\documentclass[sort&compress,preprint]{elsarticle}
\usepackage{geometry}
\usepackage{setspace}
\usepackage[absolute,overlay]{textpos}
\usepackage{amsmath,graphicx} 
\usepackage{siunitx}
\usepackage{comment}
\usepackage{braket}
\usepackage{amssymb}
\usepackage{gensymb}
\usepackage{mathtools}
\usepackage[thinc]{esdiff}
\usepackage[capitalise]{cleveref}
\usepackage{xfrac}
\usepackage[dvipsnames]{xcolor}
\usepackage[normalem]{ulem}
\usepackage{multirow}
\usepackage{leftidx}
\usepackage{mathrsfs}

\usepackage{float}
\usepackage{siunitx}

\journal{Journal of the Electrochemical Society}

\usepackage{subfloat}
\usepackage{caption}
\usepackage{subcaption}
\usepackage{graphicx}
\graphicspath{{}}

\begin{document}

\begin{frontmatter}
\onecolumn
\title{A diffuse-interface model for predicting the evolution of metallic negative electrodes and interfacial voids in solid-state batteries with homogeneous and polycrystalline solid electrolyte separators.}
\author[UF,LLNL]{Sourav Chatterjee \corref{cor1}}
\ead{chatterjee5@llnl.gov}
\author[UF]{Michael Tonks \corref{cor1}}
\ead{michael.tonks@ufl.edu}
\author[QuantumScape]{William Gardner}
\ead{WGardner@quantumscape.com}
\author[QuantumScape]{Marina Sessim}
\ead{MSessim@quantumscape.com}

\cortext[cor1]{Sourav Chatterjee}
\cortext[cor1]{Michael Tonks}

\address[UF]{Department of Materials Science and Engineering, University of Florida, Gainesville, FL 32611, USA}
\address[LLNL]{Materials Science Division, Lawrence Livermore National Laboratory, Livermore, CA 94550, USA}
\address[QuantumScape]{QuantumScape Battery, Inc.,1730 Technology Drive, San Jose, CA 95110, USA}

\begin{abstract}
This paper presents a novel diffuse-interface electrochemical model that simultaneously simulates the evolution of the metallic negative electrode and interfacial voids during the stripping and plating processes in solid-state batteries. The utility and validity of this model are demonstrated for the first time on a cell with a sodium (Na) negative electrode and a Na-$\beta^{\prime\prime}$-alumina ceramic solid electrolyte (SE) separator. Three examples are simulated. First, stripping and plating with a perfect electrode/electrolyte interface; second, stripping and plating with a single interfacial void at the electrode/electrolyte interface; third, stripping with multiple interfacial voids. Both homogeneous SE properties and polycrystalline SEs with either low or high conductivity grain boundaries (GBs) are considered for all three examples. Heterogeneous GB conductivity has no significant impact on the behavior with a perfect electrode/electrolyte interface. However, it does result in local changes to void growth due to interactions between the void edge and the GBs. The void growth rate is a linear function of the flux of Na atoms at the void edge, which in turn depends on the applied current density. We also show that the void coalescence rate increases with applied current density and can be marginally influenced by GB conductivity.
\end{abstract}


\end{frontmatter}

%%%%%%%%%%%%%%%%%%%%%%%%%%%%%%%%%%%%%%%%%%%%
\section{Introduction}
\label{sec:introduction}
Since the 1990s, lithium (Li)-ion batteries (LIBs) have been the prevalent form of energy-storage technology used in portable devices and electric vehicles \cite{takada2013progress, wang2016general}. However, conventional LIBs with a graphite negative electrode are soon expected to reach their specific energy limit ($\sim 250$ Wh/kg) \cite{janek2016solid,barchasz2012lithium, liu2019pathways}. By replacing graphite with Li metal as the negative electrode, cells with higher specific energy ($>$ $500$ Wh/Kg) and voltage could be potentially realized because Li has $\sim 10$ times higher specific capacity ($3861$ mAh/g) than graphite ($372$ mAh/g); and has one of the lowest standard reduction potentials ($-3.04$ V vs standard hydrogen electrode) \cite{liu2019pathways,feng2022review, park2017dendrite}. However, the global reserves of Li are limited, and the future cost of Li is expected to increase given the rising energy-storage demands \cite{vaalma2018cost, battistel_electrochemical_2020}. Batteries with metallic sodium (Na) negative electrodes are an attractive alternative to Li metal batteries because Na is more abundant and has a lower cost than Li and has a higher specific capacity ($1166$ mAh/g) than graphite. 

Batteries with metallic negative electrodes and ionically-conducting liquid electrolytes fail due to dendrite-induced cell shortcircuiting \cite{zhao2018solid, lee2019sodium, park2017dendrite, han2019high}. Another disadvantage of liquid electrolytes is that they are highly flammable \cite{takada2013progress}. Substituting liquid electrolytes with ionically-conducting ceramic-based solid electrolytes is a promising strategy to enable metallic negative electrodes and mitigate the fire risk associated with LIBs. Consequently, all-solid-state batteries (ASSBs) with metallic negative electrodes and ceramic-based solid-state electrolytes are considered a safer and promising alternative to LIBs \cite{lu_electrolyte_2018}.

Despite these advantages, ASSBs can fail due to void-induced dendrite formation at relatively high cycling current densities \cite{kazyak2020li, spencer2022high}, making them unsuitable for fast-charging applications. Several researchers have reported Na or Li dendrite penetration through the ceramic-based solid electrolyte separators at cycling current densities that are well below the desired current densities required for fast-charging applications \cite{ECheng2017, sharafi2017controlling,wang2019characterizing,kasemchainan2019critical, XKe2020,raj2022direct} \cite{hong2018operando} \cite{lee2019sodium}, \cite{bay2020sodium}. Recently, Kasemchainan \textit{et al.} \cite{kasemchainan2019critical} pointed out two critical current densities that are relevant in understanding dendrite formation during the cycling of ASSBs. First is the critical current density for plating (CCP), above which cell failure invariably occurs due to dendrite formation. Second is the critical current density for stripping (CCS), above which cell failure occurs due to void-induced dendrite formation. Cycling above CCS inevitably leads to void formation at the electrode/solid electrolyte separator interface. These voids increase the interfacial resistance \cite{koshikawa2018dynamic} and cause the current to get concentrated near the void edges. During subsequent charging cycle, these current density hotspots induce dendrite formation as the local current density may well exceed CCP \cite{kasemchainan2019critical}, \cite{raj2022direct} \cite{manalastas2019mechanical,devaux2015failure,krauskopf2020physicochemical}. Consequently, the CCS determines the maximum cycling current density below which ASSBs can be safely cycled without dendrite formation \cite{spencer2019sodium}. Thus, understanding factors that mitigate void formation and increase CCS can help design practical and fast-charging ASSBs. 

Several experimental groups \cite{krauskopf2019toward, kazyak2020li, lu2022void, krauskopf2019diffusion,spencer2019sodium, spencer2022high}, including Kasemchainan \textit{et al.} \cite{kasemchainan2019critical}, have suggested factors that enhance CCS in ASSBs with metallic electrodes. One is applying an increased stack pressure on the metallic negative electrode as it suppresses void growth due to creep deformation. Another factor that helps reduce void formation is using alkali metal electrodes with lower creep resistance, such as Na or potassium \cite{park2021semi}. This explains why Na-metal ASSBs typically exhibit higher CCS than Li-metal ASSBs for a given applied stack pressure and temperature \cite{spencer2022high}. For instance, Jolly \textit{et al.} \cite{spencer2019sodium} reported CCS to be around $1.5-2.5$ mA/cm$^2$ at room temperature in an ASSB with a Na negative electrode and a Na-${\beta^{\prime\prime}}$-alumina ceramic solid electrolyte separator when cycled under a stack pressure of $4-9$ MPa. Although these factors are useful indicators in mitigating void growth, a quantitative understanding of the impact of metallic electrodes and solid electrolyte properties on void evolution is still lacking.

Moreover, atomistic \cite{MYang2021,yang2021maintaining} and continuum scale \cite{agier2022void,zhao2022phase,barai2024study} models have been proposed in the literature to understand void formation and growth in ASSBs. However, these works have been applied only to Li metal batteries. For instance, Yang \textit{et al.} studied the impact of adhesion energy and pressure on void formation in a Li-metal/Li$_7$La$_3$Zr$_2$O$_{12}$ (LLZO) cell using large-scale molecular dynamics simulations. Their results suggest strong interface adhesion and high pressure mitigate void formation during stripping \cite{MYang2021}. At the continuum scale, Zhao \textit{et al.}\ \cite{zhao2022phase} used the phase-field method to simulate the evolution of micron-sized single and multiple interfacial voids in a Li/LLZO cell. Consistent with experimental observation \cite{kasemchainan2019critical}, they showed that an interfacial void grows during stripping and shrinks during plating preferentially along the electrode/solid electrolyte interface. Agier \textit{et al.}\ \cite{agier2022void} used a sharp-interface approach to show that sufficient Li flux concentration is required for void growth to occur during stripping in a Li/LLZO cell. Barai \textit{et al.}\ \cite{barai2024study} showed the impact of both bulk and surface diffusion of Li on void growth in a Li/LLZO cell. One limitation of these works is that they all assume a homogenous solid electrolyte separator. Typical solid electrolyte separators in ASSBs are polycrystalline, consisting of grain boundaries (GBs) with an ionic conductivity higher or lower than the grain interior \cite{milan2022role, dawson2018atomic,breuer2015separating,yamada2015reduced}. These GBs are believed to provide paths for dendrite propagation in ASSBs \cite{ECheng2017, tian2019interfacial, tantratian2021unraveling}. Thus, quantifying the impact of GB transport properties on interfacial void evolution is essential. 

Another aspect missing from these continuum-scale models is the shrinkage and expansion of the negative electrode that occurs during the stripping and plating of an ASSB. The thickness of the electrode changes because of the movement of the electrode/electrolyte interface at a rate proportional to the interfacial current density \cite{mishra2021perspective}. Capturing the movement of the electrode/electrolyte interface while simultaneously considering the inhomogeneities at this moving interface is numerically challenging.  To overcome this issue, Narayan and Anand \cite{narayan2020modeling} proposed a mechanics-based model that simulates the shrinkage and growth of the negative electrode using a hypothetical interphase layer between the Li metal and the LLZO solid electrolyte with a perfect interface with no voids. Alternatively, Jang \textit{et al.}\ \cite{jang2021towards,jang2022battphase} introduced a phase-field model similar to immersed-interface approaches to simulate the evolution of Li negative electrode with a liquid electrolyte by explicitly considering the moving sharp interface condition at the electrode/electrolyte interface. Since they model a liquid electrolyte, their approach assumes that the electrode/electrolyte interface remains perfect during stripping and plating.
 
This work presents a novel phase-field model that can simultaneously simulate the evolution of metallic negative electrodes and interfacial voids in an ASSB. Furthermore, we consider homogenous and polycrystalline solid electrolytes with low-conductive and high-conductive GBs. Although phase-field models considering solid-electrolyte GB properties exist in the literature \cite{tian2019interfacial}, these models neither consider the evolution of interfacial voids nor that of the negative electrode. Moreover, contrary to previous works, we apply our model to an all-solid-state cell with a Na negative electrode and Na-$\beta^{\prime\prime}$-alumina ceramic solid electrolyte separator. To the best of our knowledge, stripping and plating of a Na metal solid-state cell has never been simulated. Our goal in this work is twofold. First, to model the shrinkage and growth of the metallic negative electrode during stripping and plating under an applied current density, assuming a perfect electrode/electrolyte interface. Second, to model the evolution of interfacial voids while simultaneously considering the volume change occurring in the negative electrode. In addition, we report on the impact of solid electrolyte GBs on the evolution of the metallic negative electrode and interfacial voids. 

This paper is organized as follows. In Section \ref{sec:formulation}, we formulate a multi-phase-field model and provide the governing equations with the interfacial and boundary conditions.  In Section \ref{numerical_method}, we parameterize the model to simulate a cell with a Na negative electrode and a Na-$\beta^{\prime\prime}$-alumina solid electrolyte separator. In Section \ref{sec:results}, we use three test cases in each subsection to demonstrate the utility and validity of our model. These subsections are subdivided into a homogeneous solid electrolyte case and polycrystalline solid electrolyte cases with low-conductive and high-conductive GBs. In Section \ref{sec:conclusions}, we finish with some concluding remarks and indicate possible future directions.

\section{Formulation}
\label{sec:formulation}
We aim to model two processes occurring in a solid-state cell with a metallic negative electrode. First, we want to simulate the shrinkage and growth of the metallic negative electrode during stripping and plating, assuming a perfect contact between the electrode and the solid electrolyte separator. Second, we want to introduce initial voids at the electrode/electrolyte interface and simulate their evolution during stripping and plating while concurrently simulating the evolution of the metallic negative electrode. We assume that void nucleation has already occurred and focus only on void growth and shrinkage. We then wish to use this model to quantify the impact of applied current density and solid-electrolyte grain boundary diffusivity/ionic conductivity on (i) deposition and depletion rates and (ii) void migration and coalescence rates during electroplating and stripping.

\begin{figure}[tbph]
\centering
\begin{subfigure}{0.6\textwidth}
	\includegraphics[trim=0 0 0 0, clip, keepaspectratio,width=\linewidth]{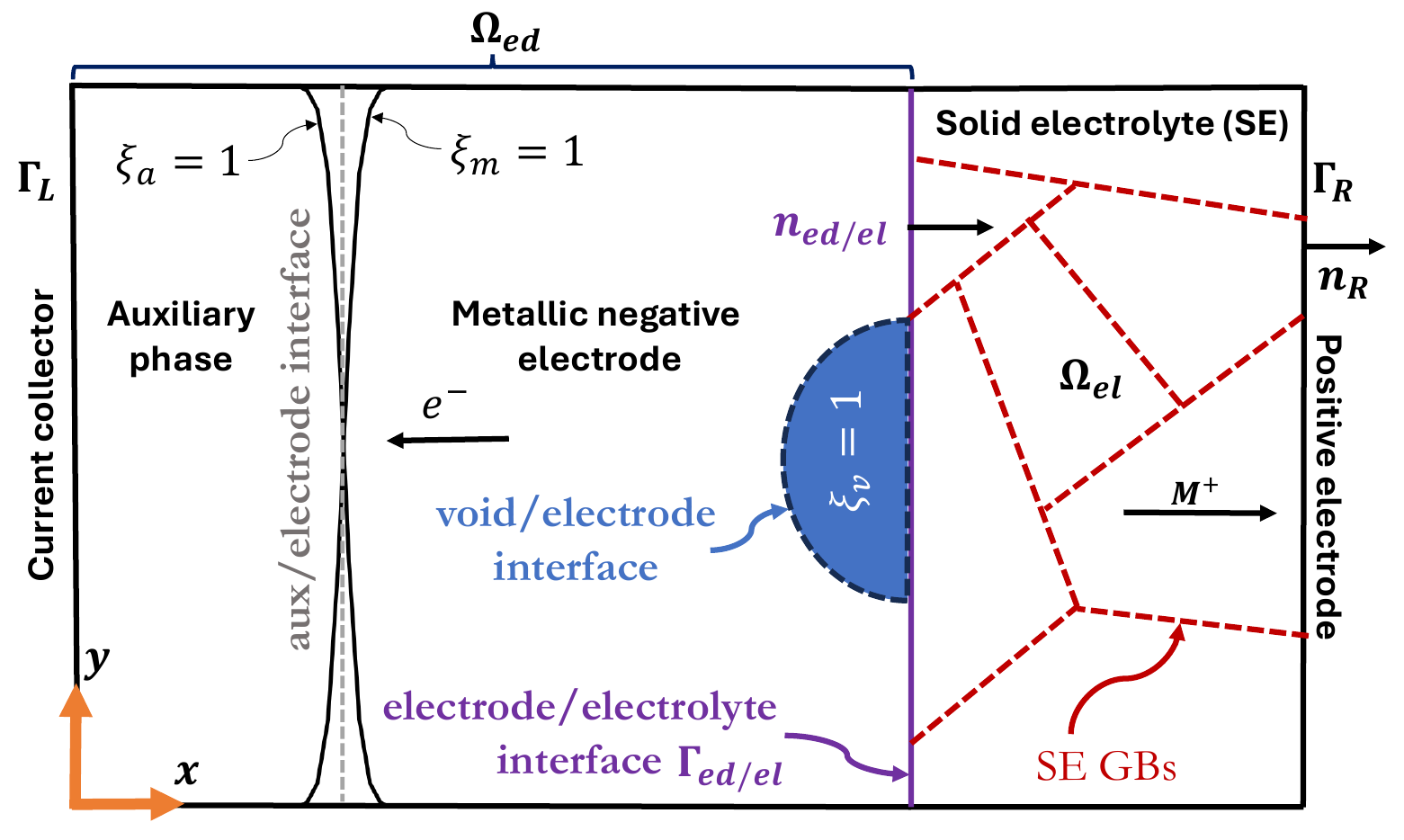}
	\subcaption{}
	\label{Fig0a}
\end{subfigure}
\begin{subfigure}{0.35\textwidth}
	\includegraphics[trim=0 0 0 0, clip, keepaspectratio,width=\linewidth]{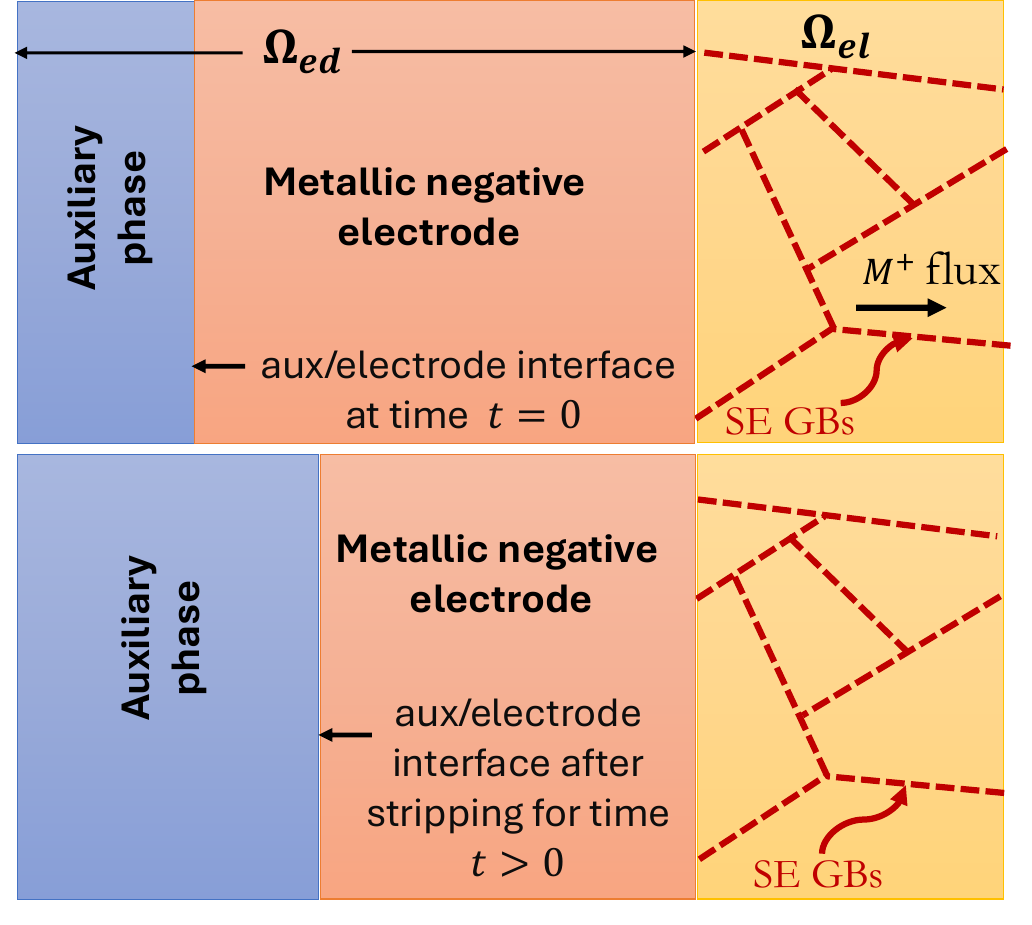}
	\subcaption{}
	\label{Fig0b}
\end{subfigure}
\caption{Schematics demonstrating the functioning of our model, where (a) shows how the domain is subdivided into two subregions: $\Omega_{ed}$ containing the negative electrode and $\Omega_{el}$ representing the solid electrolyte separator. The grey and blue dotted lines mark the diffuse aux/electrode and void/electrode interfaces, respectively. The sharp electrode/electrolyte interface is marked with a solid purple line. (b) shows the shrinkage of the negative electrode during stripping due to the movement of the aux/electrode interface under an applied flux and current.}
\label{Fig0}
\end{figure}

In this work, we use the phase-field approach to simulate the shrinkage and growth of the metallic negative electrode during stripping or plating since it does not require explicitly tracking the moving interface. Specifically, we introduce an ``auxiliary" phase into our model representing the empty space that is sandwiched between the metallic negative electrode and the current collector using a spatially-varying phase-field variable field, $\xi_a$ (Fig.~\ref{Fig0a}). Note that we do not explicitly represent the current collector for simplicity. Likewise, the negative electrode is modeled using a variable field, $\xi_m$. These variables equal $1$ in their corresponding phases and $0$ elsewhere. They smoothly vary across a submicron-thick diffuse aux/electrode interface region, loosely indicated by a grey dotted line in Fig.~\ref{Fig0a}.
On the other hand, the solid electrolyte does not change and the electrode/electrolyte interface is modeled as stationary and sharp, as indicated by a solid purple line in Fig.~\ref{Fig0a}. Consequently, the negative electrode grows or shrinks at the cost of the auxiliary phase due to the movement of this diffuse aux/electrode interface instead of the electrode/electrolyte interface, as is done in electrochemical models with liquid electrolytes \cite{mishra2021perspective,jang2021towards,jang2022battphase}. This is shown schematically in Fig.~\ref{Fig0b}.

Moreover, the interfacial voids are modelled using a variable field, $\xi_v$, that equals $1$ in a void and $0$ elsewhere. A dotted blue line indicates the diffuse void/electrode interface in Fig.~\ref{Fig0a}. Note the voids are purposefully placed at the electrode/electrolyte interface, as shown by experiments \cite{kasemchainan2019critical}, \cite{spencer2019sodium}. Thus, these voids act as barriers to the migration of metal ions across the electrode/electrolyte interface during stripping or plating. Our approach to model voids resembles that of Zhao \textit{et al.} \cite{zhao2022phase}.
Nevertheless, unlike their approach, our model can capture both void migration and the evolution of negative electrodes simultaneously during stripping or plating without explicitly tracking the electrode/electrolyte interface. Further, unlike \cite{zhao2022phase}, we also consider the impact of solid electrolyte grain boundaries on void evolution. We employ the phase-field approach to model these grain boundaries, as indicated by dotted red lines in Fig.~\ref{Fig0a}. We assume these boundaries to be static since we simulate stripping and plating near room temperature ($300$ K).

We divide our simulation domain into two subregions, as shown in Fig.~\ref{Fig0}. One region, referred to as $\Omega_{ed}$, contains the negative electrode and auxiliary phases, which we simulate with or without void(s). The other region corresponds to the solid electrolyte separator, referred to as $\Omega_{el}$. The static electrode/electrolyte interface lies between these two subregions and is indicated as $\Gamma_{ed/el}$ in Fig.~\ref{Fig0a}. The following subsections provide the electrode/electrolyte interfacial conditions and the governing equations in the two subregions. The governing equations are derived for a generic pure alkali metallic negative electrode. We hereafter refer to the diffusing metal atoms by the letter $M$. During stripping, the metal atom oxidises at the electrode/electrolyte interface as $M (ed) \rightarrow M^{^+} (el) + e^{-} (ed)$. Subsequently, the metal cation $M^{^+}$ diffuses through the solid electrolyte separator domain into the positive electrode. These cations migrate in the opposite direction during plating and reduce at the electrode/electrolyte interface. For simplicity, we use a half-cell model and do not mesh the positive electrode. 
\subsection{Negative electrode and auxiliary region with a void}
\label{subsection:electrode}
We must first define an energy functional to derive the evolution equations for the phase-field variables representing the metallic negative electrode, auxiliary phase and void. This work follows the grand-potential-based phase-field formulations \cite{plapp2011unified,Choudhary2012} and is similar to a recent electrochemical molten salt corrosion model \cite{bhave2023electrochemical}. This formulation assumes that the phases have equal diffusion potential in the diffuse interfacial regions but have different phase compositions, similar to the Kim-Kim-Suzuki model \cite{kim1999phase}. This assumption is important because it separates the interfacial and bulk driving forces required to model microstructures at the micrometer length scales \cite{plapp2011unified}. Specifically, the grand-potential functional, $\Psi^{ed}$, is defined as 
\begin{align}
\begin{split}
\Psi^{ed}= \int_{\Omega_{ed}}\left[\sum_{\substack{p=\\{a,m,v}}}\frac{\alpha}{2}\left(\nabla \xi_{p} \right)^2
+m\left\{\sum_{\substack{p=\\{a.m,v}}}\left( \frac{\xi_{p}^{4}}{4} - \frac{\xi_{p}^{2}}{2} \right) + 
1.5\left(\xi_{a}^{2}\xi_{m}^{2} + \xi_{a}^{2}\xi_{v}^{2} + \xi_{m}^{2}\xi_{v}^2\right) 
+ \frac{1}{4}\right\}
+ \omega_{bulk}^{ed} 
\right]dV\\
\quad  \text{in}  \quad \Omega_{ed},
\label{Eqn1}
\end{split}
\end{align}
where $\alpha$ is the gradient-energy coefficient, $m$ is the barrier height, and $\omega^{ed}_{bulk}$ is the bulk grand potential density. $\alpha$ and $m$ are calculated using the interfacial energy $\gamma$ and interface width $\mathit{l}_w$ according to the relations \cite{moelans2011}: $\alpha=(3/4)\gamma \mathit{l}_w$ and $m = 6.0\left(\gamma/\mathit{l}_w\right)$. The bulk grand potential density is defined as:
\begin{align}
\omega^{ed}_{bulk} = \sum_{\substack{p=\\{a.m,v}}}h^{ed}_{p}\omega_{p} = h_{a}^{ed}\omega_{a} + h_{m}^{ed}\omega_{m} + h_{v}^{ed}\omega_{v} \quad  \text{in}  \quad \Omega_{ed},
\label{Eqn2}
\end{align}
where $\omega_{a}$, $\omega_{m}$ and $\omega_{v}$ are the bulk grand potential densities of the auxiliary, negative electrode and void phases, respectively. The bulk grand potential density of a given phase is defined as \cite{plapp2011unified}:
\begin{align}
\omega_{p} (\mu_{M}^{ed}) = f_{p}(\mu_{M}^{ed}) -  \mu^{ed}_{M}\frac{c_{M}^{p}(\mu_{M}^{ed})}{v_{m}}, \quad p = a,m,v \quad  \text{in}  \quad \Omega_{ed},
\label{Eqn3}
\end{align}
where $v_{m}$ is the molar volume of the metal, $\mu_{M}^{ed}(\boldsymbol{x},t)$ is the diffusion potential of the metal $M$ in the region $\Omega_{ed}$, $f_{p=a,m,v}(\mu_{M}^{ed})$ and $c_M^{p=a,m,v}(\mu_{M}^{ed})$ are the free energy densities and phase mole fractions of the auxiliary phase, metallic negative electrode, and the void, respectively. Note that these properties are functions of the metal diffusion potential because it is the independent variable in a grand-potential-based model \cite{plapp2011unified}.  $h_{p}^{ed}$ is the interpolation function, which is defined as follows \cite{moelans2011}:
\begin{align}
h_{p}^{ed} = \frac{\xi_{p}^{2}}{\left[ \xi_{a}^{2} + \xi_{m}^{2} + \xi_{v}^{2} \right]},  \quad p = a, m, v \quad  \text{in}  \quad \Omega_{ed}
\label{Eqn4}
\end{align}
By taking the variational derivative of Eq. \eqref{Eqn1} and using the Allen-Cahn equation \cite{allen1979microscopic}, it follows that the phase-field variables $\xi_{p=a,m,v}$ evolve according to:
\begin{align}
\begin{split}
\frac{\partial \xi_{p}}{\partial t}  + L_{\phi}\left[m\left\{\xi_{p}^{3} - \xi_{p}  + 3\xi_{p}\sum_{\substack{q=a,m,v \\ q\neq p}} \xi_{q}^{2}\right\} 
- \alpha \nabla^2 \xi_{p} + 
\sum_{\substack{q=a,m,v}}\frac{\partial h^{ed}_{q}}{\partial \xi_{p}}\omega_{q}\right] =0 \quad p = a,m,v \quad \text{in}\quad  \Omega_{ed},
\label{Eqn5}
\end{split}
\end{align}
where $L_{\phi}$ denotes the phase-field kinetic coefficient assumed to be a constant and equal for all phases for simplicity.

We assume that the flux of metal atoms in the region $\Omega_{ed}$, i.e., $\boldsymbol{j}_{M}^{ed}$, is directly proportional to the gradient in the diffusion potential of the metal atoms. This yields:
\begin{align}
\begin{split}
\boldsymbol{j}_{M}^{ed} = -\left[\frac{\left(h_{a}^{ed}{M}_{a} + h_{m}^{ed}M_{m} + h_{v}^{ed}{M}_{v} \right)}{v_{m}}\nabla \mu_{M}^{ed}\right]  \quad  \text{in} \quad \Omega_{ed},
\label{Eqn6}
\end{split}
\end{align}
where ${M}_{a}$, ${M}_{m}$ and ${M}_{v}$ denote the mobilities of the metal in the auxiliary phase, negative electrode and void, respectively. Further, these mobilities are related to the diffusivities using the relation $M_{p} = D_{M}^{p}dc_{M}^{p}/d\mu_{M}^{ed}$, where $dc_{M}^{p}/d\mu_{M}^{ed}$ is the inverse of the thermodynamic factor of phase $p$. It should be noted that Eq.~\eqref{Eqn6} in the bulk of the negative electrode, i.e., $\xi_m=1$, reduces to $\left.\boldsymbol{j}_{M}^{ed}\right|_{\xi_{m}=1} = -\left(D_{M}^{p}/v_{m}\right)\left.\nabla c_{M}^{ed}\right|_{\xi_{m=1}}$ (see Appendix \ref{AppendixA1}), that is the flux of metal species is directly proportional to its concentration gradient. This shows that our flux relation is identical to other approaches if metal diffusion inside the electrode is considered (see, for instance, \cite{fabre2011charge, tian2017simulation,schmidt2023three}). Moreover, the flux in Eq. \eqref{Eqn6} must obey the mass conservation equation, which reads:
\begin{align}
\frac{1}{v_{m}}\frac{\partial c_{M}^{ed}}{\partial t} +\nabla \cdot \boldsymbol{j}_{M}^{ed}= 0 \quad  \text{in}  \quad \Omega_{ed}.
\label{Eqn7}
\end{align}
We, however, do not directly solve for Eq. \eqref{Eqn7}.  Several grand-potential-based phase-field formalisms \cite{plapp2011unified, Cogswell2015, hong2018phase, jeon2022phase} have formulated an evolution equation for diffusion potential from Eq. \eqref{Eqn7}. In this work, we take a similar approach. Thus, Eq. \eqref{Eqn7} may be equivalently written as follows (see Appendix \ref{AppendixA2} for the derivation): 
\begin{align}
\left[\sum_{\substack{p=\\a,m,v}}\frac{dc_{M}^{p}}{d \mu_{M}^{ed}}\frac{h_{p}^{ed}}{v_{m}}\right]\frac{\partial \mu^{ed}_{M}}{\partial t}
+\nabla \boldsymbol{j}_{M}^{ed}
+\sum_{\substack{p=\\a,m,v}}\left(\sum_{q=a,m,v}\frac{\partial h_{p}}{\partial \xi_{q}}\frac{\partial \xi_{q}}{\partial t}\right)  \frac{c_{M}^{p}}{v_{m}}
= 0
\quad  \text{in}  \quad \Omega_{ed},
\label{Eqn8}
\end{align}
Thus, we obtain the diffusion potential of metal in the region $\Omega_{ed}$ by solving Eq. \eqref{Eqn8} in combination with the flux relation Eq. \eqref{Eqn6}.

Moreover, we assume that electrons are the sole charge-carrying species in the auxiliary phase, negative electrode, and void region. Thus, Ohm's law gives the electric current density in the region $\Omega_{ed}$, i.e., $\boldsymbol{i}^{ed}$:
\begin{align}
\boldsymbol{i}^{ed}  = -\left(\sigma_{a}h_{a}^{ed}  + \sigma_{m}h_{m}^{ed} + \sigma_{v}h_{v}^{ed}\right)\nabla \phi^{ed} \quad  \text{in}  \quad \Omega_{ed},
\label{Eqn9}
\end{align}
where $\sigma_{a}$,  $\sigma_{m}$ and $\sigma_{v}$ are the electronic conductivities of the auxiliary phase, metallic negative electrode, and void phases, respectively, and $\phi^{ed}$ is the electric potential in the region $\Omega_{ed}$. Since charge must be conserved, the electric potential is obtained by solving the charge conservation equation assuming that there are no free charges in the region $\Omega_{ed}$, which yields \cite{schmidt2023three}:
\begin{align}
\nabla \cdot \boldsymbol{i}^{ed} = 0 \quad  \text{in}  \quad \Omega_{ed}.
\label{Eqn10}
\end{align}
\subsection{Solid electrolyte separator region}
We solve only for the diffusion of metal ions and the electric potential inside the solid electrolyte separator. As indicated before, we use the phase-field approach to investigate the role of grain boundaries in the transport of metal ions through the solid electrolyte. Thus, we represent each grain with a single-order parameter, which assumes a value of $1$ inside the grain and $0$ elsewhere. We denote this variable as $\xi_{gi}^{el}$, where the subscript $i$ indicates a unique grain. We also assume these grain boundaries to be stationary, i.e., $\partial \xi_{gi}^{el}/\partial t=0$ for all $i$, since we aim to study stripping and plating near room temperature. 

We assume that the flux of metal ions in the solid electrolyte separator depends on diffusion and migration. This yields:
\begin{align}
\boldsymbol{j}_{M^{z^+}}^{el} = -\left\{\frac{M^{g,el}_{M^{z^+}} (h^{el}_{g} - h^{el}_{gb})  + M^{gb,el}_{M^{z^+}}h^{el}_{gb}}{v_{m}}\right\}\left(\nabla \mu^{el}_{M^{z^+}} + zF\nabla \phi^{el}\right) \quad  \text{in}  \quad \Omega_{el},
\label{Eqn11}
\end{align}
where $\mu^{el}_{M^{z^+}}$ is the diffusion potential of metal cations $M_{z^{+}}$ in the solid electrolyte, $\phi^{el}$ is the electric potential in the solid electrolyte, $z$ denotes the charge on the metal cation, $F$ is the Faraday constant, $M^{g,el}_{M^{z^+}}$ and $M^{gb,el}_{M^{z^+}}$ respectively denote the mobilities of metal cation in the solid electrolyte grains and grain boundaries, and $h^{el}_{g}$ and $h^{el}_{gb}$, respectively, denote the interpolation functions representing the solid electrolyte grains and grain boundaries. These interpolation functions depend on the order parameters, $\xi_{gi}$, representing the grains and grain boundaries. Specifically, we assume that these functions take the following form \cite{muntaha2023impact}:
\begin{align}
h_{g}^{el} = \frac{\left(\xi_{gi}^{el}\right)^2}{\sum_{i=1}^{N_{g}}\left(\xi_{gi}^{el}\right)^{2}}, \quad h_{gb}^{el} = 16\sum_{i=1}^{N_{g}}\sum_{\substack{j>i}}^{N_{g}} \left(\xi_{gi}^{el}\right)^{2}\left(\xi_{gj}^{el}\right)^{2} \quad  \text{in}  \quad \Omega_{el},
\label{Eqn12}
\end{align}
where $N_{g}$ indicates the number of solid electrolyte grains. It should be noted that $h_{g}^{el}=1$ only inside the solid electrolyte grains, whereas $h_{gb}^{el}$ is non-zero within the diffuse solid-electrolyte grain boundaries.

Further, we assume that the composition of the metal cations in the solid electrolyte $c_{M^{z^+}}^{el}$ can be analytically expressed as a function of metal cation diffusion potential $\mu_{M^{z^+}}^{el}$, i.e., $c_{M^{z^+}}^{el} = \hat{c}^{el}\left(\mu_{M^{z^+}}^{el}\right)$. This is generally true provided the form of the solid electrolyte-free energy is parabolic, dilute solution, or ideal solution \cite{plapp2011unified}. Consequently, the composition gradient may be expressed as $\nabla c_{M^{z^+}}^{el} =\partial \hat{c}^{el}/{\partial \mu_{M^{z^+}}}\nabla \mu_{M^{z^+}}$, where $\partial \hat{c}^{el}/{\partial \mu_{M^{z^+}}}$ is the inverse of the thermodynamic factor of the solid electrolyte. This factor relates the mobilities in Eq. \eqref{Eqn11} to the solid electrolyte grain and grain boundary diffusivities by the relation $L_{M^{z^+}}^{p,el} = D_{M^{z^+}}^{p,el}d\hat{c}^{el}/d\mu_{M^{z^+}}^{el}$ \cite{balluffi2005kinetics},  where $p=\{g,gb\}$; it thus follows that Eq. \eqref{Eqn11} may equivalently be written as $\boldsymbol{j}^{M^{z^+}}_{el} = -\left[D^{g,el}_{M^{z^+}} (h^{el}_{g} - h^{el}_{gb})  + D^{gb,el}_{M^{z^+}}h^{el}_{gb}\right]/v_{m}\left(\nabla c^{el}_{M^{z^+}} + zF\nabla \phi^{el}\right)$, which is the Nernst-Planck equation \cite{fabre2011charge, tian2017simulation,schmidt2023three}.

The flux of metal cations must satisfy the conservation of mass equation. This yields:
\begin{align}
\frac{1}{v_{m}}\frac{\partial c_{M^{z^+}}^{el}}{\partial t} + \nabla \boldsymbol{j}^{M^{z^+}}_{el} =0 \quad  \text{in}  \quad \Omega_{el}.
\label{Eqn13}
\end{align}
Similar to Eq. \eqref{Eqn8}, we rewrite Eq. \eqref{Eqn13} as a diffusion potential evolution equation by taking the time derivative of the constitutive relation between composition and diffusion potential, i.e.,  $c_{M^{z^+}}^{el} = \hat{c}^{el}\left(\mu_{M^{z^+}}^{el}\right)$. This yields:
\begin{align}
\frac{1}{v_{m}}\frac{\partial \hat{c}}{\partial \mu_{M^{z^+}}}\frac{\partial \mu_{M^{z^+}}}{\partial t} + \nabla \boldsymbol{j}^{M^{z^+}}_{el} = 0 \quad  \text{in}  \quad \Omega_{el}.
\label{Eqn14}
\end{align}
Thus, we obtain the diffusion potential of metal cations in the solid electrolyte domain by solving Eq. \eqref{Eqn14} in combination with the flux relation from Eq.~\eqref{Eqn11}. Once the diffusion potential is known, the constitutive relation $c_{M^{z^+}}^{el} = \hat{c}^{el}\left(\mu_{M^{z^+}}^{el}\right)$ gives the composition of metal cations in the solid electrolyte separator.

The electric potential in the solid electrolyte $\phi^{el}$ is obtained by solving the charge conservation equation assuming no free charges, similar to Eq. \eqref{Eqn10}. This yields:
\begin{align}
\nabla \cdot \boldsymbol{i}^{el} = 0 \quad  \text{in}  \quad \Omega_{el},
\label{Eqn15}
\end{align}
where $\boldsymbol{i}^{el}$ is the ionic current density in the solid electrolyte. We further assume $\boldsymbol{i}^{el}$ is given by Ohm's law according to: 
\begin{align}
\boldsymbol{i}^{el}= -\left[\kappa_{g}^{el}(h_{g}^{el} -h_{gb}^{el}) + \kappa_{gb}^{el}h_{gb}^{el}\right] \nabla \phi^{el} \quad  \text{in}  \quad \Omega_{el},
\label{Eqn16}
\end{align}
 where $\kappa_{g}^{el}$ and $\kappa_{gb}^{el}$ are the ionic conductivities of the grain and grain boundary, respectively. This relation also implies that we assume that no concentration gradient of the metal cations exists in the solid electrolyte, which is reasonable because the solid electrolyte is a single-ion conductor. 

\subsection{Interfacial and boundary conditions}
At the electrode/electrolyte interface, the mass flux of metal species on the electrode side equals the mass flux of metal cations on the solid electrolyte side. This ensures mass conservation at the interface, which yields
\begin{align}
\left.\boldsymbol{j}_{M}^{ed}\right|_{\Gamma_{ed/el}}\boldsymbol{n}_{ed/el} = \left.\boldsymbol{j}_{M^{z^+}}^{el}\right|_{\Gamma_{ed/el}}\boldsymbol{n}_{ed/el},
\label{Eqn17}
\end{align}
where $\boldsymbol{n}_{ed/el}$ is the unit normal at the electrode/electrolyte interface that points from the electrode to the electrolyte (Fig.~\ref{Fig0a}). In addition, we assume that the diffusion potential of the metal species equals the diffusion potential of the metal cation at the electrode/electrolyte interface, which implies the system is at a local chemical equilibrium at this interface. Specifically,
\begin{align}
 \left.\mu_{M}^{ed}\right|_{\Gamma_{ed/el}} = \left.\mu^{el}_{M^{z^{+}}}\right|_{\Gamma_{ed/el}}.
\label{Eqn18}
\end{align}
Likewise, we assume the current density and electric potential are continuous at the electrode/electrolyte interface, similar to Zhao \textit{et al.} \cite{zhao2022phase}. This yields the following equations:
\begin{align}
\left.\boldsymbol{i}^{ed}\right|_{\Gamma_{ed/el}}\boldsymbol{n}_{ed/el} &= \left.\boldsymbol{i}^{el}\right|_{\Gamma_{ed/el}}\boldsymbol{n}_{ed/el}, \label{Eqn19}\\
\left.\phi^{ed}\right|_{\Gamma_{ed/el}} &=  \left.\phi^{el}\right|_{\Gamma_{ed/el}}.
\label{Eqn20}
\end{align}
Equation \eqref{Eqn20} means that our model does not consider charge transfer or activation overpotential at the electrode/electrolyte interface. 
Further, at the current collector/electrode interface on the leftmost boundary $\Gamma_{L}$ (Fig.~\ref{Fig0a}), we assume the electric potential is zero. This yields: 
\begin{align}
\left.\phi^{ed}\right|_{\Gamma_{L}} = 0.  \label{Eqn21}
\end{align}
At the rightmost boundary  $\Gamma_{R}$ (Fig.~\ref{Fig0a}), we apply natural boundary conditions on the electric potential and the diffusion potential variables defined in the region $\Omega_{el}$, i.e., $\phi^{el}$ and $\mu_{M^{z^+}}^{el}$, respectively. Specifically, we simulate galvanostatic charging/discharging conditions by applying a current density $i_{app}$. During stripping, this current leaves through the rightmost solid-electrolyte boundary into the positive electrode, whereas it moves into the solid electrolyte from the positive electrode during plating.
The applied flux is related to the applied current density by Faraday's law. This yields:
\begin{align}
\left.\boldsymbol{i}^{el}\right|_{\Gamma_{R}} \boldsymbol{n}_{R}&= i_{app}, \label{Eqn22}\\
\left.\boldsymbol{j}_{M^{z^+}}^{el}\right|_{\Gamma_{R}}\boldsymbol{n}_{R} &= \frac{i_{app}}{zF},
\label{Eqn23}
\end{align}
where $\boldsymbol{n}_{R}$ is the unit normal at the rightmost solid electrolyte separator boundary. Except for these non-homogenous natural boundary conditions, we assume zero flux boundary conditions at all external surfaces for all variables.

\section{Numerical method and model parameters}
\label{numerical_method}
Equations \eqref{Eqn5}, \eqref{Eqn8}, \eqref{Eqn10}, \eqref{Eqn13} and \eqref{Eqn16} are the governing equations of the model in the two subregions $\Omega^{ed}$ and $\Omega^{el}$. The variables in the two subregions are connected through the interfacial conditions at the electrode/electrolyte interface, i.e., Eqs. \eqref{Eqn17}, \eqref{Eqn18}, \eqref{Eqn19}, \& \eqref{Eqn20}. We have implemented these equations in the Multiphysics Object-Oriented Simulation Environment (MOOSE) framework, an open-source finite-element-based software \cite{giudicelli2024moose}. We use the phase-field module in MOOSE \cite{tonks2012object,schwen2017rapid} to implement the governing equations. We apply a Newton solver with a LU preconditioner to solve these equations implicitly in MOOSE. An adaptive time-stepping scheme is employed to reduce the computation time. The relative and absolute tolerances for convergence are set at $10^{-8}$ and $10^{-9}$, respectively. For numerical convergence reasons, the equations are non-dimensionalized. The dimensionless forms of these equations are provided in Appendix \ref{nondimensional_form}. 

To illustrate the utility of our model and verify its implementation, we parameterize the model for a pure metallic Na negative electrode and a Na-$\beta^{\prime\prime}$-alumina (NBA) ceramic solid electrolyte separator. The properties and parameters we use in the model are shown in Table \ref{tab:my_table} and we discuss them, below.

We first start from the chemical grand potential densities defined in Eq. \eqref{Eqn3}.  For simplicity, we assume that the chemical free energies of the auxiliary phase, Na negative electrode, and void have parabolic forms, i.e., $f_{p} = A^{p}/2(c_{Na}^{p}- c_{Na}^{eq,p})^2$ for $p=a,m,v$. Using this expression, the chemical grand-potential densities in Eq. \eqref{Eqn3} may be expressed as: $\omega_{p} = -\left[\left(\mu_{Na}^{ed}\right)^2/\left(v_{m}^2A^{p}\right) + \left(\mu_{Na}^{ed}c_{Na}^{eq,p}\right)/v_{m}\right]$ (see Appendix \ref{AppendixA3}). A similar expression is required for the electrolyte phase since we need a constitutive relation between $c_{Na^+}^{el}$ and $\mu_{Na^+}^{el}$, as previously suggested. For simplicity we also assume that the free energy of the electrolyte phase is also parabolic, i.e., $f_{el} = (A^{el}/2)(c_{Na^{^{+}}}^{el} - c_{Na^{^{+}}}^{eq,el})^{2}$. Consequently, the constitutive relation in the electrolyte is $c_{Na^{^{+}}}^{el} =  \mu_{Na^{^{+}}}^{el}/\left(v_{m}A^{el}\right) + c_{Na^{^{+}}}^{eq,el}$. For each phase in $\Omega^{ed}$, two parameters are required to calculate these grand potential densities: i) the equilibrium mole fractions of Na, $c^{eq,p}_{Na}$, and ii) the parabolic coefficient $A^{p}$. Likewise, to construct the constitutive relation for the electrolyte, we need the equilibrium mole fractions of $Na^+$ $c^{eq,el}_{Na^+}$ and the parabolic coefficient $A^{el}$. Note that dilute or ideal solution-free energy expressions could also be used to model these phases. However, employing logarithmic functions for phases with dilute concentrations leads to numerical convergence issues \cite{hu2007thermodynamic}, and therefore we use parabolic expressions.

We assume that the Na concentrations in the auxiliary phase and void are negligible compared to the electrode. Specifically, the equilibrium mole fractions of Na in the auxiliary phase and void $c_{Na}^{eq,a/v}$ are both chosen to be $10^{-8}$. The equilibrium mole fraction of Na in the metal phase $c_{Na}^{eq,m}$ is obtained from the equilibrium vacancy concentration $c_{v}^{eq}$ using $c_{Na}^{eq,m} = 1-c_{v}^{eq}$. The equilibrium concentration of $Na^+$ in the solid electrolyte $c_{Na^{+}}^{eq,el}$ is selected as $0.0563$ based on the chemical formula of NBA, Na$_{1.67}$Mg$_{0.67}$Al$_{10.33}$O$_{17}$ \cite{heinz2021grain}. Moreover, since there is no experimental data to fit the parabolic free energies, the parabolic coefficients $A^p$ of the auxiliary phase, negative electrode, and void are chosen to be high enough to ensure Na concentrations inside these phases do not deviate significantly from their equilibrium values during stripping or plating. The parabolic coefficient for the electrolyte phase $A^{el}$ is selected in the same manner. This assumption is similar to modelling stoichiometric compound phases in phase-field models \cite{hu2007thermodynamic, kellner2022modeling}. 

The 0 V boundary conditions should be applied at the auxiliary/electrode interface. However, applying a boundary condition at the moving diffuse interface requires changes to the model formulation \cite{yu2012extended}. To simplify the model, we apply the 0 V boundary condition at the left boundary and assume that the electronic conductivity in the auxiliary phase equals the Na negative electrode, i.e., $\sigma_{a}=\sigma_{m}$. The electronic conductivity of the void phase is taken to be nearly $10^{-14}$ times smaller than that of the Na negative electrode. The ionic conductivities of the NBA solid electrolyte grains and grain boundaries are calculated from an Arrhenius-type equation \cite{heinz2021grain}, i.e., $\kappa_{p}^{el} = (\mathcal{K}_{p}/T)e^{-E_{p}/k_{B}T}$, where $p=\{g,gb\}$. It has been found that the ionic conductivity of NBA grain boundaries is usually lower than that in the grain interior \cite{heinz2021grain}. Nevertheless, it has been reported that the grain boundary conductivity of ceramic solid electrolytes may be enhanced by surface modification \cite{yamada2015reduced, wang2022optimizing}. Thus, we also consider a scenario where the grain boundary conductivity is $10$ times higher than in the grain interior. Moreover, for simplicity, the Na diffusivity in the auxiliary phase is assumed to be equal to that in the negative electrode, i.e., $D_{Na}^{a}= D_{Na}^{m}$. In the void regions, the Na diffusivity is set to 0 (Table \ref{tab:my_table}). Lastly, the diffusivities of Na$^+$ in the NBA grains and grain boundaries are obtained from the ionic conductivities using the Nerst-Einstein relation, i.e., $D_{p} = \kappa_{p}RT/(c^{eq, el}_{Na^{^{+}}}F^2 )$, where $p=\{g,gb\}$.

The aux/electrode and void/electrode interfaces have a fixed width $l_{w}$ of $0.5$ $\mu$m. Note that this is also the assumed grain boundary width of the static solid electrolyte grain boundaries. The assumed width is significantly smaller than the mean solid electrolyte grain size, yet almost $10^3$ times higher than the physical interface width. This results in significantly faster GB transport in the model than would occur in the real system. To compensate for the wide GBs, the GB conductivity can be reduced, as shown in Ref. \cite{muntaha2023impact}. 

We model the behavior in 2D to reduce the computational cost, though the actual void and GB structures are 3D in a real material. The grid spacing $\Delta x$ is one-third of the interface width to resolve the interface regions accurately. The simulation domain is $50$ $\mu$m thick, while the solid electrolyte separator is $18.5$ $\mu$m thick along the x-axis in all the simulations. The domain is $80$ $\mu$m wide along the y-axis. The simulations are performed on a uniform finite element mesh generated within MOOSE. The 2D mesh is discretized using four-noded quadrilateral elements. The solid electrolyte microstructure is synthetically generated in MOOSE using a Voronoi tessellation and is shown in Fig.~\ref{Fig_SI_secS5} of the Supporting Information.

\begin{table}[tbp]
\caption{Model and numerical parameters used to perform the simulations.}  
  \centering
    \begin{tabular}{lccc}
        \hline
        Parameter & Value & Units & Refs.\\
        \hline
        \multicolumn{4}{l}{\textbf{Constants}}\\
        $T$ &300 & K & ---\\
        $R$ &8.314& J/(mol K) & ---\\
        $F$& 96487& C/mol &---\\
        $z$ &1& --- &---\\ 
        $k_B$ & $8.62\times 10^{-5}$  &  eV/K &---\\
        $v_{m}$& 23.78 &cm$^3$/mol&---\\
        & & &\\ % Just a gap
        \multicolumn{4}{l}{\textbf{Interfacial properties}}\\
        $\gamma$ & 0.22 & J/m$^2$ & \cite{jain2013commentary}\\
        $\mathit{l}_w$ &0.5& $\mu m$&---\\
        $\alpha$& $(3/4)\gamma \mathit{l}_w$&J/m&\cite{moelans2011}\\
        $m$ & $6.0\left(\gamma/\mathit{l}_w\right)$&J/m$^3$&\cite{moelans2011}\\
        $L_{\phi}$  &$10^{-4}$& m$^3$/Js & ---\\
        %%%%
        & & &\\ % Just a gap
        \multicolumn{4}{l}{\textbf{Auxiliary-Na-electrode-void}}\\
         %Thermodynamic properties
         $c_v^{eq}$ & $e^{-2.0}e^{-0.157/k_{B}T}$ & & \cite{sullivan1964measurement}\\
         $c_{Na}^{eq,a}, c_{Na}^{eq,m}, c_{Na}^{eq,v}$& $10^{-8}, 1-c_v^{eq}, 10^{-8}$ & ---&---\\
        $A^{a}/f_c, A^{m}/f_c, A^{v}/f_c$& $310, 11, 310 $ & --- &---\\
        %%%%%%%%
        %Diffusivity
        %%%%%%%
        $D_{Na}^{m}$ & $6.33\times 10^{-13} $ & m$^{2}$/s& \cite{mundy1971effect} \\
        $D_{Na}^{v}$ & $0 $ & m$^{2}$/s& ---\\
        $D_{Na}^{a}$ & $6.33\times 10^{-13}$ & m$^{2}$/s& ---\\
        %%%%%%%%
        %Electronic conductivity
        %%%%%%%
        $\sigma_{m}$ & $2.1 \times 10^{5}$ & S/cm& \cite{west2022solid} \\
        $\sigma_{v}$ & $2.1 \times 10^{-9}$ & S/cm & ---\\
        $\sigma_{a}$ & $2.1 \times 10^{5}$  & S/cm & ---\\
        %%%%%%%%%%%%%%
        & & &\\ % Just a gap
        \multicolumn{4}{l}{\textbf{Na-$\beta^{\prime\prime}$-alumina solid electrolyte separator}}\\
        %%%%%%%%%%%
        $c^{eq,el}_{Na^{^{+}}}$ & $0.0563$ & ---&\cite{heinz2021grain}\\
        $A^{el}/f_c$ & $4\times 10^{3}$ & --- &---\\
        $E_{g}$  & $0.20$ & eV & \cite{heinz2021grain}\\
 	$E_{gb}$  & $0.35$ & eV & \cite{heinz2021grain}\\
	$\mathcal{K}_{g}$ & $8.537 \times 10^{3}$ & (S K)/cm & \cite{heinz2021grain}\\
        $\mathcal{K}_{gb}$ & $4.786 \times 10^{3}$ & (S K)/cm & \cite{heinz2021grain}\\
        $\kappa^{el}_{(g\, \text{or} \, gb)}$&$\left[\mathcal{K}_{(g\, \text{or} \, gb)}/T\right]e^{-E_{(g\, \text{or} \, gb)}/k_{B}T}$&S/cm&\\
        $D_{Na^{+}}^{(g\, \text{or}\, gb)}$& $\left[\kappa_{(g\, \text{or}\, gb)}^{el}RT\right]/(c^{eq, el}_{Na^{^{+}}}F^2)$& m$^{2}$/s& ---\\
        %%%%
         & & &\\ % Just a gap
        \multicolumn{4}{l}{\textbf{Non-dimensionalization and mesh}}\\
        $\Delta x$ &$\mathit{l}_w/3.0$&$\mu m$&---\\
        $l_c$& $1\times 10^{-9}$ & m& ---\\
        $\tau$ &$l_{c}^2/D_{Na}^{m}$ &s&---\\
        $f_c$& $RT/v_{m}$ & J/m$^{3}$& ---\\
         %%%%
        \hline
    \end{tabular}
    \label{tab:my_table}
\end{table}

%%%%%%%%%%%%%%%%%%%%%%%%%%%%%%%%%%%%%%%%%%%%
\section{Results and discussion}
\label{sec:results}
In this section, we show the utility of our model by simulating three cases. First, we show the evolution of the negative electrode during stripping and plating, assuming a perfect electrode/electrolyte interface. Second, we simulate stripping and plating, assuming a single interfacial void exists at the electrode/electrolyte interface. Finally, we consider the case in which multiple interfacial voids form at the electrode/electrolyte interface. To assess the impact of GBs, each of these cases is subdivided into two subcases: one with a homogenous solid electrolyte (SE) and another with polycrystalline SEs with low-conductivity and high-conductivity grain boundaries (GBs).

\subsection{Perfect electrode/electrolyte interface}
\subsubsection{Stripping and plating simulations with a homogeneous SE}
\label{RS1.1}
We first simulate the shrinkage of a metallic sodium (Na) negative electrode in perfect contact with a homogeneous SE during stripping for different applied current densities. For our purposes here, we define the applied current density during stripping or discharge as positive, indicating an outgoing flux of Na$^+$ at the rightmost SE separator boundary. The starting thickness of the negative electrode is $28.6$ $\mu$m for all cases. The simulations are run for a stripping duration of $3$ h.

\begin{figure}[tbp]
\centering

\begin{subfigure}{0.58\textwidth}
	\includegraphics[trim=0 0 0 0, clip, keepaspectratio,width=\linewidth]{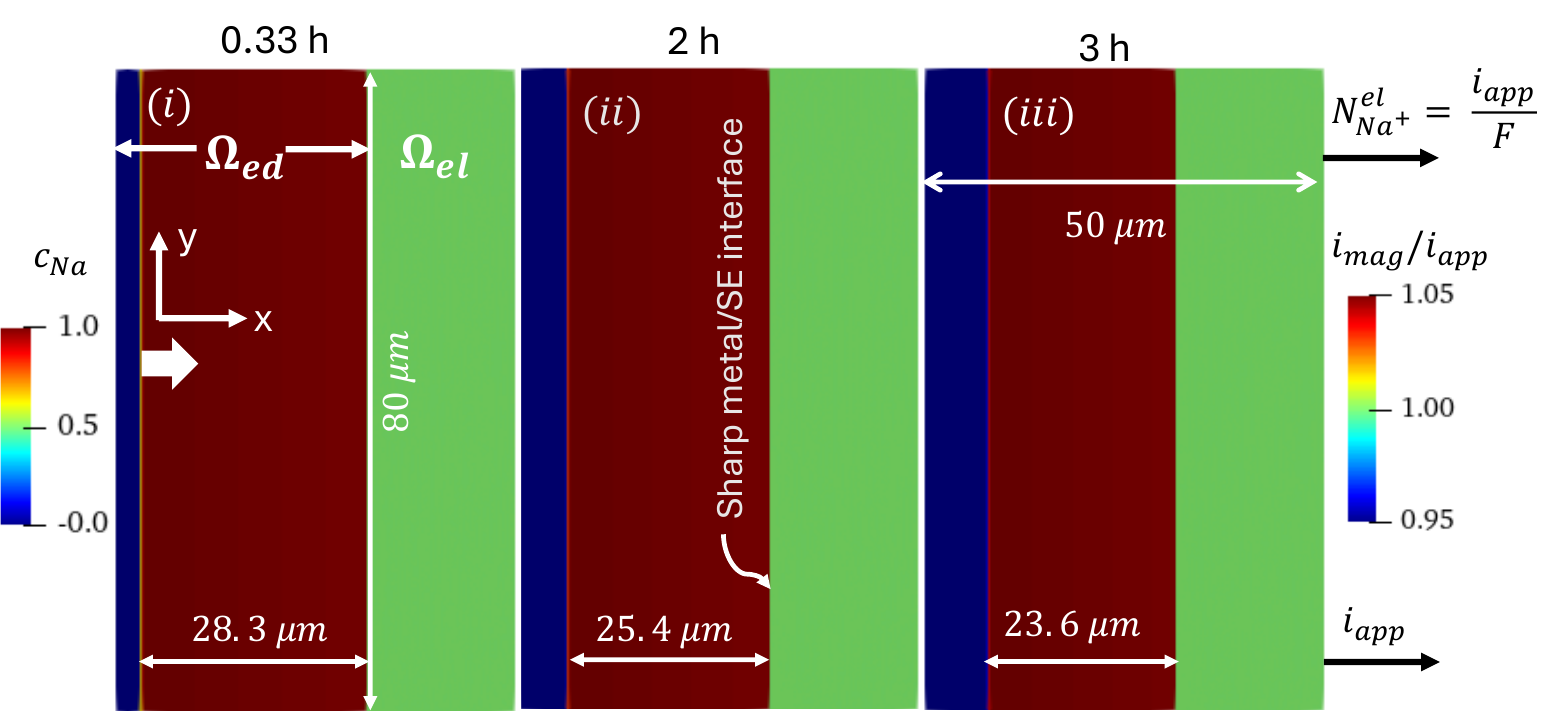}
	\subcaption{}
	\label{FigR1a}
\end{subfigure}
\begin{subfigure}{0.41\textwidth}
	\includegraphics[trim=0 0 0 0, clip, keepaspectratio,width=\linewidth]{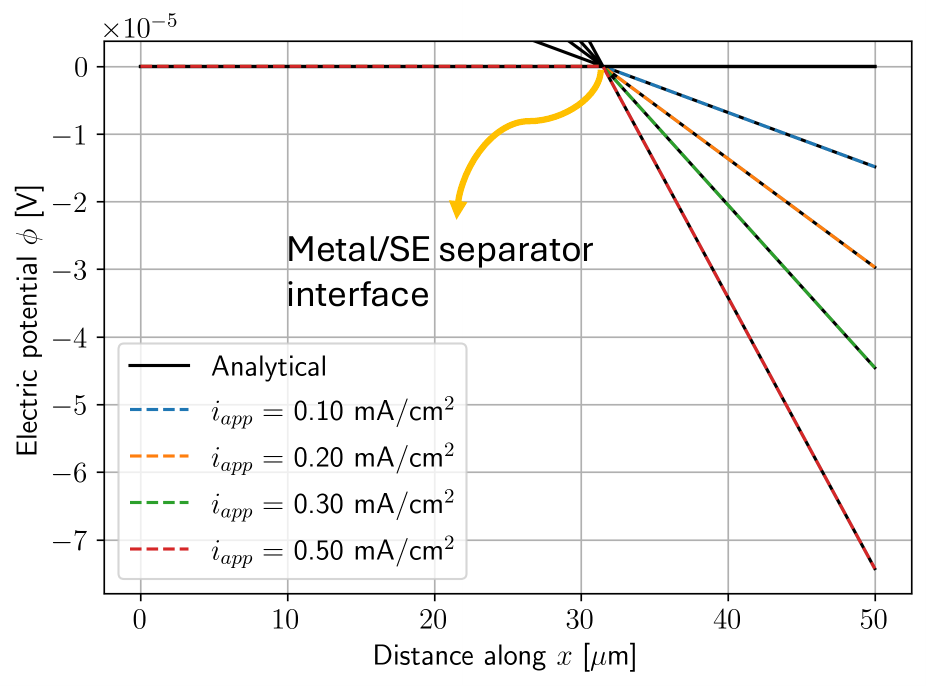}
	\subcaption{}
	\label{FigR1b}
\end{subfigure}
\begin{subfigure}{0.44\textwidth}
	\includegraphics[trim=0 0 0 0, clip, keepaspectratio,width=\linewidth]{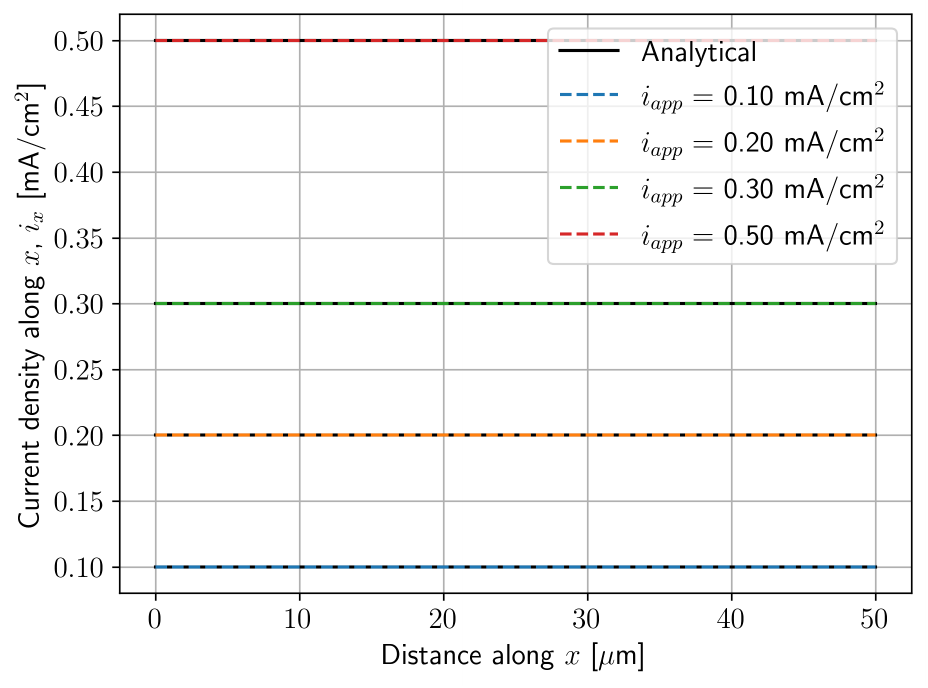}
	\subcaption{}
	\label{FigR1c}
\end{subfigure}
\begin{subfigure}{0.44\textwidth}
	\includegraphics[trim=0 0 0 0, clip, keepaspectratio,width=\linewidth]{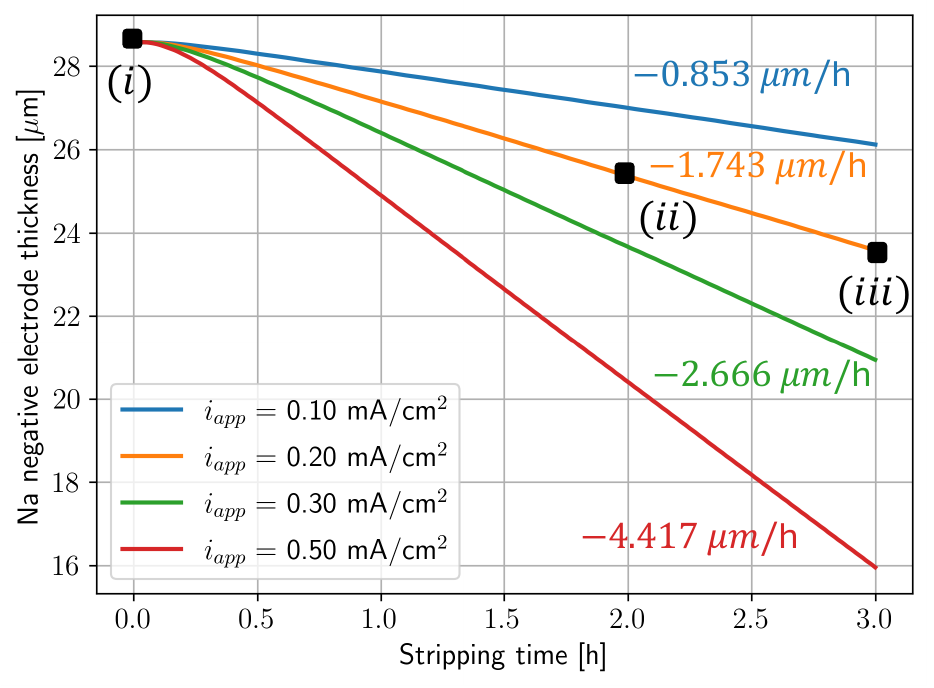}
	\subcaption{}
	\label{FigR1d}
\end{subfigure}
\begin{subfigure}{0.44\textwidth}
	\includegraphics[trim=0 0 0 0, clip, keepaspectratio,width=\linewidth]{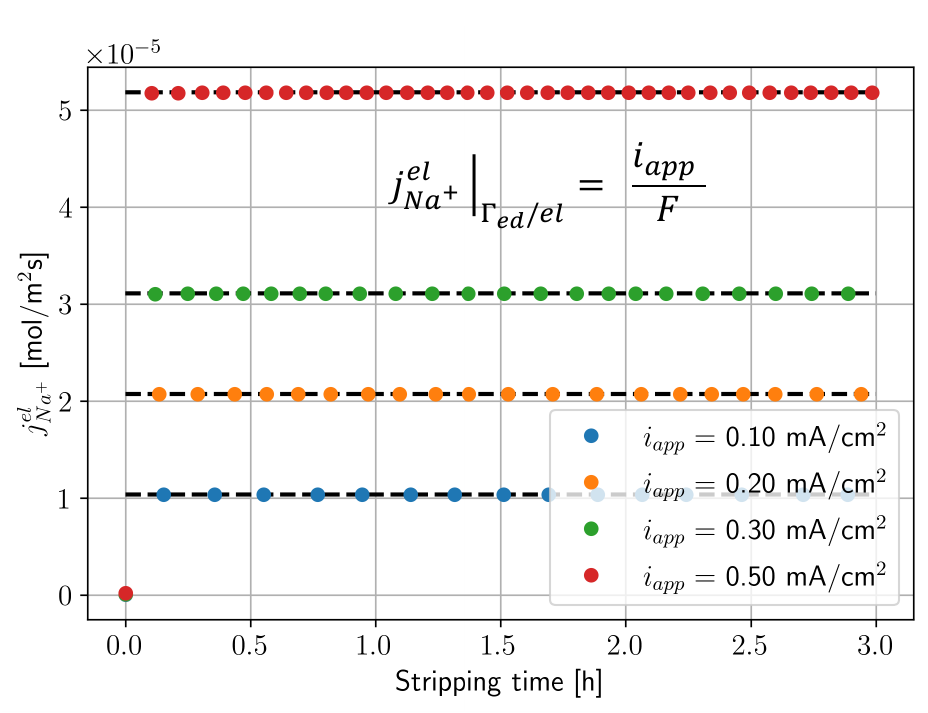}
	\subcaption{}
	\label{FigR1e}
\end{subfigure}
\begin{subfigure}{0.44\textwidth}
	\includegraphics[trim=0 0 0 0, clip, keepaspectratio,width=\linewidth]{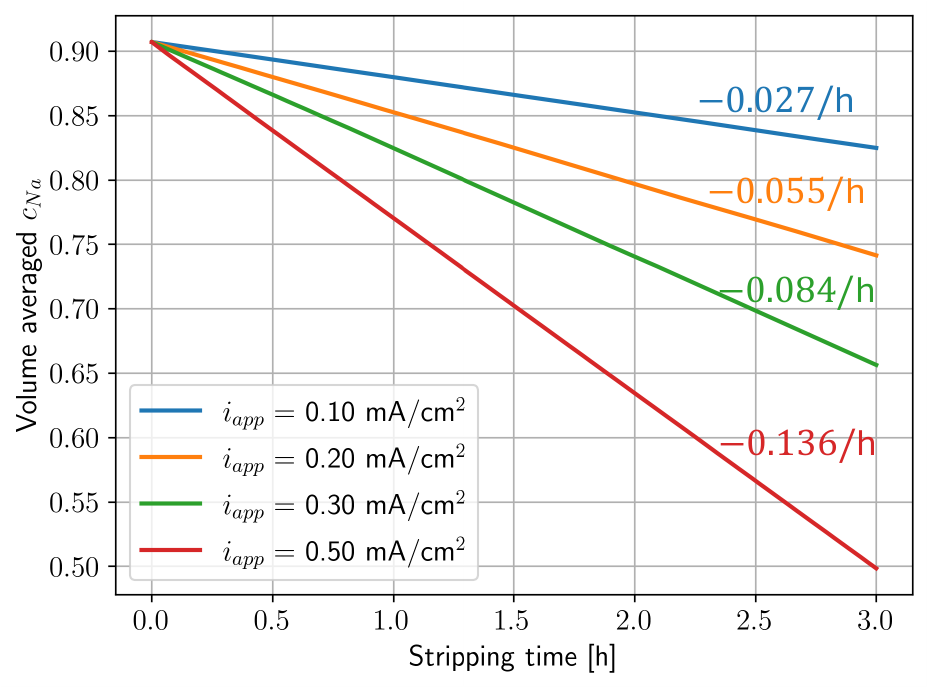}
	\subcaption{}
	\label{FigR1f}
\end{subfigure}
\caption{Stripping simulations with a perfect electrode/electrolyte interface and a homogeneous SE separator for different applied current densities. (a) Examples of the shrinkage of the Na negative electrode at three times with an applied current density of $0.2$ mA/cm$^2$. $\Omega_{ed}$ is shaded by the Na concentration and $\Omega_{el}$ by the ratio of the local current density to the applied current density. The spatial distribution along the x-axis of (b) electric potential and (c) current density. The variation with time in the (d) Na electrode thickness and (e) area-averaged mass flux of Na${^+}$ along the electrode/electrolyte interface $j_{Na^+}^{el}$. In (d), the black markers labelled $(i)$--$(iii)$ correspond to the microstructures shown in (a). The dotted lines in (e) indicate the ratio of applied current density to the Faraday constant. (f) Volume-averaged Na mole fraction $c_{Na}$ with stripping time. Values of the slopes obtained by linear fitting the curves are shown in (d) and (f).}
\end{figure}

Figure \ref{FigR1a} depicts the shrinkage of the Na negative electrode during stripping for an applied current density of $0.2$ mA/cm$^2$ at $0.33$ h, $2$ h, and $3$ h. The diffuse aux/electrode interface moves to the right, causing the negative electrode to shrink during stripping (see also Supplementary Video S$1$ in the Supporting Information), while the sharp electrode/electrolyte interface remains stationary. 
%This is contrary to sharp-interface models for cells with metallic electrodes \cite{mishra2021perspective}, \cite{jang2021towards}, where the negative electrode shrinks at the expense of the SE separator during stripping due to the movement of the electrode/electrolyte interface. 
In addition, a spatially constant concentration of Na is maintained across each of the three phases. 

Independent of time, the electric potential is zero in the x-direction across the auxiliary and Na electrode phases and varies linearly across the SE separator, as shown in Fig.~\ref{FigR1b}. The magnitude of the negative slope varies with the applied current density and is consistent with the analytical solution (see Section \ref{SI_secS1} of the Supporting Information for details). The current density is constant in the x-direction and equal to the applied current density (see Fig.~\ref{FigR1c}). Since the domain and boundary conditions are homogeneous in the y-direction, the electric potential is constant in the y-direction, and the current density in the y-direction is zero (Fig.~\ref{Fig_SI_secS1}). These results verify that charge is conserved in our model.
   
\begin{figure}[tbp]
\centering
\begin{subfigure}{0.4\textwidth}
	\centering
	\includegraphics[trim=0 0 0 0, clip, keepaspectratio,width=\linewidth]{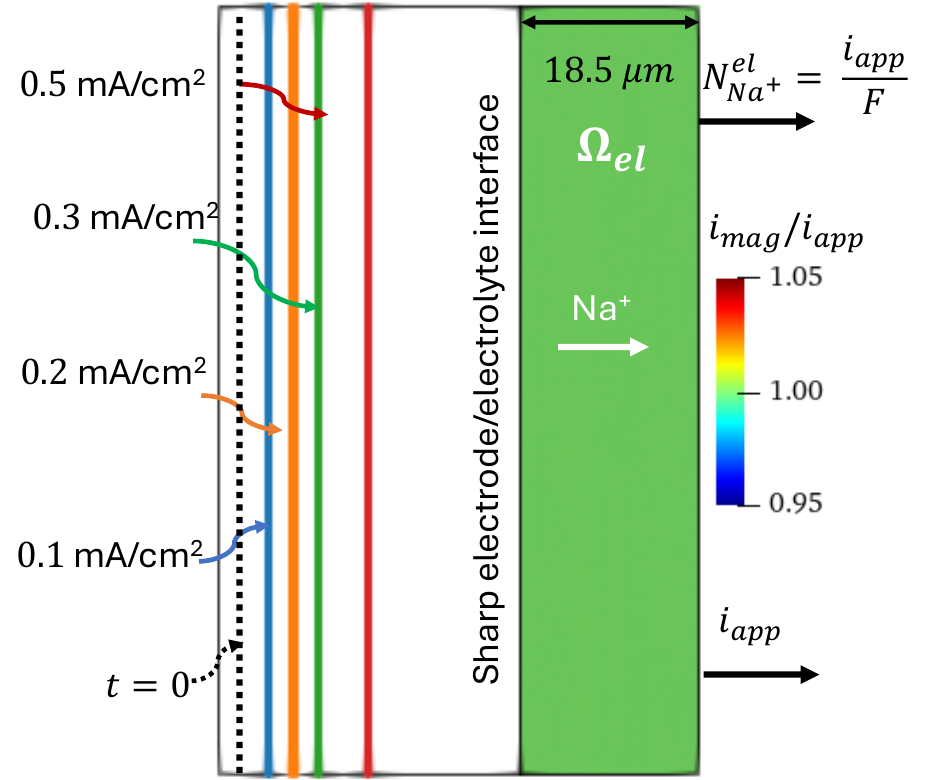}
	\subcaption{}
	\label{FigR1g}
\end{subfigure}
\begin{subfigure}{0.4\textwidth}
	\centering
	\includegraphics[trim=0 0 0 0, clip, keepaspectratio,width=\linewidth]{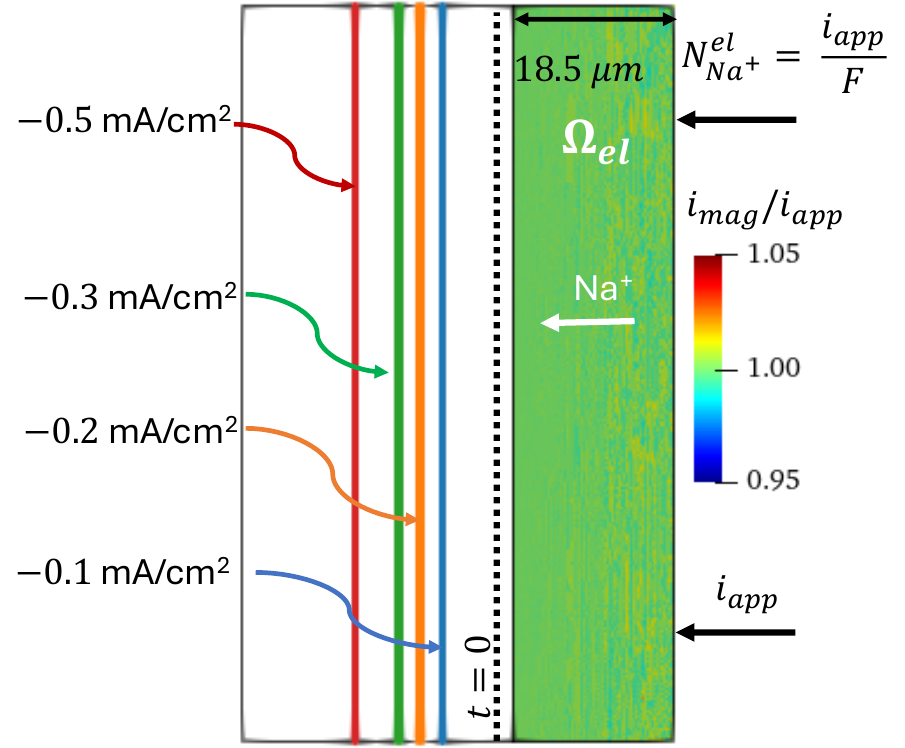}
	\subcaption{}
	\label{FigR2g}
\end{subfigure}
\caption{Impact of applied current density on Na electrode thickness during stripping and plating. The aux/electrode interface contour for $\xi_{a}=0.5$ after stripping (a) and plating (b) for $3$ h with different applied current densities. The dotted black lines indicate the initial position of the aux/electrode interface. In (a) and (b), the region $\Omega_{el}$ is shaded by the current density normalized by the applied current density and the arrows indicate the direction of Na$^{+}$ flux.}
\end{figure}

In addition, we determine the position in the x-direction of the moving aux/electrode interface (where $\xi_m=0.5$) with time to quantify the electrode shrinkage rate, as shown in Fig.~\ref{FigR1d}. The negative electrode thickness decreases linearly with stripping time for any applied current density. The slopes, i.e., the depletion rates, obtained after linear fitting are shown in Fig.~\ref{FigR1d} and Table \ref{TabR1}. These depletion rates are roughly proportional to the applied current density. In Table \ref{TabR1}, we also show depletion rates from a sharp-interface model \cite{mishra2021perspective} (see Section \ref{SI_secS2} of the Supplementary Information for details), and they are similar to those from our model. 
The impact of applied density on the depletion rate means that the electrode thickness after 3 h decreases with increasing applied current density, as shown in Fig.~\ref{FigR1g}. 

\begin{table}[btp]
\caption{Depletion and deposition rates during stripping and plating calculated using the sharp-interface electrochemical model \cite{mishra2021perspective} and our predicted values. 
The predicted depletion and deposition rates were obtained by linearly fitting the simulated curves shown in Figs. \ref{FigR1d} and \ref{FigR2d}, respectively.}
\centering
\begin{tabular}{l|cccc}
 \hline
$i_{app}$ [mA/cm$^2$] & $\pm 0.1$ & $\pm 0.2$ &  $\pm 0.3$ & $\pm 0.5$ \\ 
\hline \hline
Sharp[$\mu$m/h] & $\mp0.885$  & $\mp1.770$  & $\mp 2.655$ & $\mp 4.425$  \\ \hline
Stripping$^{\text{This paper}}$  [$\mu$m/h] & $-0.853$  & $-1.743$ & $-2.666$ &$-4.417$  \\ \hline
Plating$^{\text{This paper}}$  [$\mu$m/h] & $+0.878$ & $+1.738$  & $+2.620$ &$+4.278$ \\ \hline
\end{tabular}
\label{TabR1}
\end{table}

In our model, the velocity of the aux/electrode interface and, consequently, the electrode shrinkage rate depends on the average normal flux of Na/Na$^+$ leaving the electrode/electrolyte interface, which was calculated using Eq.~\eqref{EqnS16a} provided in Section \ref{SI_secS4} of the Supporting Information. Figure \ref{FigR1e} depicts the area-averaged normal flux of Na$^+$ leaving the electrode/electrolyte interface as a function of time for different applied current densities. The average interfacial flux remains constant and is equal to the ratio of the applied current density to the Faraday constant, i.e., $i_{app}/F$. This ratio is indicated by dotted black lines in Fig.~\ref{FigR1e}. It is also equal to the applied flux at the right SE separator boundary because the Na$^{+}$ concentration in the SE separator during stripping remains spatially uniform and constant with time (see Fig.~\ref{Fig_SI_secS1} and \ref{Fig_SI_secS2a}, respectively, of the Supporting Information). 

Furthermore, as the Na negative electrode gets consumed, the overall Na concentration in the electrode-auxiliary region decreases during stripping, which was calculated using Eq.~\eqref{EqnS14} provided in Section \ref{SI_secS4} of the Supporting Information. Figure \ref{FigR1f} shows that the overall Na concentration decreases linearly with stripping time for different applied current densities. Note that the growth of the auxiliary phase does not contribute significantly to the overall Na concentration because the Na mole fraction in the auxiliary phase is negligible ($\approx 10^{-6}$), as shown in Fig.~\ref{Fig_SI_secS1}. The slopes, i.e., the Na loss rates, obtained after linear fitting are also shown in Fig.~\ref{FigR1f} and are proportional to the applied current density. We also analytically calculate the Na loss rates (see Section \ref{SI_secS4} of the Supporting Information). Table \ref{TabR2} shows that the calculated Na loss rates agree with the analytical values. This verifies that mass is conserved in our simulations. It is important to point out that mass conservation in moving sharp-interface electrochemical models is often an issue, especially during cycling simulations, and can require a reformulation of the interfacial conditions at the moving electrode/electrolyte interface \cite{mishra2021perspective,jang2021towards}. Our approach naturally conserves mass and does not require interfacial conditions at the moving interface.
\begin{table}[btp]
\caption{Na loss and gain rates during stripping and plating calculated using an analytical approach and our predicted values. The analytical approach is described in Section \ref{SI_secS4} of the Supporting Information. The predicted loss and gain rates were obtained by linearly fitting the simulated curves shown in Figs. \ref{FigR1f} and \ref{FigR2f}, respectively.}
\centering
\begin{tabular}{l|cccc}
 \hline
$i_{app}$ [mA/cm$^2$] & $\pm 0.1$  & $\pm0.2$ &  $\pm0.3$  & $\pm0.5$ \\ 
\hline \hline
Analytical [h$^{-1}$] & $\mp 0.028$  & $\mp0.056$ & $\mp0.084$ & $\mp 0.141$ \\ \hline
Stripping$^{\text{This paper}}$  [h$^{-1}$]& $-0.027$ & $-0.055$   & $-0.084$ &$-0.136$  \\ \hline
Plating$^{\text{This paper}}$  [h$^{-1}$] & $+0.028$ & $+0.056$   & $+0.084$ & $+0.139$ \\ \hline
\end{tabular}
\label{TabR2}
\end{table}
We also investigate the growth of a Na-negative electrode during plating with different applied current densities. The applied current density is negative during plating, indicating an incoming flux of Na$^+$ at the rightmost SE separator boundary. Like stripping, we initialize all our plating simulations with a $5.6$ $\mu$m wide negative Na electrode and simulate $3$ h. 

\begin{figure}[tbp]
\centering
\begin{subfigure}{0.55\textwidth}
	\includegraphics[trim=0 0 0 0, clip, keepaspectratio,width=\linewidth]{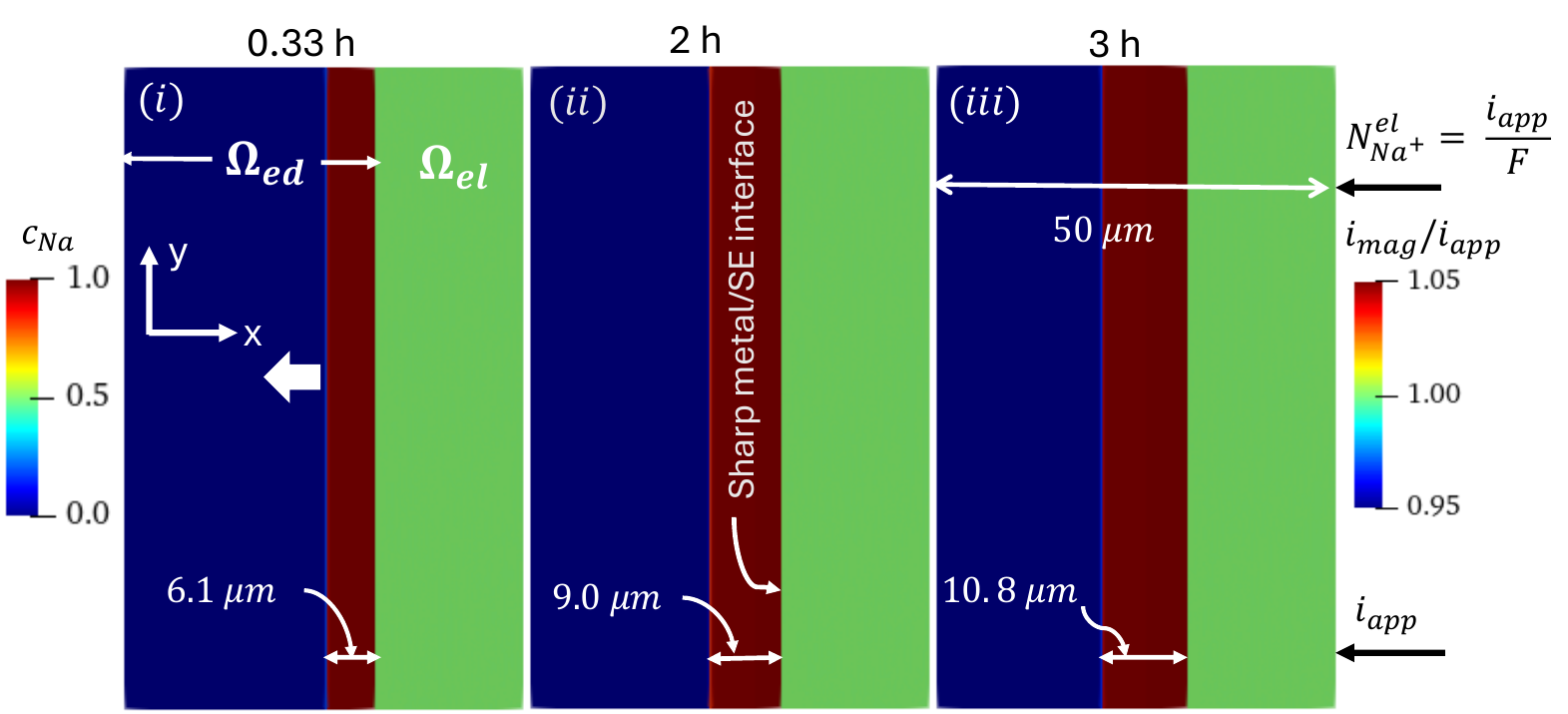}
	\subcaption{}
	\label{FigR2a}
\end{subfigure}
\begin{subfigure}{0.41\textwidth}
	\includegraphics[trim=0 0 0 0, clip, keepaspectratio,width=\linewidth]{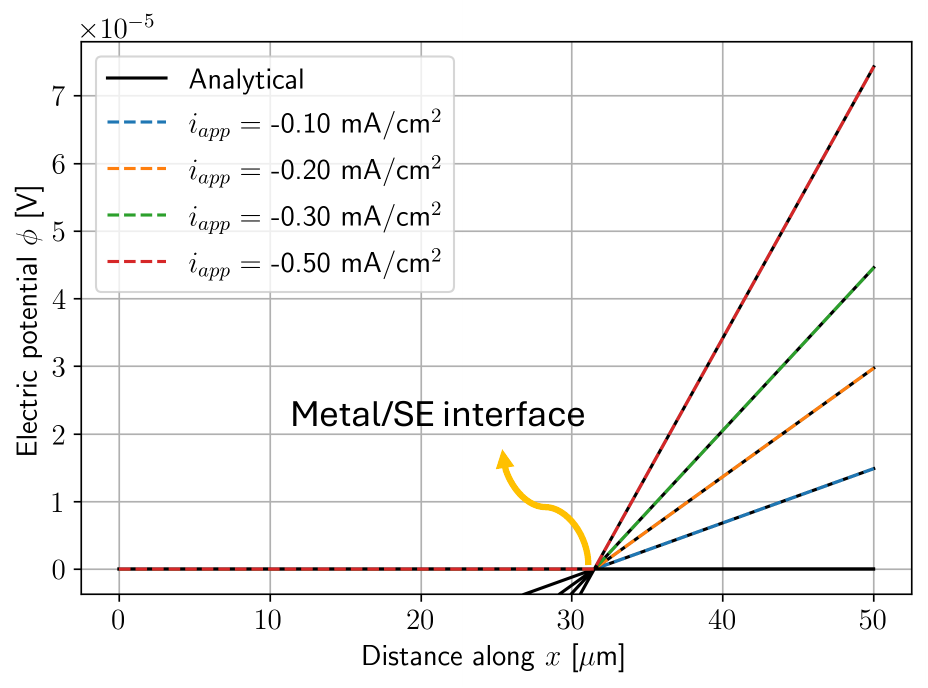}
	\subcaption{}
	\label{FigR2b}
\end{subfigure}
\begin{subfigure}{0.44\textwidth}
	\includegraphics[trim=0 0 0 0, clip, keepaspectratio,width=\linewidth]{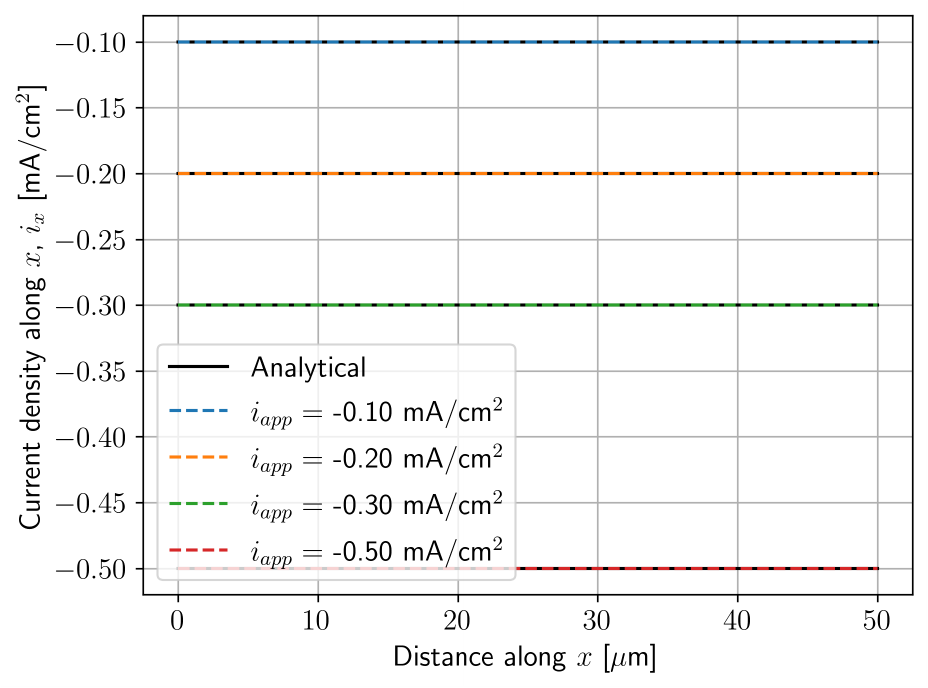}
	\subcaption{}
	\label{FigR2c}
\end{subfigure}
\begin{subfigure}{0.44\textwidth}
	\includegraphics[trim=0 0 0 0, clip, keepaspectratio,width=\linewidth]{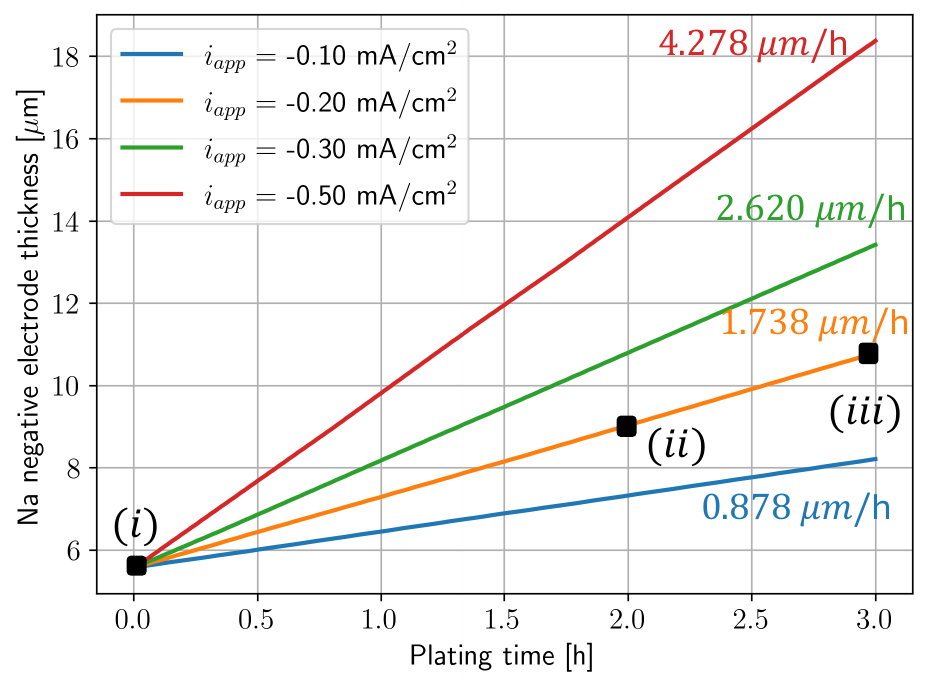}
	\subcaption{}
	\label{FigR2d}
\end{subfigure}
\begin{subfigure}{0.44\textwidth}
	\includegraphics[trim=0 0 0 0, clip, keepaspectratio,width=\linewidth]{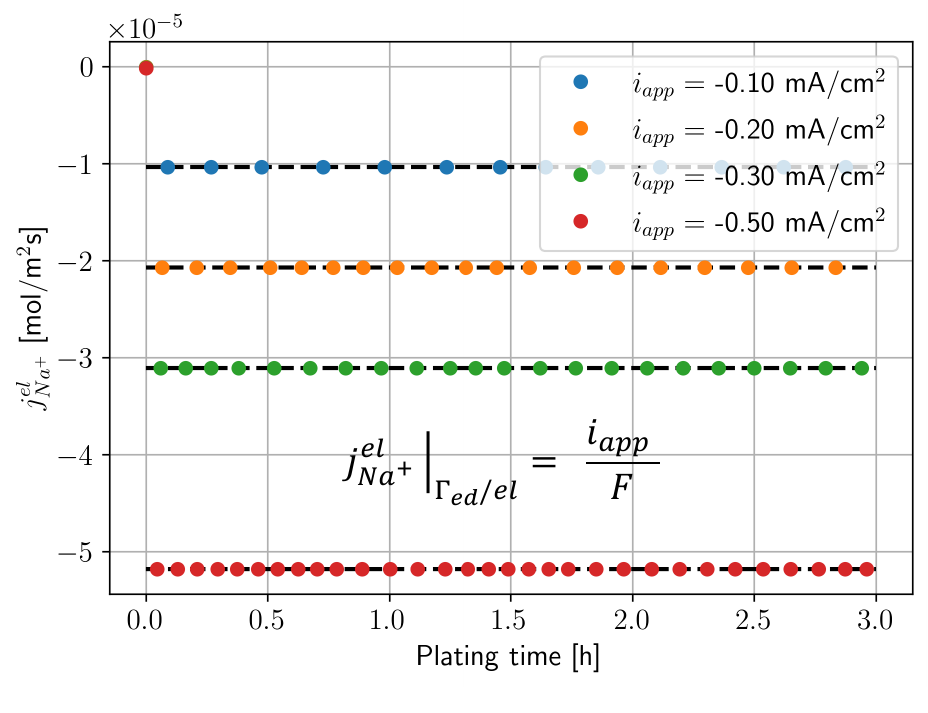}
	\subcaption{}
	\label{FigR2e}
\end{subfigure}
\begin{subfigure}{0.44\textwidth}
	\includegraphics[trim=0 0 0 0, clip, keepaspectratio,width=\linewidth]{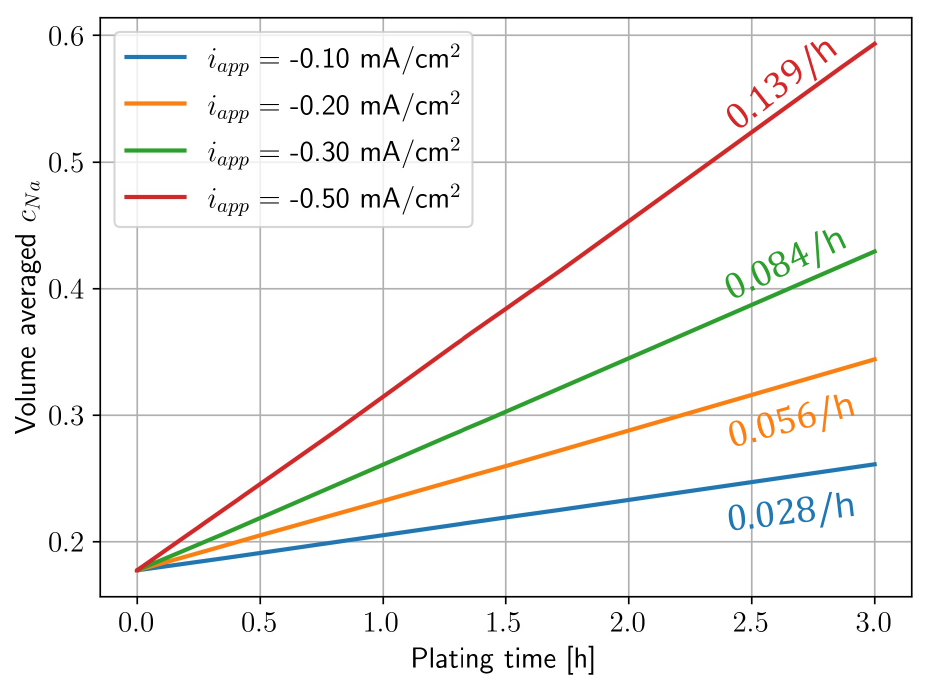}
	\subcaption{}
	\label{FigR2f}
\end{subfigure}
\caption{Plating simulations assuming a perfect electrode/electrolyte interface and a homogeneous solid electrolyte separator. (a) Examples of the growth of the Na negative electrode at three times with an applied current density of $-0.2$ mA/cm$^2$. $\Omega_{ed}$ is shaded by the Na concentration and $\Omega_{el}$ by the ratio of the local current density to the applied current density. The spatial distribution along the x-axis of (b) electric potential and (c) current density. The variation with time in the (d) Na electrode thickness and (e) area-averaged mass flux of Na${^+}$ along the electrode/electrolyte interface $j_{Na^+}^{el}$. In (d), the black markers labeled $(i)$--$(iii)$ correspond to the microstructures shown in (a). The dotted lines in (e) indicate the ratio of applied current density to the Faraday constant. (f) Variation in the volume-averaged Na mole fraction $c_{Na}$ with stripping time. Values of the slopes obtained by linear fitting the curves are shown in (d) and (f).\label{Fig4}}
\end{figure}

Figure \ref{FigR2a} shows the growth of the Na negative electrode during plating or charging for an applied current density of $-0.2$ mA/cm$^2$ again at $0.33$ h, $2$ h, and $3$ h. As expected, the diffuse aux/electrode interface moves incrementally to the left, causing the growth of the negative electrode during plating (see also Supplementary Video S$2$ in the Supporting Information). We have again verified that charge is conserved by comparing the electric potential and current density distributions against the analytical solution previously discussed in our stripping simulations, as shown in Figs.~\ref{FigR2b} and \ref{FigR2c}. 

Further, to compare our deposition rates with sharp-interface models \cite{mishra2021perspective}, we again calculate the Na negative electrode thickness with time, as shown in Fig.~\ref{FigR2d}. The electrode thickness increases linearly with plating time, and the slopes, i.e., the deposition rates, are shown in the figure and Table \ref{TabR1}. Like stripping, Table \ref{TabR1} shows that the deposition rate increases proportionally with the magnitude of the applied current density. The difference in the deposition and depletion rates is negligible, $\leq 2\%$, since the applied current densities in the plating and stripping simulations have the same magnitude, just the opposite sign. The final electrode thickness increases with increasing applied current density after plating for 3 h, as shown in Fig.~\ref{FigR2g}. roportional to the applied current density, as shown in Fig.~\ref{FigR2e}. Also, the steady-state flux values have the same magnitude but opposite sign since they are equal to $|\boldsymbol{i}_{app}|/F$. Since this interfacial flux controls the velocity of the aux/electrode interface, the deposition rates are proportional to the applied current density, as shown in Table \ref{TabR1}. In addition, Fig.~\ref{FigR2f} shows that the overall Na concentration in the electrode-auxiliary region increases linearly with time during plating for different applied current densities. The slopes, i.e., the Na gain rates calculated after linear fitting, are shown in the figure and Table \ref{TabR2}. The difference between the Na loss and gain rates is negligible ($\leq 1\%$).

\subsubsection{Cyclic simulations with a polycrystalline solid electrolyte (SE)}
\label{RS1.3}
In this subsection, we investigate the role of the SE GB properties on the deposition and depletion kinetics of the Na-negative electrode during cycling. Specifically, we consider two SEs, one with low-conductivity GBs and another with high-conductivity GBs. The spatial distributions of ionic conductivities along the GBs and grains for these two SEs are presented in Fig.~\ref{Fig_SI_secS5} in Section \ref{SI_secS5} of the Supporting Information. Previously, in Section \ref{numerical_method}, we noted that the diffusivities along the GBs were obtained from the ionic conductivities; thus, they also differ from the SE grain diffusivity. For the purpose of discussion below, we will hereafter refer only to the GB-to-grain conductivity ratios, i.e., $\kappa_{gb}^{el}/\kappa_{g}^{el}$, and it should be understood that the GB-to-grain diffusivity ratios are the same as the conductivity ratios. We compare the behavior for cases in which the conductivity of the GBs is the same as the grains ($\kappa_{gb}^{el}/\kappa_{g}^{el}=1$), is higher than the grains ($\kappa_{gb}^{el}/\kappa_{g}^{el}=10$), and is lower than the grains ($\kappa_{gb}^{el}/\kappa_{g}^{el}=0.67$).

\begin{figure}[tbp]
\centering
\begin{subfigure}{0.94\textwidth}
	\includegraphics[trim=0 0 0 0, clip, keepaspectratio,width=\linewidth]{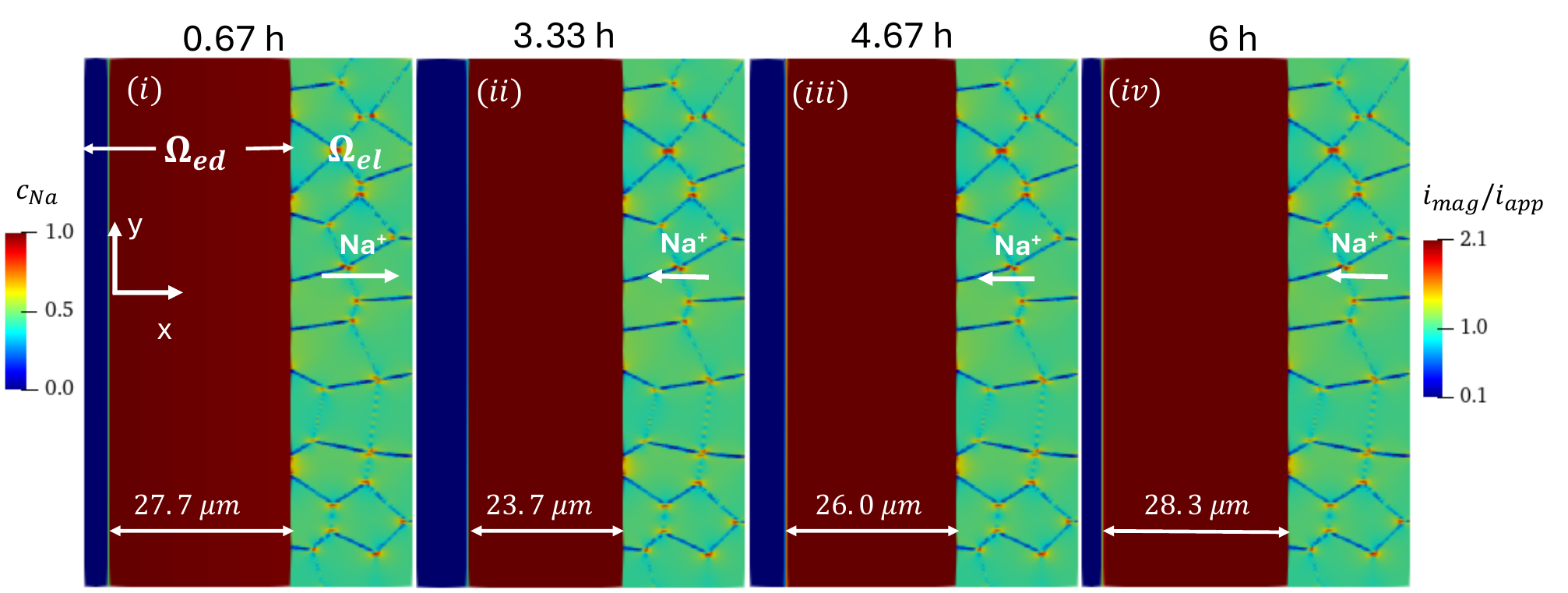}
	\subcaption{}
	\label{FigR5a}
\end{subfigure}
\begin{subfigure}{0.94\textwidth}
	\includegraphics[trim=0 0 0 0, clip, keepaspectratio,width=\linewidth]{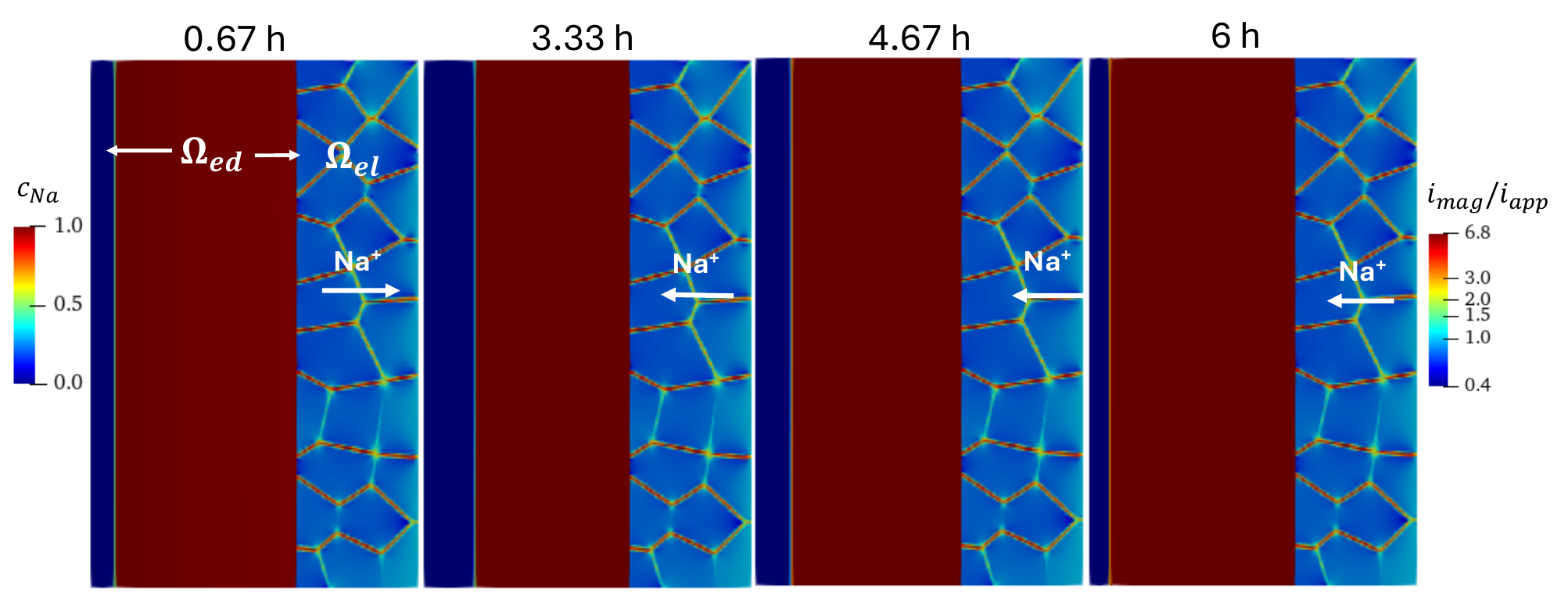}
	\subcaption{}
	\label{FigR5b}
\end{subfigure}
\caption{Cycling simulations assuming a perfect electrode/electrolyte interface and polycrystalline SE separator. Evolution of the Na negative electrode during cycling with an applied current of $0.2$ mA/cm$^2$ with (a) $\kappa_{gb}^{el}/\kappa_{g}^{el}=0.067$ and (b) $\kappa_{gb}^{el}/\kappa_{g}^{el}=10$. $\Omega_{ed}$ is shaded by the Na concentration and $\Omega_{el}$ by the ratio of the local current density to the applied current density. The arrows indicate the direction of Na$^+$ flux.}
\end{figure}

In our cyclic simulations, we simulate stripping for $3$ h and then plating for $3$ h under an applied cycling current density of $0.2$ mA/cm$^2$. Figures \ref{FigR5a} and \ref{FigR5b} depict the evolution of the Na negative electrode during cycling in contact with a SE having low conductivity and high conductivity GBs, respectively (see also Supplementary Videos S$3$ and S$4$). The Na negative electrode evolution is similar in both cases, i.e., it shrinks during stripping and grows during plating, as discussed in Section \ref{RS1.1}. As before, no concentration gradient of Na occurs in the electrode nor in the auxiliary phase during cycling. However, in contrast to the simulations discussed in Section \ref{RS1.1}, the current density distribution in the SE separator is heterogeneous due to the low or high conductivity GBs (Figs. \ref{FigR5a} and \ref{FigR5b}), respectively.  

\begin{figure}[tbp]
\centering
\begin{subfigure}{0.49\textwidth}
	\includegraphics[trim=0 0 0 0, clip, keepaspectratio,width=\linewidth]{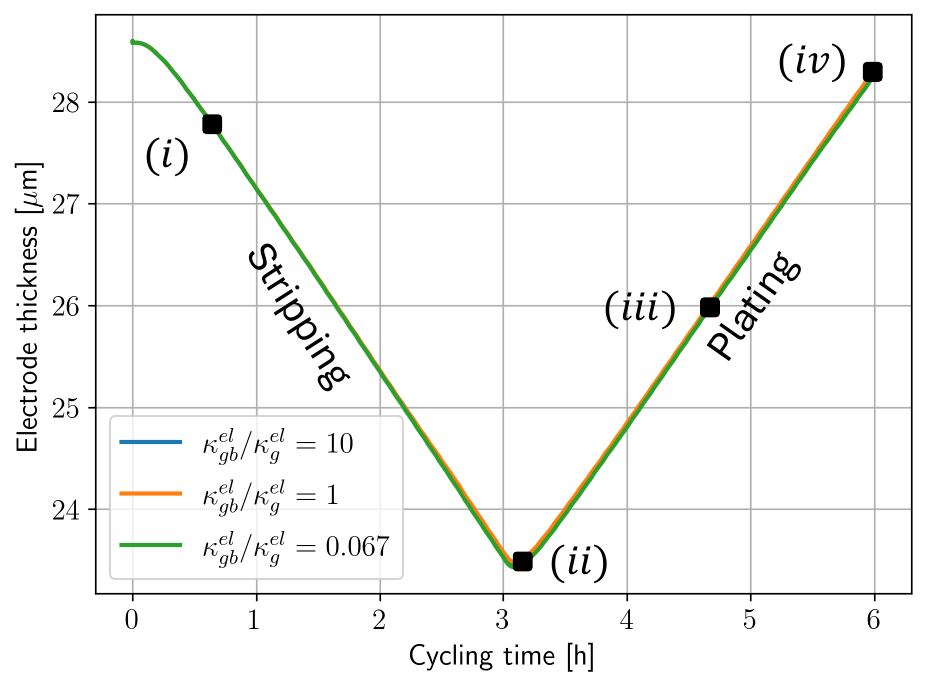}
	\subcaption{}
	\label{FigR5c}
\end{subfigure}
\begin{subfigure}{0.49\textwidth}
	\includegraphics[trim=0 0 0 0, clip, keepaspectratio,width=\linewidth]{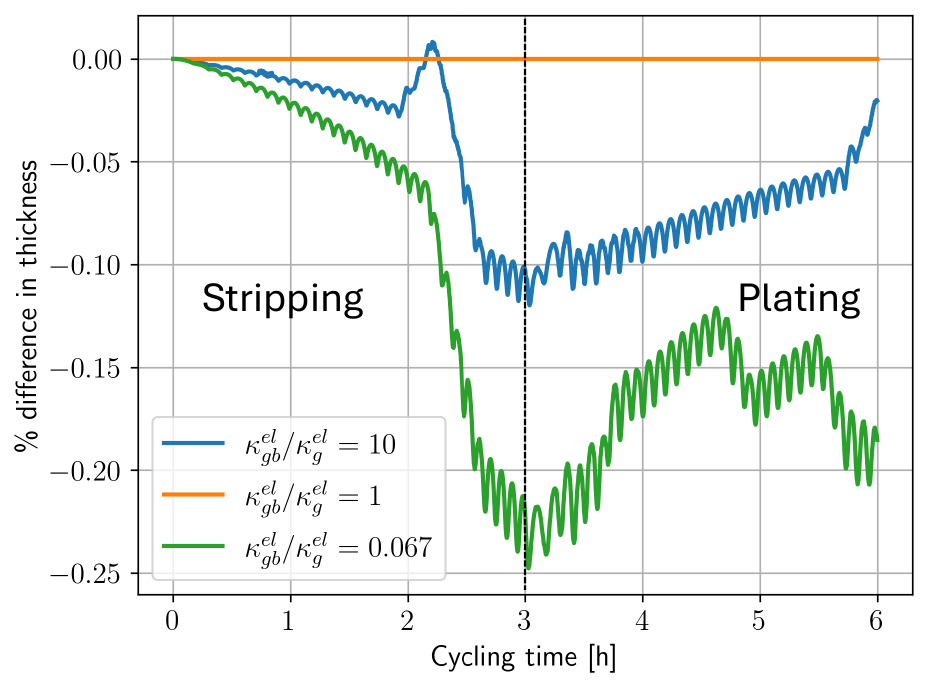}
	\subcaption{}
	\label{FigR5d}
\end{subfigure}
\begin{subfigure}{0.49\textwidth}
	\includegraphics[trim=0 0 0 0, clip, keepaspectratio,width=\linewidth]{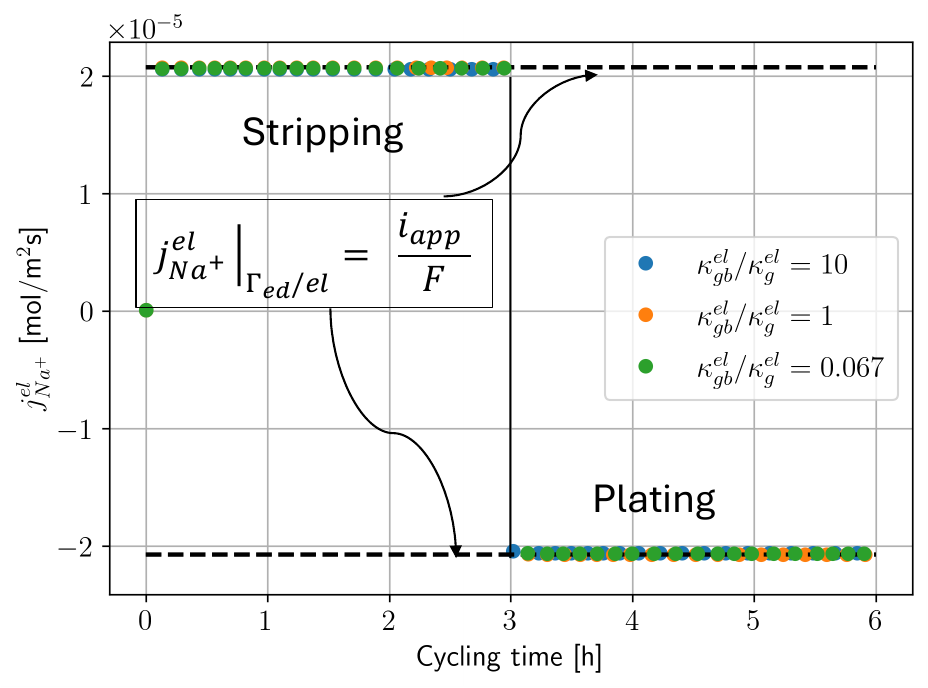}
	\subcaption{}
	\label{FigR5e}
\end{subfigure}
\begin{subfigure}{0.49\textwidth}
	\includegraphics[trim=0 0 0 0, clip, keepaspectratio,width=\linewidth]{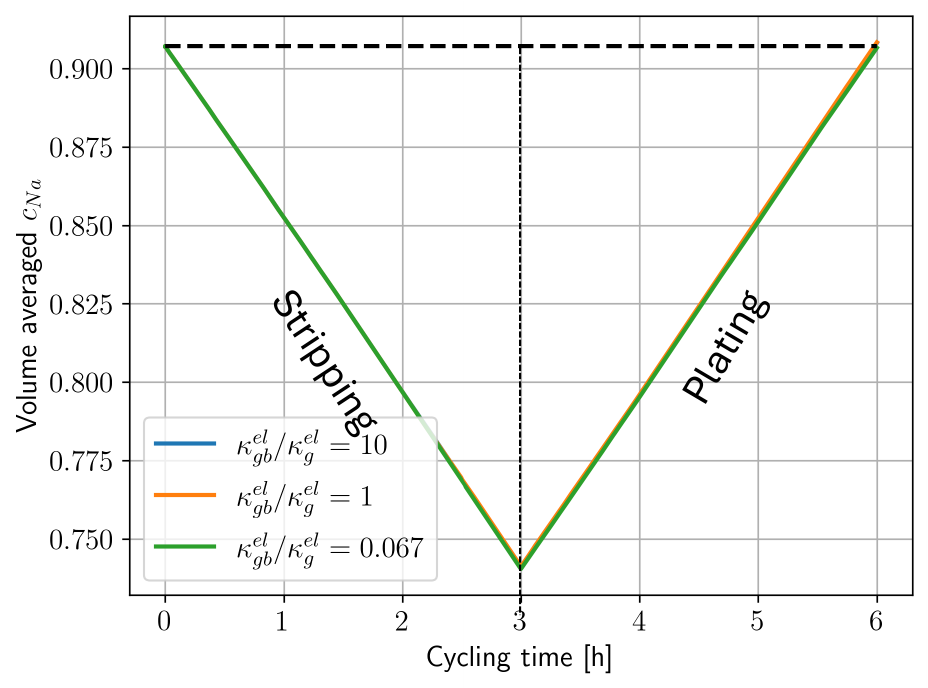}
	\subcaption{}
	\label{FigR5f}
\end{subfigure}
\caption{Analysis of the cycling simulations assuming a perfect electrode/electrolyte interface and polycrystalline SE separator. The variation with time of the (a) Na negative electrode thickness, (b) percentage difference in the electrode thickness relative to the homogeneous SE case, (c) area-averaged mass flux of Na${^+}$ at the electrode/electrolyte interface $j^{el}_{Na^+}$, and (d) volume-averaged Na mole fraction. In (a), the black markers labelled $(i)$--$(iv)$ correspond to the microstructures shown in Fig.~\ref{FigR5a}.  In (d), the black dotted line indicates the initial concentration of Na in the system.}
\end{figure}

The Na electrode thickness decreases linearly with time during stripping and then increases linearly during plating, as shown in Fig.~\ref{FigR5c}. The thickness after six hours is close to, but not equal to the original thickness. For the homogeneous SE case, this difference ($<1 \%$) is due to the small nonlinear regions at the start of the simulation and when the system cycles from stripping to plating. The change in the thickness with time is independent of the conductivity of the GBs; the behavior is identical for all three $\kappa_{gb}^{el}/\kappa_{g}^{el}$ ratios. We confirm this by calculating the percentage difference in the Na electrode thickness relative to the homogeneous SE case, shown in Fig.~\ref{FigR5d}. The maximum percentage difference never exceeds a magnitude of $0.25\%$.

This is expected because the deposition and depletion rates in our model depend on the applied flux of Na/Na$^+$ at the electrode/electrolyte interface, and this flux is directly proportional to the applied current density and is identical in the case of polycrystalline and homogenous SEs. To verify this, we calculate the average flux of Na$^{+}$ at the electrode/electrolyte interface as a function of time for different GB-to-grain conductivity ratios, shown in Fig.~\ref{FigR5e}. The steady-state fluxes are independent of the GB-to-grain conductivity ratios and are equal to the ratio of applied cycling current density divided by the Faraday constant, $i_{app}/F$. This confirms our assertion that the deposition and depletion rates only depend on the magnitude of the applied current density.

Finally, we ensure that Na is conserved during the cycling. Figure \ref{FigR5f} shows the total Na concentration during cycling for different GB-to-grain conductivity ratios. The overall Na loss and gain rates during cycling are independent of the GB-to-grain conductivity ratios. This is also consistent with our previous analysis in Section \ref{RS1.1}, which shows that the Na loss or gain rates depend only on the applied flux at the electrode/electrolyte interface. Since this flux is identical in the case of polycrystalline and homogenous SEs, as shown in Fig.~\ref{FigR5e}, the Na loss and gain rates during cycling are identical. The initial and final total Na concentrations are equal, indicating that the mass of Na is conserved in our cyclic simulations regardless of the SE GB properties.

%
%\begin{comment}

\subsection{Single void at the electrode/solid electrolyte interface}
\subsubsection{Stripping and plating simulations with a single-crystal solid electrolyte (SE)}
\label{secR2.1}
Next, we simulate the evolution of a single void at the electrode/electrolyte interface during stripping for different applied current densities. Similar to Zhao \textit{et al.} \cite{zhao2022phase}, we begin all our simulations with a semi-circular void of radius $4.8$ $\mu$m at the electrode/electrolyte interface. However, unlike Zhao \textit{et al.} \cite{zhao2022phase}, the thickness of the negative electrode in our model varies during stripping. The initial Na negative electrode thickness is assumed to be $28.6$ $\mu$m in all simulations. The simulations run for a stripping time of $1$ h, or until the void completely spans the electrode/electrolyte interface, i.e., complete contact loss occurs. Once complete contact loss occurs, the flux of Na/Na$^+$ at the electrode/electrolyte interface becomes effectively zero, and further stripping cannot occur.

\begin{figure}[tbp]
\centering
\begin{subfigure}{0.86\textwidth}
	\includegraphics[trim=0 0 0 0, clip, keepaspectratio,width=\linewidth]{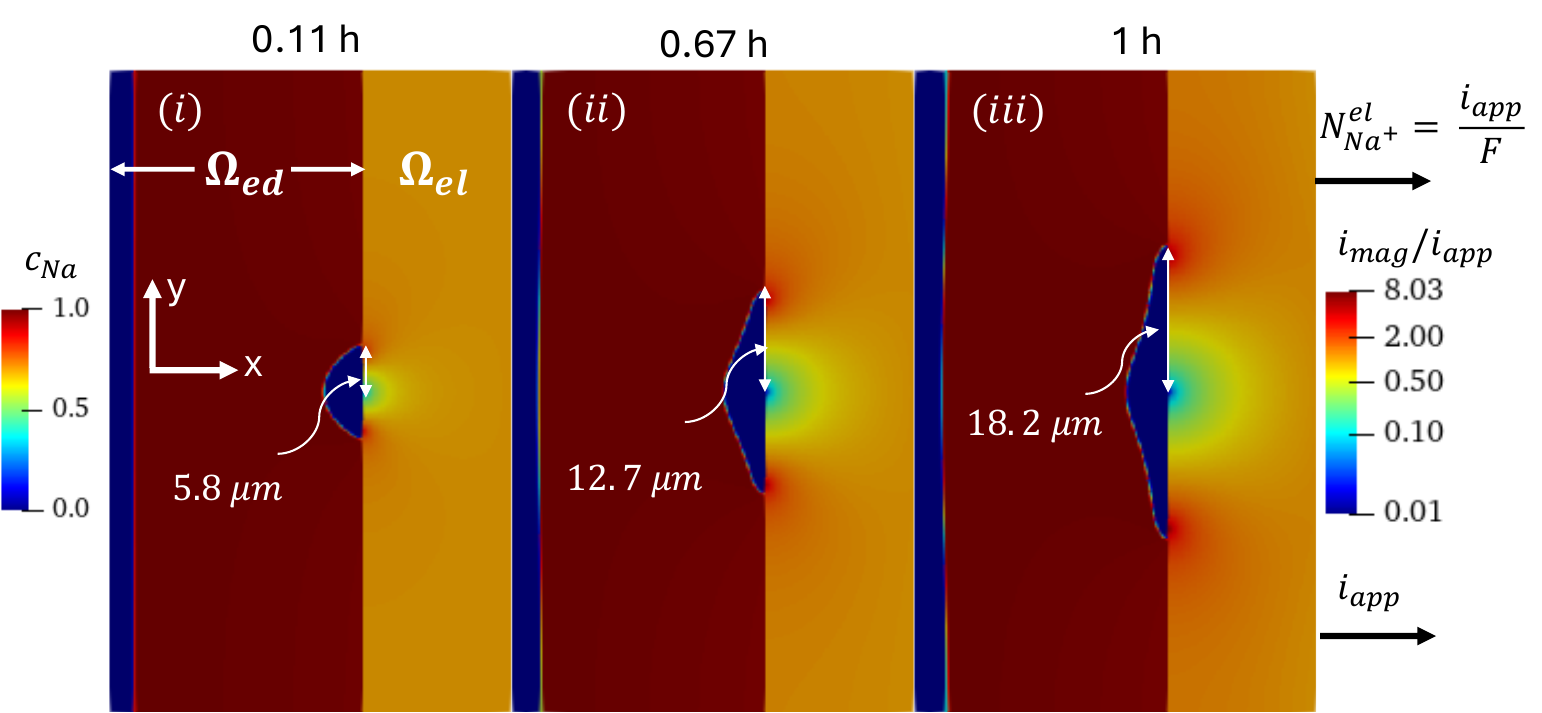}
	\subcaption{}
	\label{FigR6a}
\end{subfigure}
\begin{subfigure}{0.86\textwidth}
	\includegraphics[trim=0 0 0 0, clip, keepaspectratio,width=\linewidth]{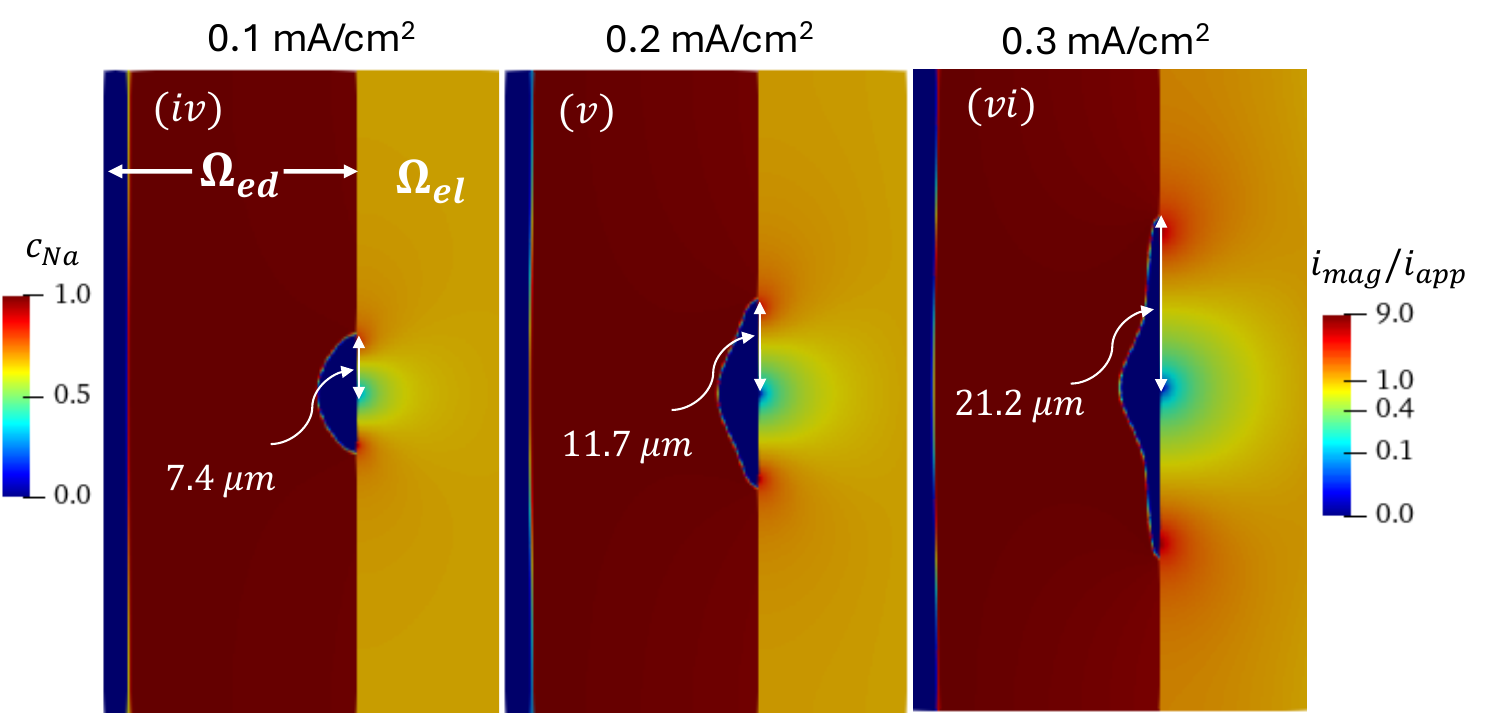}
	\subcaption{}
	\label{FigR6b}
\end{subfigure}
\caption{Stripping simulations with a single interfacial void at the electrode/electrolyte interface and a homogeneous solid electrolyte separator with different applied current densities. (a) Void evolution with an applied current density of $0.2$ mA/cm$^2$. (b) Void morphology for different applied current densities after stripping for 0.6 h. $\Omega_{ed}$ is shaded by the Na concentration and $\Omega_{el}$ is shaded by the ratio of the local current density to the applied current density.}
\end{figure}

Figure~\ref{FigR6a} shows the evolution of a single void during stripping under an applied current density of $0.2$ mA/cm$^2$. The void grows due to the formation of vacancies at the electrode/electrolyte interface as Na atoms diffuse into the electrolyte during stripping. These vacancies cause the void to change its shape from a semi-circle to a pancake, which qualitatively agrees with the void shapes observed in experiments \cite{kasemchainan2019critical,spencer2019sodium} and void simulations in Li/LLZO system \cite{zhao2022phase,agier2022void}. The void growth normal to the electrode/electrolyte interface is small compared to its growth along the interface. While the negative electrode shrinks during stripping, it shrinks more on the top and the bottom than it does in the center because the void blocks the flux of Na$^+$ across the electrode/electrolyte interface. Thus, the aux/electrode interface is no longer flat (see Supplementary Video S$5$ in the Supporting Information). Also, the current density in the SE separator is higher near the void edges and lower near the void center due to contact loss. 

The magnitude of the applied current density has a large impact on the void growth, as shown in Fig.~\ref{FigR6b} after stripping for $0.6$ h. The void shapes and electrode thickness are similar in all cases, but the void size and current constriction increase with increasing applied current density, which is similar to Zhao's analysis for the Li/LLZO system \cite{zhao2022phase}.

\begin{figure}[tbp]
\centering
\begin{subfigure}{0.6\textwidth}
	\includegraphics[trim=0 0 0 0, clip, keepaspectratio,width=\linewidth]{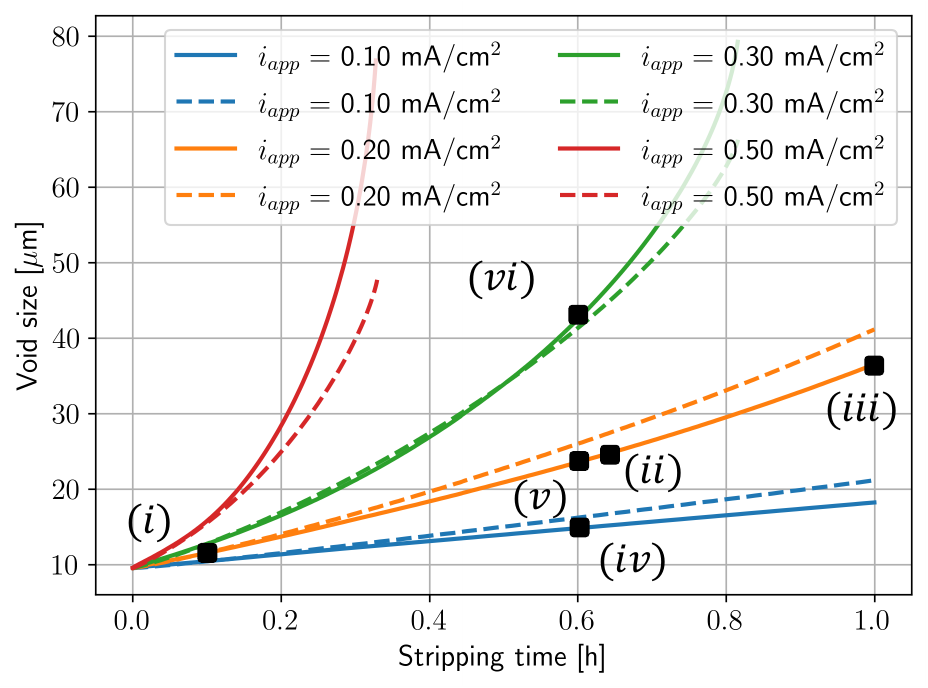}
	\subcaption{}
	\label{FigR6c}
\end{subfigure}
\begin{subfigure}{0.49\textwidth}
	\includegraphics[trim=0 0 0 0, clip, keepaspectratio,width=\linewidth]{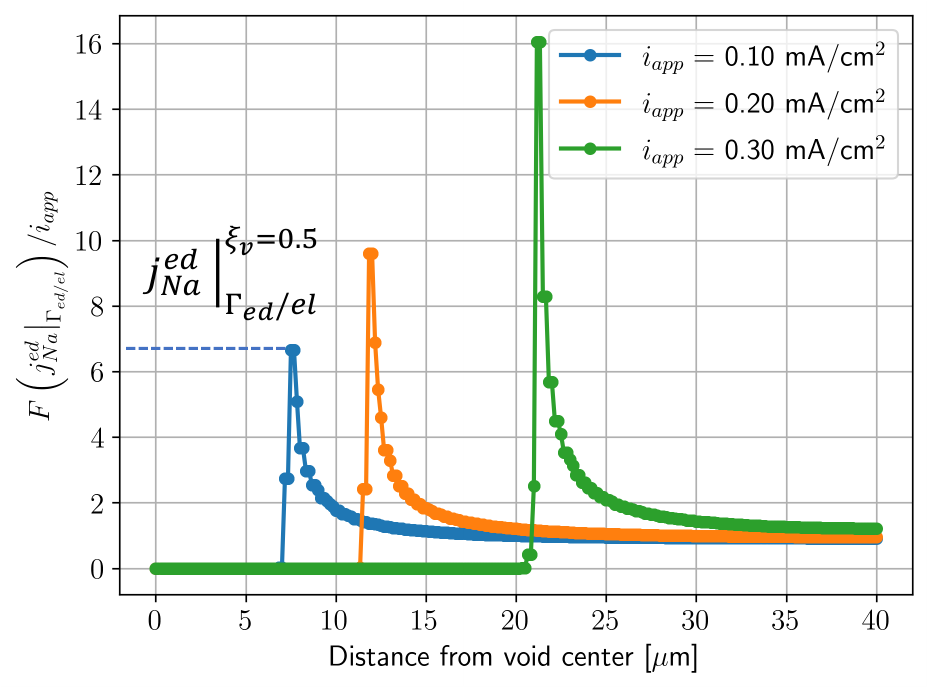}
	\subcaption{}
	\label{FigR6e}
\end{subfigure}
\begin{subfigure}{0.49\textwidth}
	\includegraphics[trim=0 0 0 0, clip, keepaspectratio,width=\linewidth]{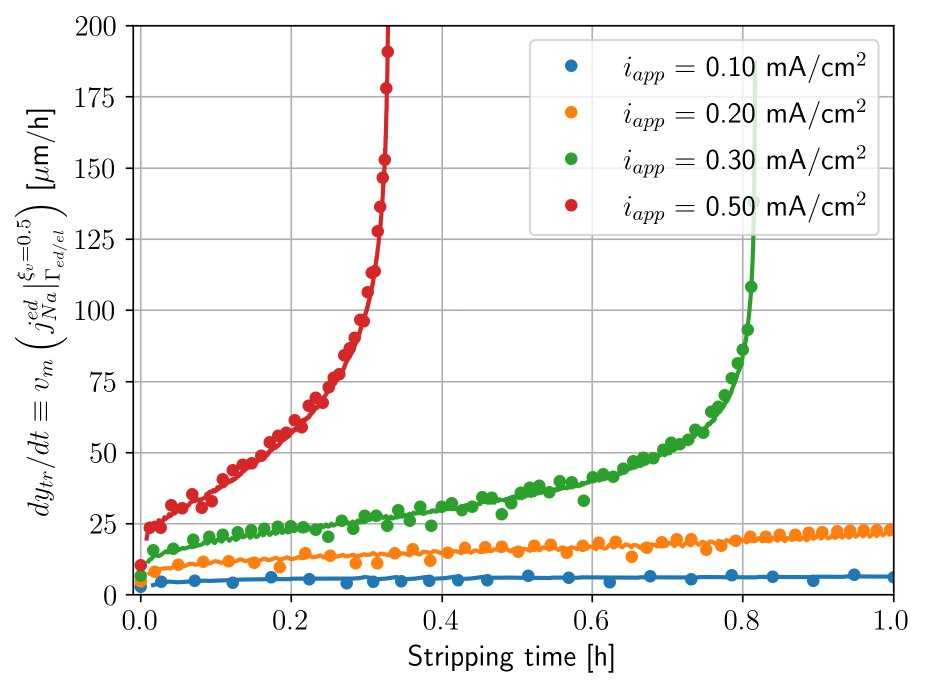}
	\subcaption{}
	\label{FigR6f}
\end{subfigure}
\caption{Analysis of stripping simulations with a single interfacial void at the electrode/electrolyte interface and a homogeneous solid electrolyte separator with different applied current densities. (c) Void size (twice the distance between the void center and edge along the electrode/electrolyte interface) over time. The black markers labelled $(i)$--$(vi)$ correspond to the microstructures shown in Figs.~\ref{FigR6a} and \ref{FigR6b}. (b) The normal component of the Na mass flux along the electrode/solid electrolyte interface normalized with the applied flux, $\left.j_{Na}^{ed}\right|_{\Gamma_{ed/el}}/(i_{app}/F)$, with distance from the void center after $0.6$ h. The dotted horizontal line indicates the void edge flux for $i_{app} = 0.1$ mA/cm$^2$. (c) Void edge velocity calculated using Eq.~\eqref{EqnR5} as a function of stripping time. The solid lines indicate the moving average value over $15$ time steps. In (a), the solid lines indicate the actual void size obtained by explicitly tracking the void edge, while the dotted lines are obtained by time integrating the data in (c).}
\end{figure}

To quantify the void growth rate during stripping, we approximate the void size by doubling the distance from the void edge (the position where $\xi_{v}=0.5$ along the electrode/electrolyte interface) to the void center. Figure \ref{FigR6c} shows the void size with time for different applied current densities. The growth of the void size increases with increasing applied current density, as shown qualitatively in Fig.~\ref{FigR6b}. The void size grows almost linearly with time at first but then accelerates with time. This acceleration is most clear for the larger applied current densities, i.e., $0.3$ and $0.5$ mA/cm$^2$. The theoretical model of Lu \textit{et al.} \cite{lu2022void} assumes a linear void growth rate. However, our results indicate that this assumption may not be accurate. 

To investigate what controls the void growth rate, we plot the spatial distribution of the normal component of the interfacial flux of Na along the electrode/electrolyte interface, i.e., $\left.j^{ed}_{Na}\right|_{x=\Gamma_{ed/el}}$, normalized by the applied flux, i.e., $i_{app}/F$, for three selected applied current densities after stripping for $0.6$ h (see Fig.~\ref{FigR6e}). The interfacial flux is zero along the void, jumps to a maximum at the void edge, and gradually decreases until it equals the applied flux at the electrode/electrolyte interface. The void size and the maximum flux value at the edge increase with increasing applied current density. Based on these results, we assume that the maximum interfacial flux at the void edge controls the rate at which the void elongates along the electrode/electrolyte interface, i.e., $x=\Gamma_{ed/el}$. Specifically,
\begin{align}
\left.\frac{dy_{tr}}{dt}\right|_{\Gamma_{ed/el}} = v_{m}j_{tr}(t),
\label{EqnR5}
\end{align} 
where $\left.dy_{tr}/dt\right|_{\Gamma_{ed/el}}$ is the velocity of the void edge along the electrode/electrolyte interface, $v_m$ is the molar volume of Na, $j_{tr}(t)= \left.j^{ed}_{Na}\right|_{\Gamma_{ed/el}}^{\xi_v=0.5}$ is the flux at the edge. The superscript $\xi_v=0.5$ indicates the value of the void phase-field variable used to approximate the position of the diffuse void/electrode interface.

Figure~\ref{FigR6f} depicts the velocity of the void edge obtained using Eq.~\eqref{EqnR5} as a function of time for different applied current densities. The velocities are close to linear for a time and then rapidly accelerate for the larger applied current densities, and we expect similar behavior would be seen with the lower applied current densities if longer times were simulated. We integrate this data with time to estimate the change in the void size and compare it with the actual void size, as shown in Fig.~\ref{FigR6c}. The approximate void sizes obtained by integrating Eq.~\eqref{EqnR5} compare reasonably well with the actual sizes, though the differences are larger for high applied current densities and stripping durations $\geq0.4$ h. One possible explanation for this deviation could be that Eq.~(\ref{EqnR5}) does not consider the contact angle between the void, metal electrode and the SE separator. Thus, these results suggest an improved approach to predicting the void growth than that used by Lu \emph{et al.} \cite{lu2022void}, but a more quantitative analytical model is still needed to predict the void growth rate, which is beyond the scope of this work. 

We also simulate the shrinkage of a single void at the electrode/electrolyte interface during plating for 1 h (or once the void completely disappears) for different applied current densities. Like stripping, a semi-circular-shaped void of radius $4.8$ $\mu$m is placed initially at the electrode/electrolyte interface. The initial width of the Na negative electrode is assumed to be $28.6$ $\mu$m in all our simulations. 

\begin{figure}[tbp]
\centering
\begin{subfigure}{0.86\textwidth}
	\includegraphics[trim=0 0 0 0, clip, keepaspectratio,width=\linewidth]{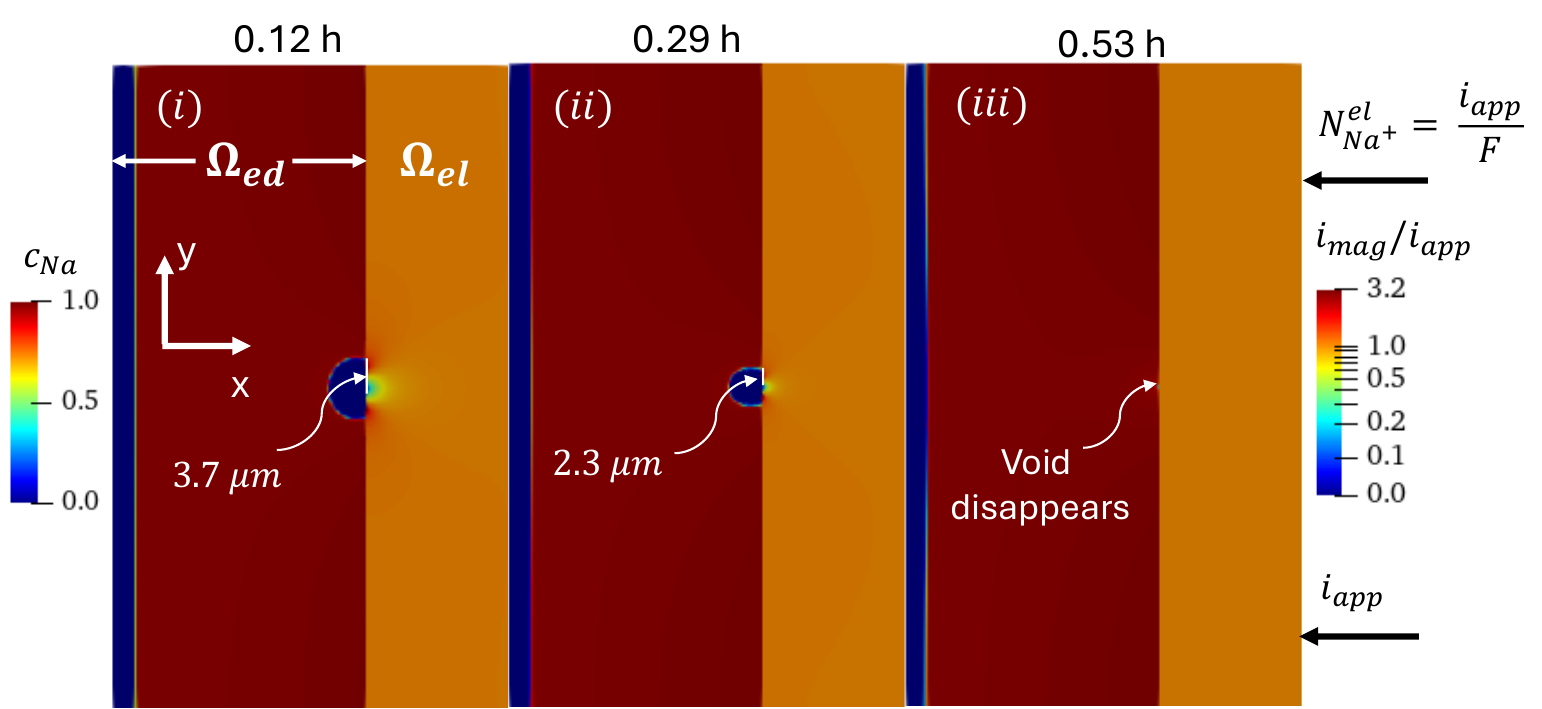}
	\subcaption{}
	\label{FigR7a}
\end{subfigure}
\begin{subfigure}{0.86\textwidth}
	\includegraphics[trim=0 0 0 0, clip, keepaspectratio,width=\linewidth]{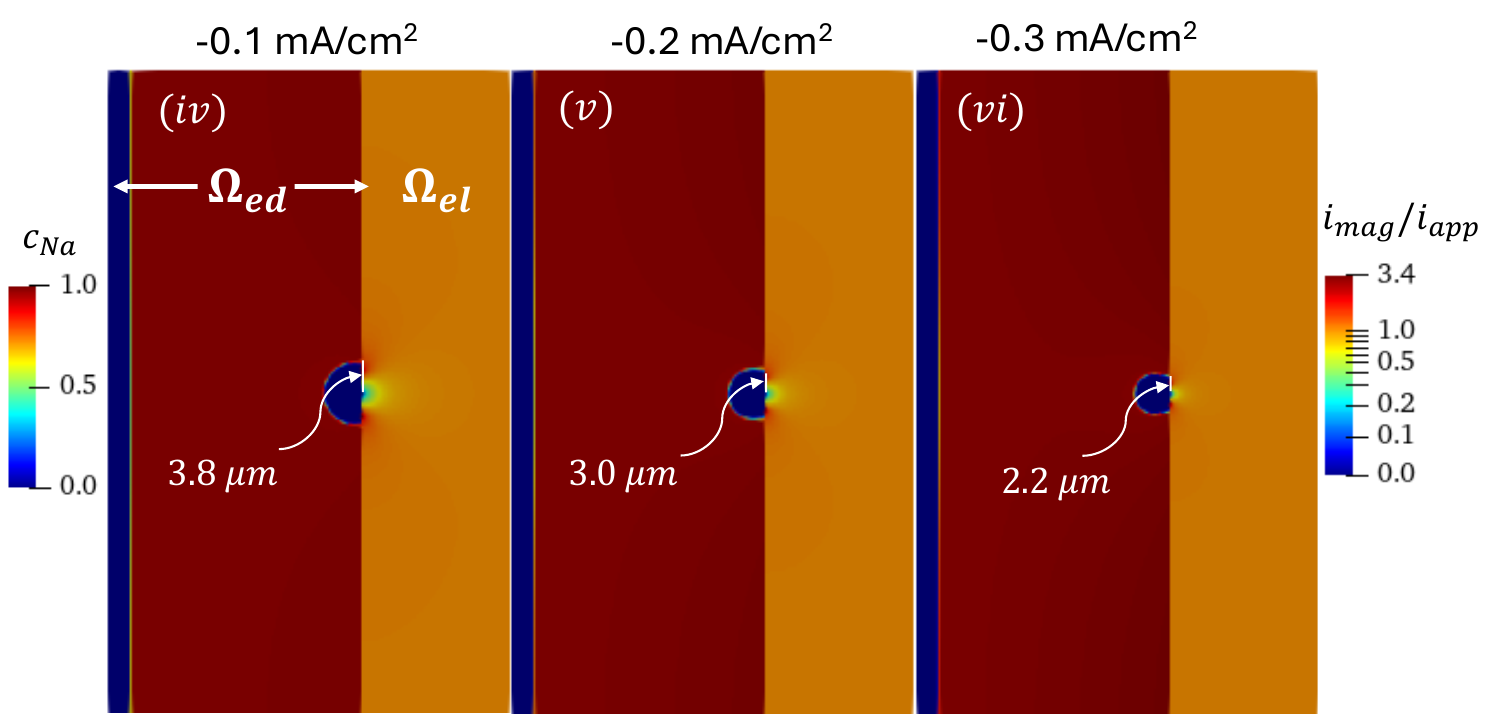}
	\subcaption{}
	\label{FigR7b}
\end{subfigure}
\caption{Plating simulations with a single interfacial void at the electrode/electrolyte interface and a homogeneous solid electrolyte separator. (a) Void evolution with an applied current density of $-0.2$ mA/cm$^2$. (b) Void morphology for different applied current densities after plating for 0.2 h. $\Omega_{ed}$ is shaded by the Na concentration and $\Omega_{el}$ is shaded by the ratio of the local current density to the applied current density.}
\end{figure}

Figure \ref{FigR7a} shows that the interfacial void preferentially shrinks along the electrode/electrolyte interface during plating for an applied current density of $-0.2$ mA/cm$^2$, which is consistent with experiments \cite{kasemchainan2019critical}, \cite{spencer2019sodium} and simulations in Li/LLZO systems \cite{zhao2022phase}. However, unlike Zhao \textit{et al.} \cite{zhao2022phase}, our plating simulations indicate that a single void disappears at the end of the plating cycle instead of forming occluded voids, as also observed in the experiments \cite{spencer2019sodium}. One possible reason could be that our system and initial void sizes are smaller than those from Zhao's simulations \cite{zhao2022phase} (system size and initial void radius in \cite{zhao2022phase} is 80 $\times$ 250 $\mu$m$^2$ and 10 $\mu$m, respectively). In addition to void shrinkage, Fig.~\ref{FigR7a} shows that the negative electrode grows during plating (see also Supplementary Video S$6$ in the Supporting Information). This growth, however, is limited because the simulation ends after a short time, approximately in $\leq 0.6$ h, i.e., when the void disappears. Once the void disappears, the behavior should be identical to the plating case without a void (Fig.~\ref{Fig4}). 

The applied current density has a large impact on the void evolution rate during plating, as shown in Fig.~\ref{FigR7b}. Although the void evolution is qualitatively similar in all cases, the final void size after plating for $0.2$ h decreases with increasing magnitude of the applied current density. This suggests that the void shrinkage rate increases with increasing magnitude of the applied current density.

\begin{figure}[tbp]
\centering
\begin{subfigure}{0.6\textwidth}
	\includegraphics[trim=0 0 0 0, clip, keepaspectratio,width=\linewidth]{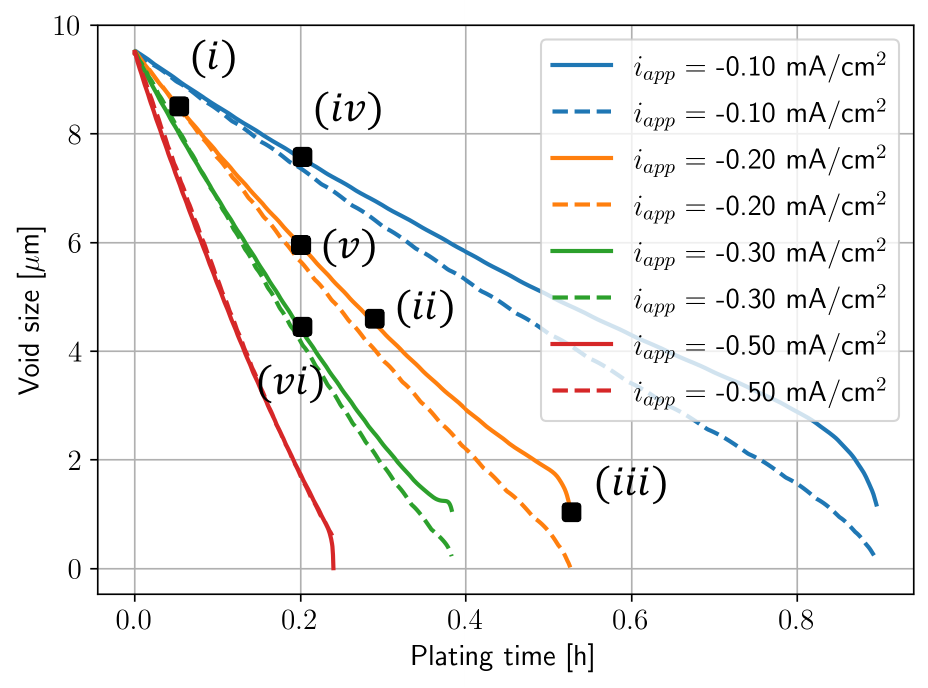}
	\subcaption{}
	\label{FigR7c}
\end{subfigure}
\begin{subfigure}{0.49\textwidth}
	\includegraphics[trim=0 0 0 0, clip, keepaspectratio,width=\linewidth]{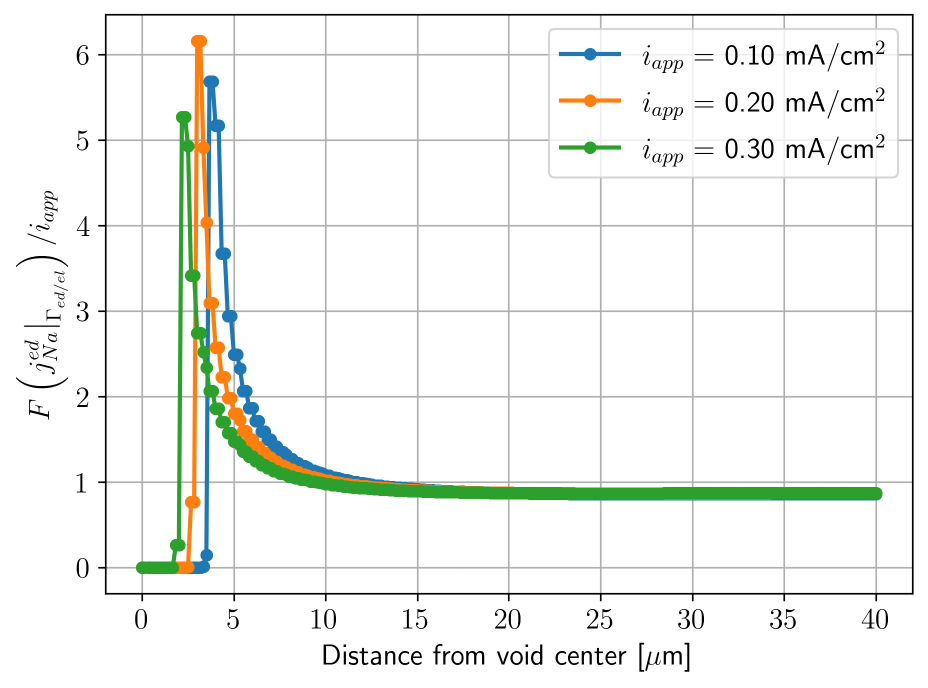}
	\subcaption{}
	\label{FigR7d}
\end{subfigure}
\begin{subfigure}{0.49\textwidth}
	\includegraphics[trim=0 0 0 0, clip, keepaspectratio,width=\linewidth]{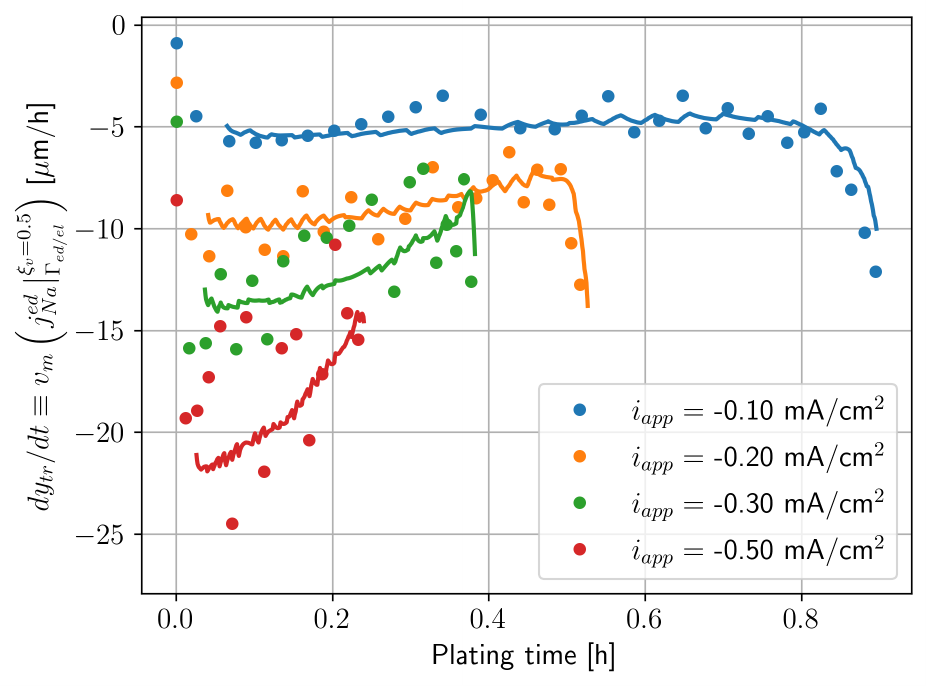}
	\subcaption{}
	\label{FigR7e}
\end{subfigure}
\caption{Analysis of plating simulations with a single interfacial void at the electrode/electrolyte interface and a homogeneous solid electrolyte separator. (a) Void size (twice the distance between the void center and edge along the electrode/electrolyte interface) over time. The black markers labelled $(i)$--$(vi)$ correspond to the microstructures shown in Fig.~\ref{FigR7a} and ~\ref{FigR7b}. (b) The normal component of the Na mass flux along the electrode/solid electrolyte interface normalized with the applied flux, $\left.j_{Na}^{ed}\right|_{\Gamma_{ed/el}}/(i_{app}/F)$, with distance from the void center after $0.2$ h. (c) Void edge velocity calculated using Eq.~\eqref{EqnR5} as a function of plating time. The solid lines indicate the moving average value over $40$ time steps. In (a), the solid lines indicate the actual void size obtained by explicitly tracking the void edge, while the dotted lines are obtained by time integrating the data in (c).}
\end{figure}

We further analyse the void shrinkage during plating by plotting the void size with time, similar to our void growth analysis during stripping (see Fig.~\ref{FigR7c}). The void sizees linearly with time until the void disappears, and the shrinkage rate increases with the magnitude of the applied current density. As with stripping, we attribute this behaviour to the rate at which the flux of Na at the void edge varies with time. Figure \ref{FigR7d} shows the normalized flux of Na along the electrode/electrolyte interface for three applied current densities after plating for $0.2$ h. Like with stripping, the Na flux is zero along the void, it then peaks at the edge, and gradually decreases to the applied flux near the top of the electrode/electrolyte interface. Note that after normalization, the flux at the edge does not increase proportionally with the applied current density. 

We again calculate the velocity of the void edge using Eq.~(\ref{EqnR5}) by determining the Na flux at the edge, $j_{tr} (t)$, as shown in Fig \ref{FigR7e}. The velocity is negative because the void size decreases with plating time. The magnitude of the edge velocity, and thus, the Na flux at the edge, increases with increasing magnitude of the applied current density. Subsequently, we estimate the void size over time by integrating velocity and the results are shown in Fig.~\ref{FigR7c} with dotted lines. Our estimated void sizes are in reasonable agreement with the actual void sizes, though the deviation increases with increasing applied current density. This result further supports our argument that the void migration rate is directly proportional to the Na flux at the edge.
\subsubsection{Cyclic simulations with polycrystalline solid electrolyte (SE) separators}
\label{secR2.2}
Next, we perform cyclic simulations to investigate the impact of SE GBs on void evolution. The cyclic simulations involve stripping for $1$ h and then plating for $1$ h at an applied cycling current density of $0.2$ mA/cm$^2$. Similar to Section \ref{RS1.3}, polycrystalline SEs with low-conductivity and high-conductivity GBs are considered. The initial interfacial void size and thickness of the negative electrode are identical to the cases discussed in Section \ref{secR2.1} for all simulations.

\begin{figure}[tbp]
\centering
\begin{subfigure}{0.88\textwidth}
	\includegraphics[trim=0 0 0 0, clip, keepaspectratio,width=\linewidth]{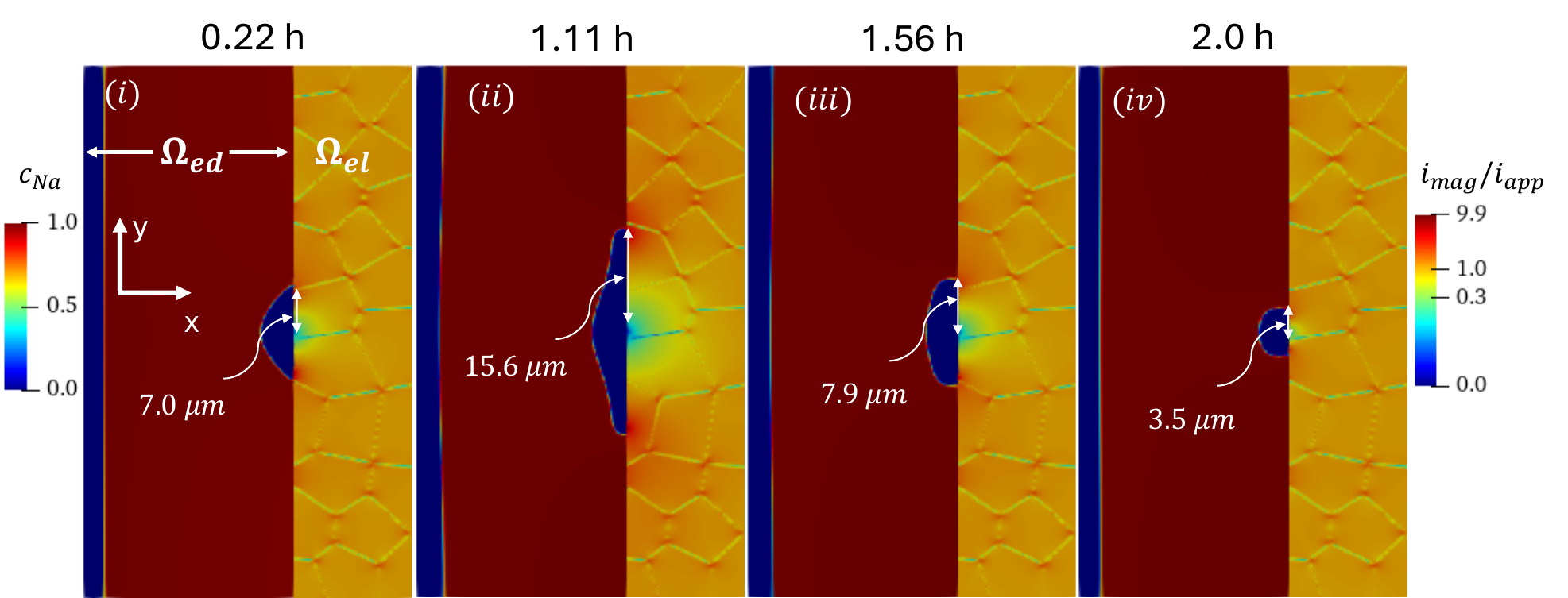}
	\subcaption{}
	\label{FigR8a}
\end{subfigure}
\begin{subfigure}{0.88\textwidth}
	\includegraphics[trim=0 0 0 0, clip, keepaspectratio,width=\linewidth]{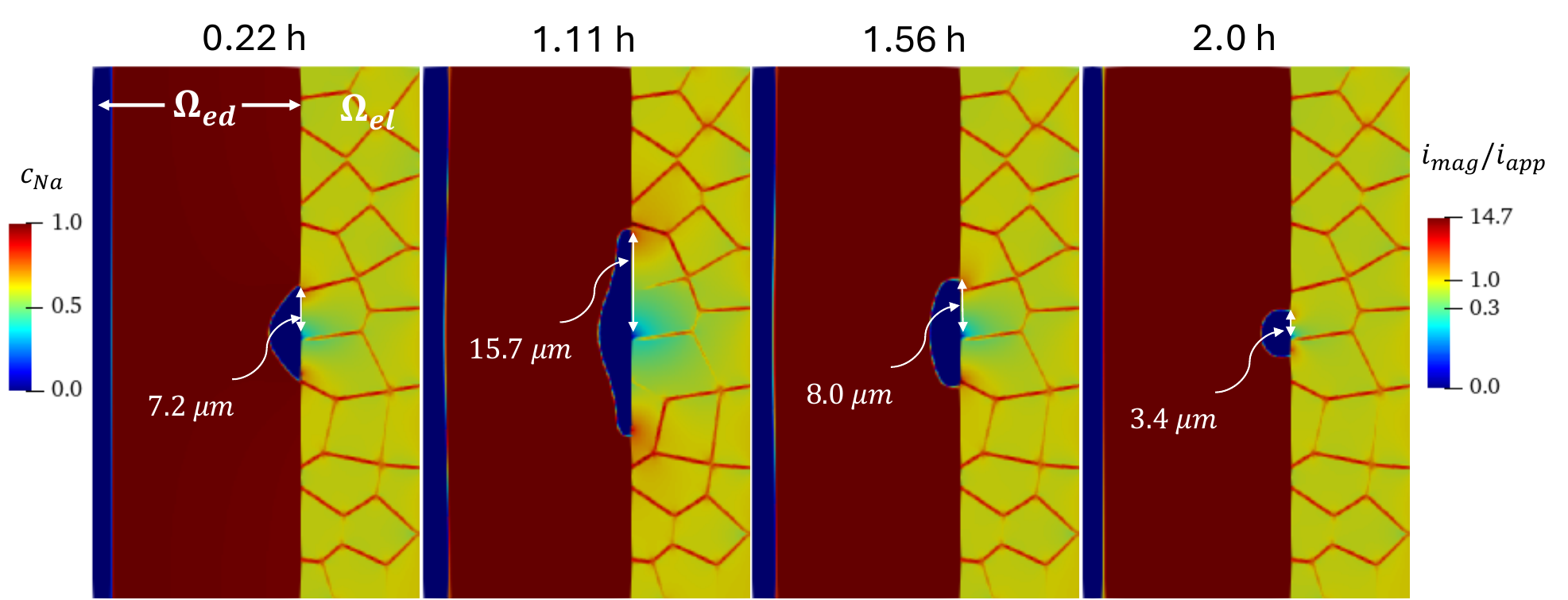}
	\subcaption{}
	\label{FigR8b}
\end{subfigure}
\caption{Cycling simulations with a single interfacial void at the interface between the negative electrode and a polycrystalline SE separator. Evolution of the void and the negative electrode during cell cycling in solid electrolytes with (a) low-conductivity and (b) high-conductivity GBs under an applied current density of $0.2$ mA/cm$^2$. $\Omega_{ed}$ is shaded by the Na concentration and $\Omega_{el}$ is shaded by the ratio of the local current density to the applied current density.}
\end{figure}

\begin{figure}[tbp]
\centering
\begin{subfigure}{0.48\textwidth}
	\includegraphics[trim=0 0 0 0, clip, keepaspectratio,width=\linewidth]{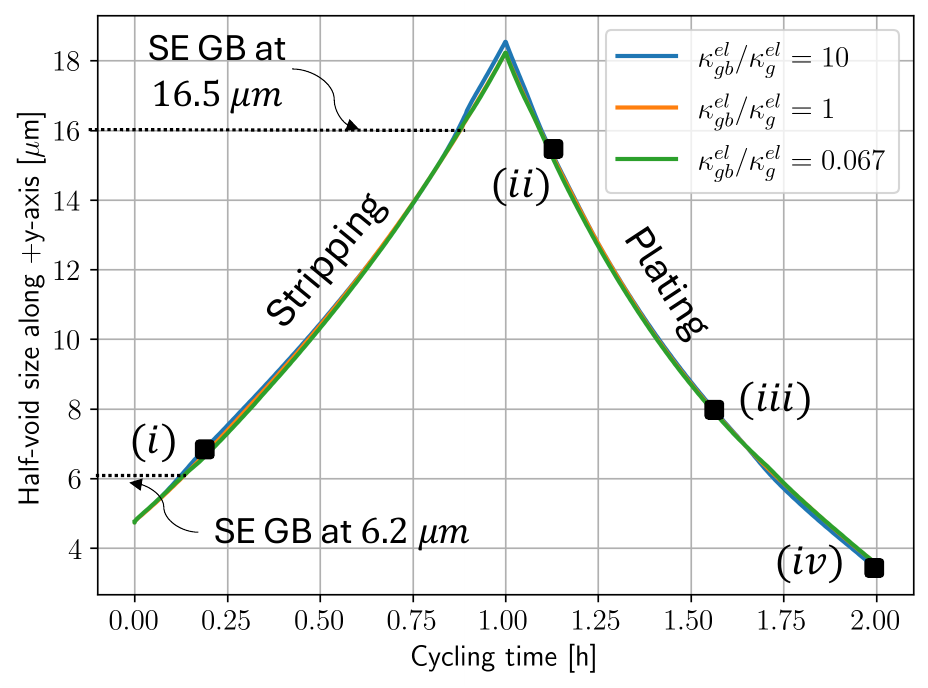}
	\subcaption{}
	\label{FigR8c}
\end{subfigure}
\begin{subfigure}{0.48\textwidth}
	\includegraphics[trim=0 0 0 0, clip, keepaspectratio,width=\linewidth]{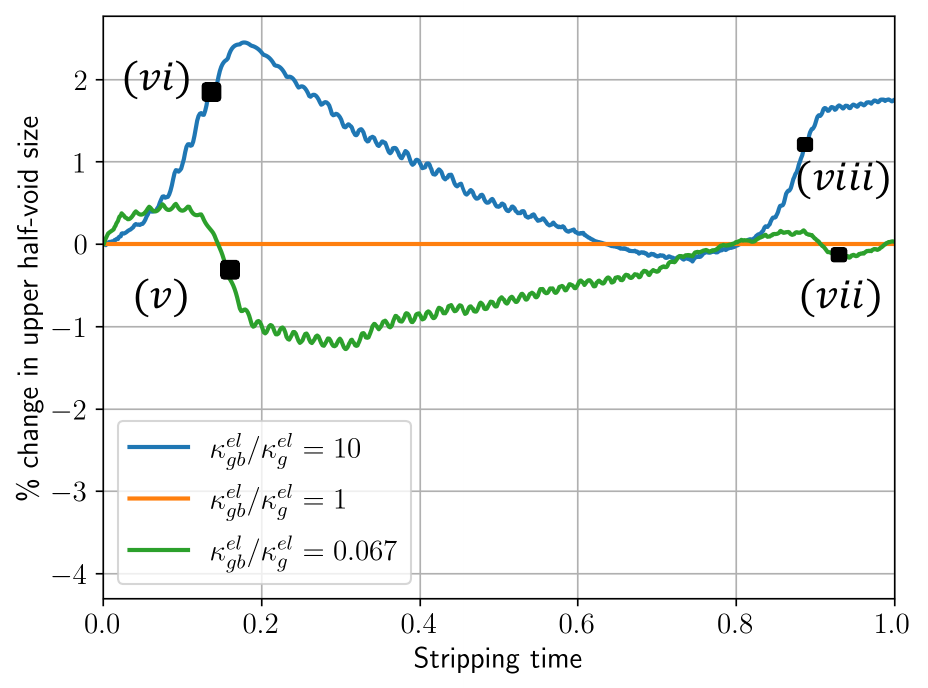}
	\subcaption{}
\label{FigR8d}
\end{subfigure}
\begin{subfigure}{0.48\textwidth}
	\includegraphics[trim=0 0 0 0, clip, keepaspectratio,width=\linewidth]{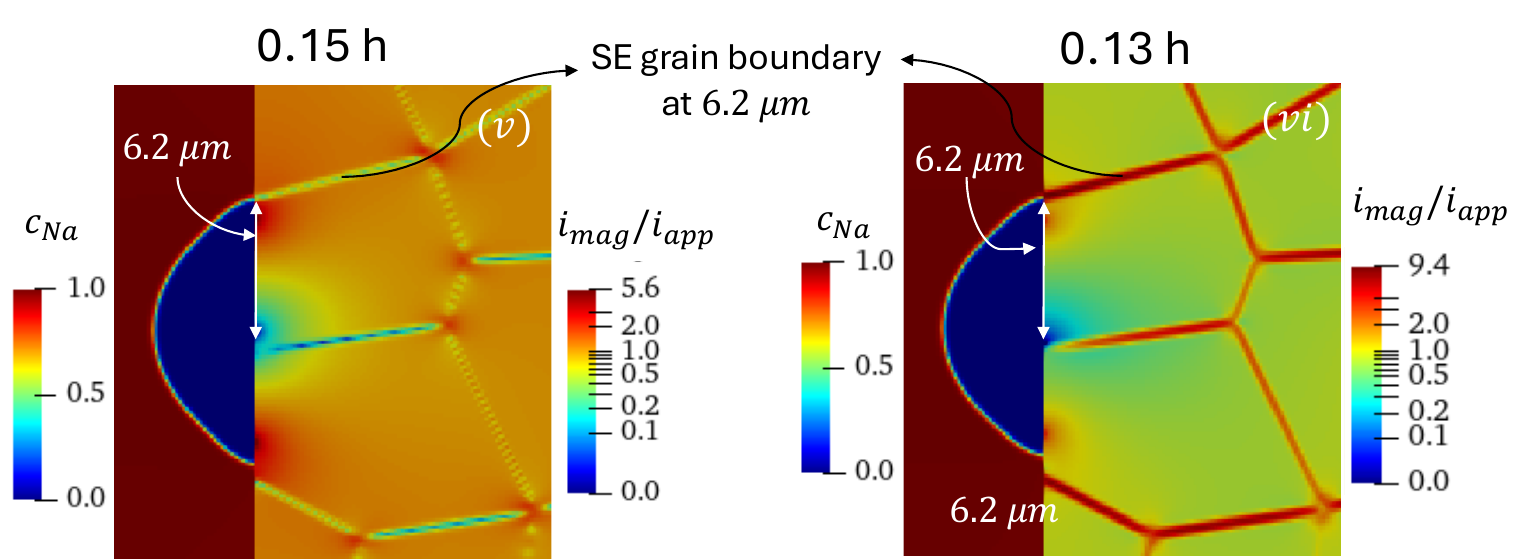}
	\subcaption{}
	\label{FigR8e}
\end{subfigure}
\begin{subfigure}{0.48\textwidth}
	\includegraphics[trim=0 0 0 0, clip, keepaspectratio,width=\linewidth]{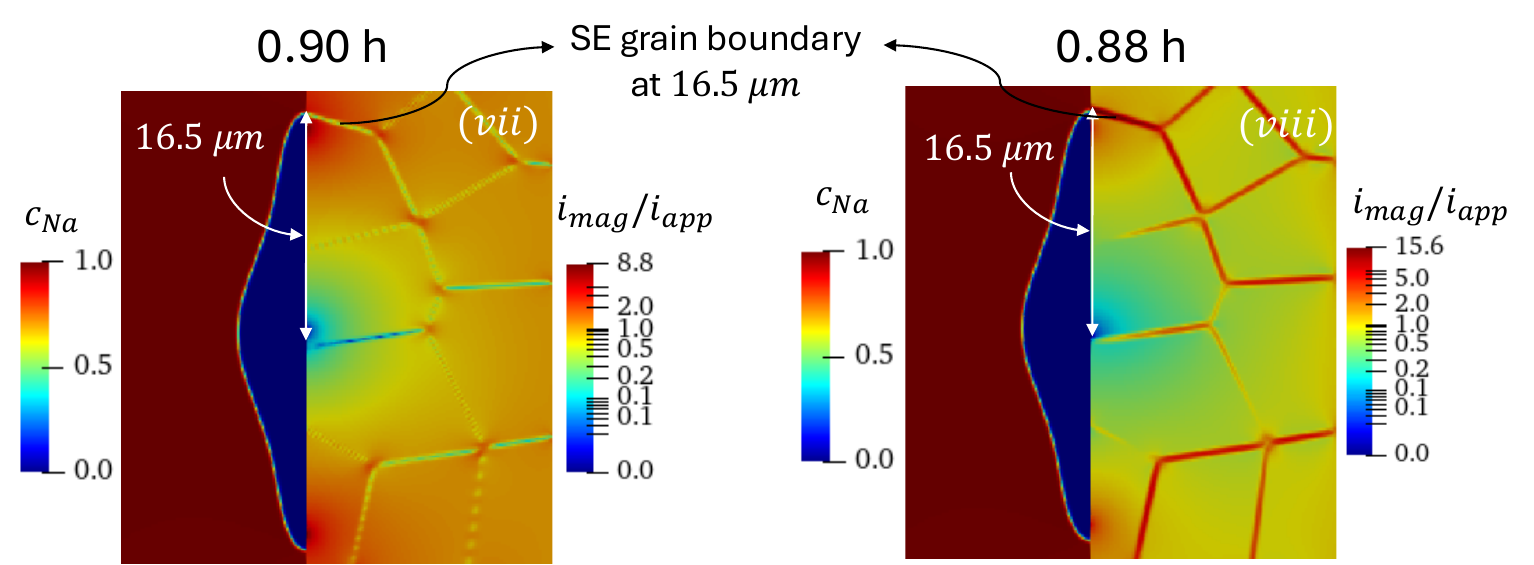}
	\subcaption{}
	\label{FigR8f}
\end{subfigure}
\begin{subfigure}{0.48\textwidth}
	\includegraphics[trim=0 0 0 0, clip, keepaspectratio,width=\linewidth]{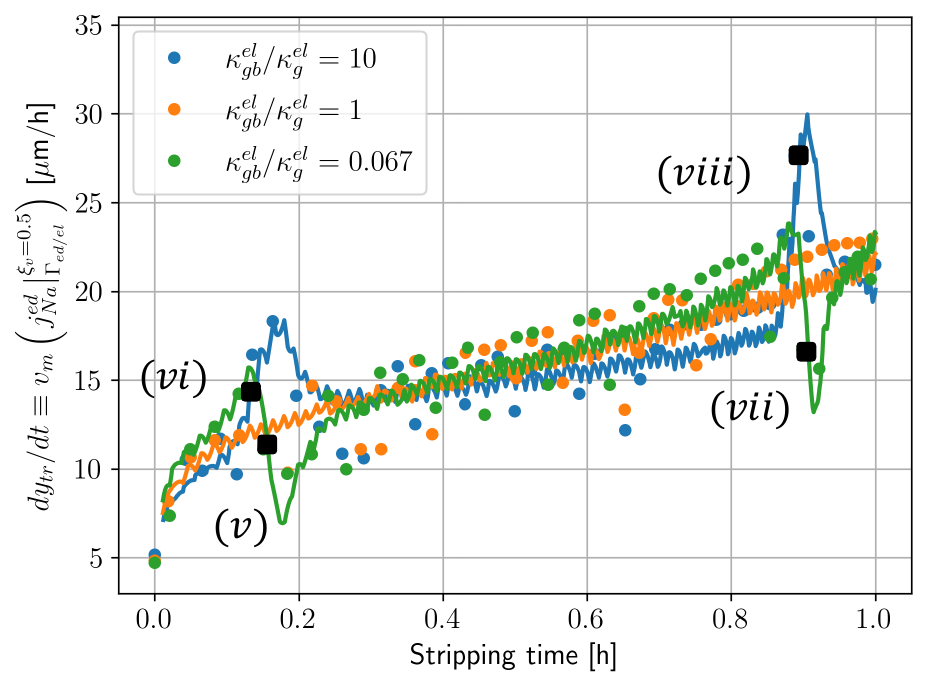}
	\subcaption{}
	\label{FigR8_velocity}
\end{subfigure}
\begin{subfigure}{0.48\textwidth}
	\includegraphics[trim=0 0 0 0, clip, keepaspectratio,width=\linewidth]{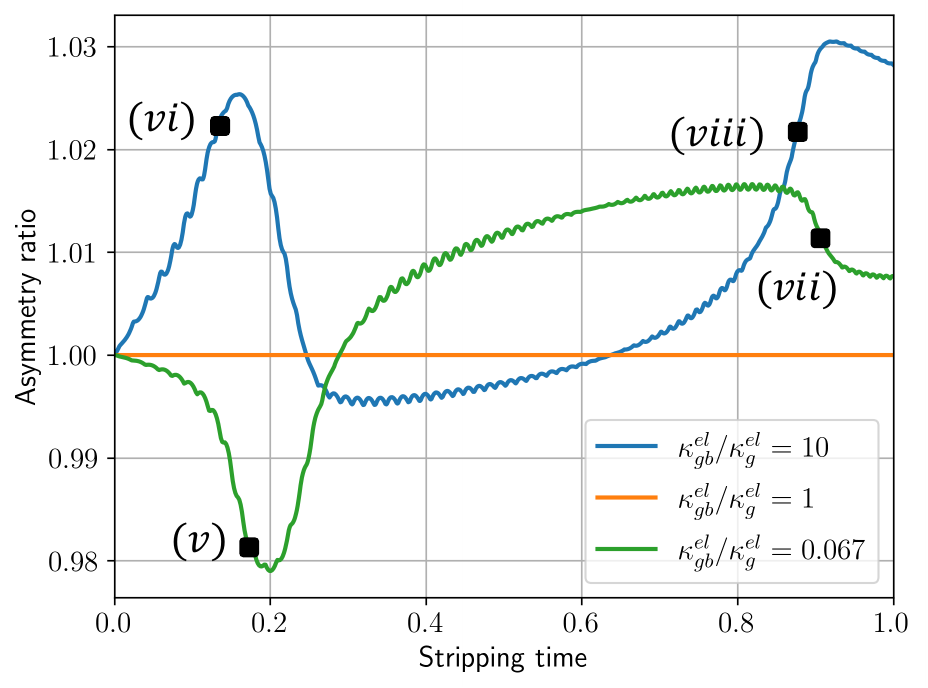}
	\subcaption{}
	\label{FigR8_asymmetry}
\end{subfigure}
\caption{Analysis of cycling simulations with a single interfacial void at the interface between the negative electrode and a polycrystalline SE separator. (a) Half-void size above the void center with cycling time with low-conductivity ($\kappa_{gb}^{el}/\kappa_{g}^{el}=0.067$), homogeneous ($\kappa_{gb}^{el}/\kappa_{g}^{el}=1.0$), and high-conductivity ($\kappa_{gb}^{el}/\kappa_{g}^{el}=10$) GBs. The black markers labelled as $(i)$--$(iv)$ correspond to the microstructures shown in Fig.~\ref{FigR8a}. (b) Percentage change in upper half-void size relative to the homogeneous SE as a function of stripping time for $\kappa_{gb}^{el}/\kappa_{g}^{el}=0.067$ and $10$. The black markers labelled $(v)$--$(viii)$ in this figure correspond to the microstructures shown in Figs. \ref{FigR8e} and \ref{FigR8f}. Zoomed-in image of the simulation domain when the void edge touches a SE GB at a distance of (c) $6.2$ $\mu$m and (d) 
$16.5$ $\mu$m above the void centre. (e) Void edge velocity using Eq.~\eqref{EqnR5} as a function of stripping time for different GB-to-grain conductivity ratios. The solid lines indicate the moving average value of the last $15$ time steps. (f) The asymmetry ratio, calculated from the ratio of the distances between the void center and the upper and lower void edges, as a function of stripping time.}
\end{figure}

Figures \ref{FigR8a} and \ref{FigR8b} show the evolution of a single interfacial void during cycling in contact with a SE having low-conductivity and high-conductivity GBs, respectively. The void evolution is almost identical in both cases, i.e., the interfacial void elongates along the electrode/electrolyte interface during stripping for $1$ h, which leads to current constriction in the SE separator near the void edges. Subsequently, the void shrinks during plating. Although the Na electrode thickness also varies during cycling, the net change in thickness is small and not apparent in Figs. \ref{FigR8a} and \ref{FigR8b}. The void size increases during stripping and decreases during plating, and is not significantly changed by the GB conductivity (see Fig.~\ref{FigR8c}). This is most clear when we plot the percentage change in the void size compared to the case with a homogeneous SE, shown in Fig.~\ref{FigR8c} for the stripping half (see Fig.~\ref{Fig1a_SI_secS6} in Section \ref{SI_secS6} of the Supporting Information for the full cycling plot). The largest deviation is just over $2\%$. Thus, the impact of SE GB conductivity and diffusivity on the overall void evolution is negligible. 

While the impact on the overall void evolution is negligible, there are local effects of the GB conductivity on the void evolution. Figure \ref{FigR8d} shows that the void size evolution tends to be faster than the homogeneous case for high conductivity GBs and slower for low conductivity GBs. In addition, the deviation from the homogeneous case changes as the void edges move past GBs. Thus, the deviations in the void growth result from interactions between the void edge and SE GBs that intersect the electrode/electrolyte interface. Figures \ref{FigR8e} and \ref{FigR8f} show the two instances when a void edge interacts with SE GBs for the low-conductivity and high-conductivity GB cases, respectively. The high current density region at the edges of the void interact with the GBs. For low conductivity GBs, the velocity of the void edge increases as it approaches and then quickly drops to significantly below the velocity of the homogeneous case as it nears and then passes the GB (see Fig.~\ref{FigR8_velocity}). The velocity then increases as it moves away from the GB. For high conductivity GBs, the velocity increases as the void edge approaches the GB and then decreases as it moves away (again, see Fig.~\ref{FigR8_velocity}). These changes in velocity occur at both the upper and lower edges of the GB and result in slight asymmetry in the shape of the void, as shown in a plot of the ratio between the distance from the void center to the upper void edge and the distance from the void center to the lower void edge from Fig.~\ref{FigR8_asymmetry}.

To further support our argument that GB conductivity locally affects void evolution, we have performed additional simulations with bi-crystal and tri-crystal SEs (see Section \ref{SI_secS7} of the Supporting Information). These simulations also suggest that the maximum deviation observed is around $6\%$.

\subsection{Multiple voids at the electrode/solid electrolyte interface}
\subsubsection{Stripping simulations with a homogeneous solid electrolyte (SE)}
\label{secR3.1}
\begin{figure}[tbp]
\centering
\begin{subfigure}{0.49\textwidth}
	\includegraphics[trim=0 0 0 0, clip, keepaspectratio,width=\linewidth]{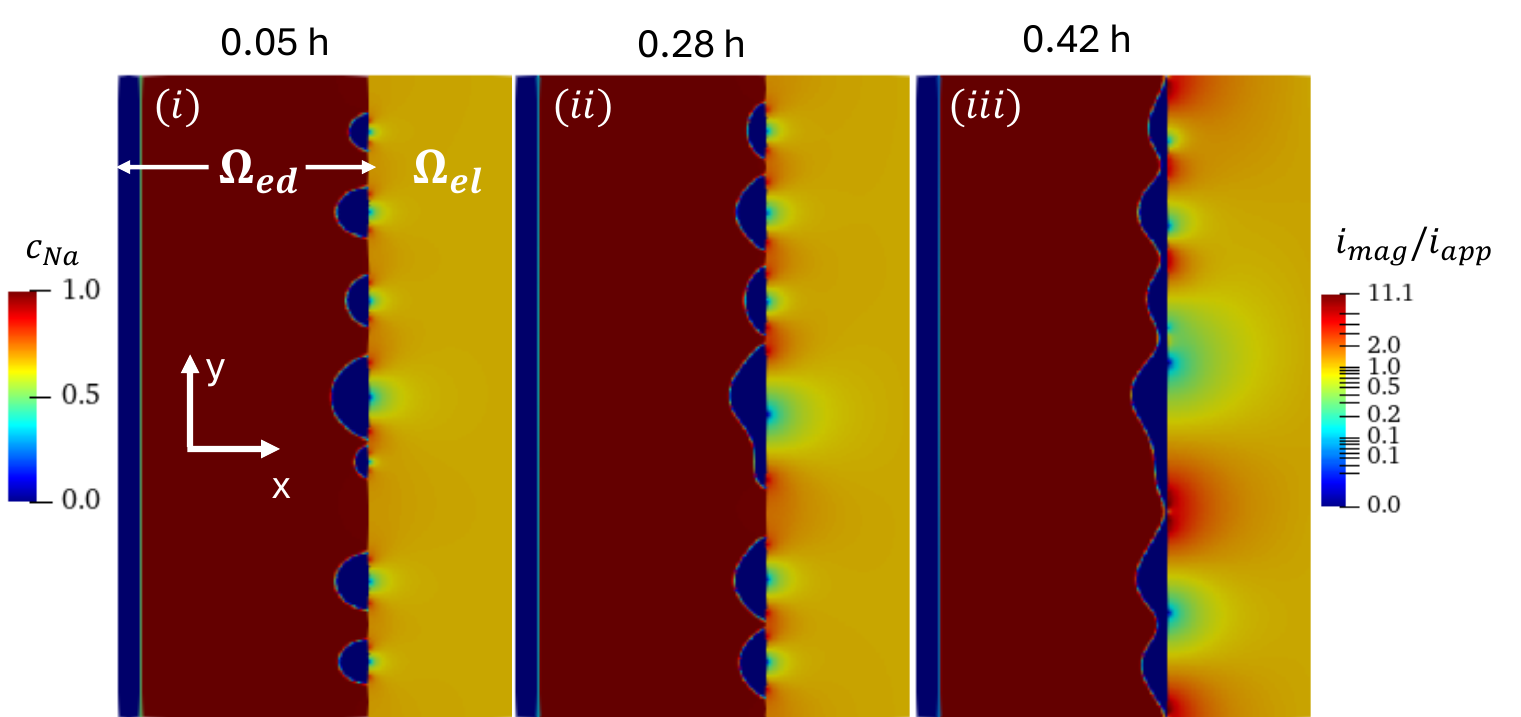}
	\subcaption{}
	\label{FigR10a}
\end{subfigure}
\begin{subfigure}{0.49\textwidth}
	\includegraphics[trim=0 0 0 0, clip, keepaspectratio,width=\linewidth]{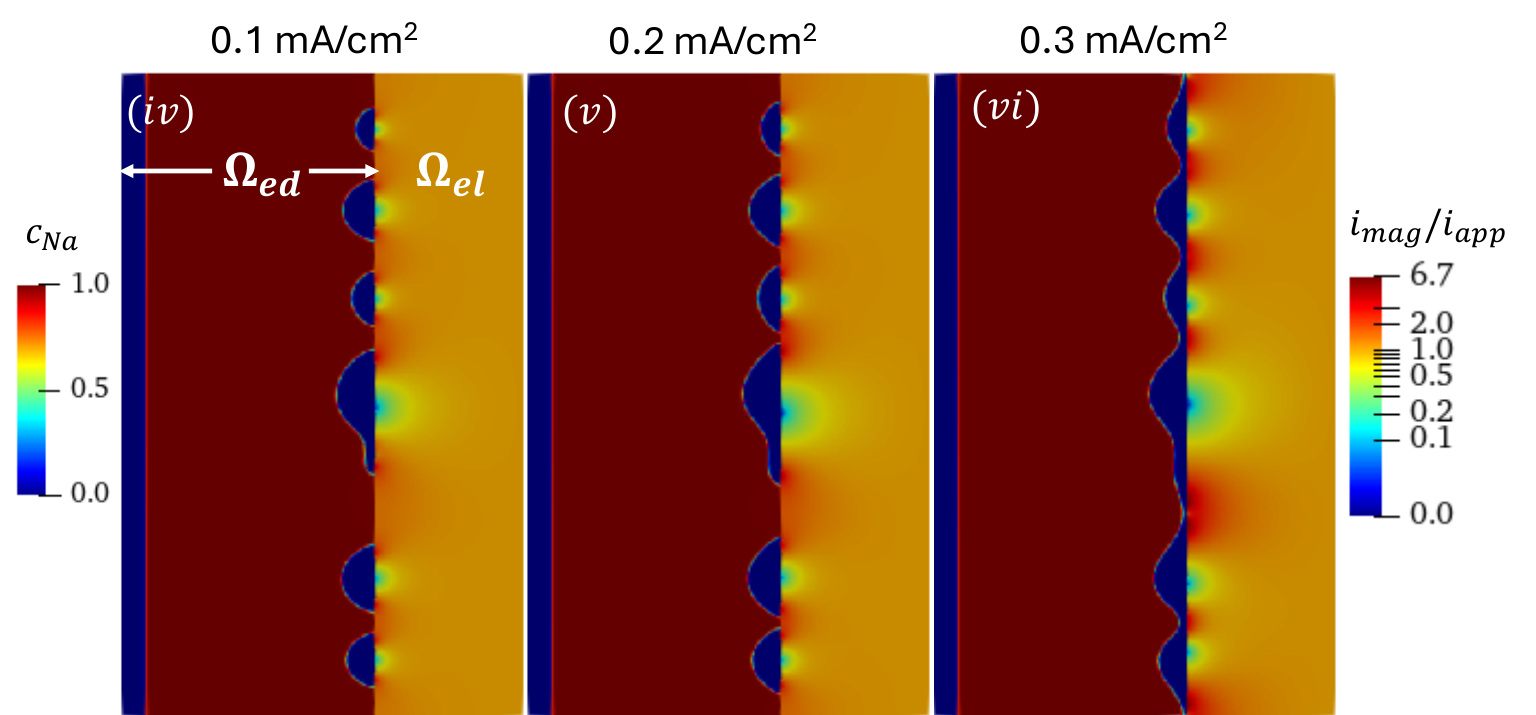}
	\subcaption{}
	\label{FigR10b}
\end{subfigure}
\begin{subfigure}{0.49\textwidth}
	\includegraphics[trim=0 0 0 0, clip, keepaspectratio,width=\linewidth]{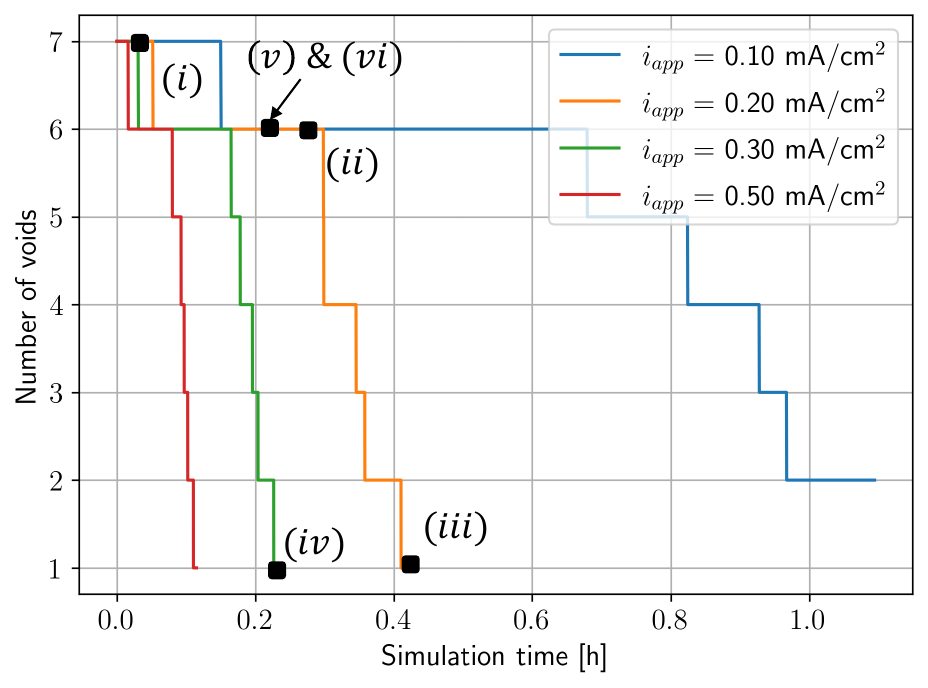}
	\subcaption{}
	\label{FigR10c}
\end{subfigure}
\begin{subfigure}{0.49\textwidth}
	\includegraphics[trim=0 0 0 0, clip, keepaspectratio,width=\linewidth]{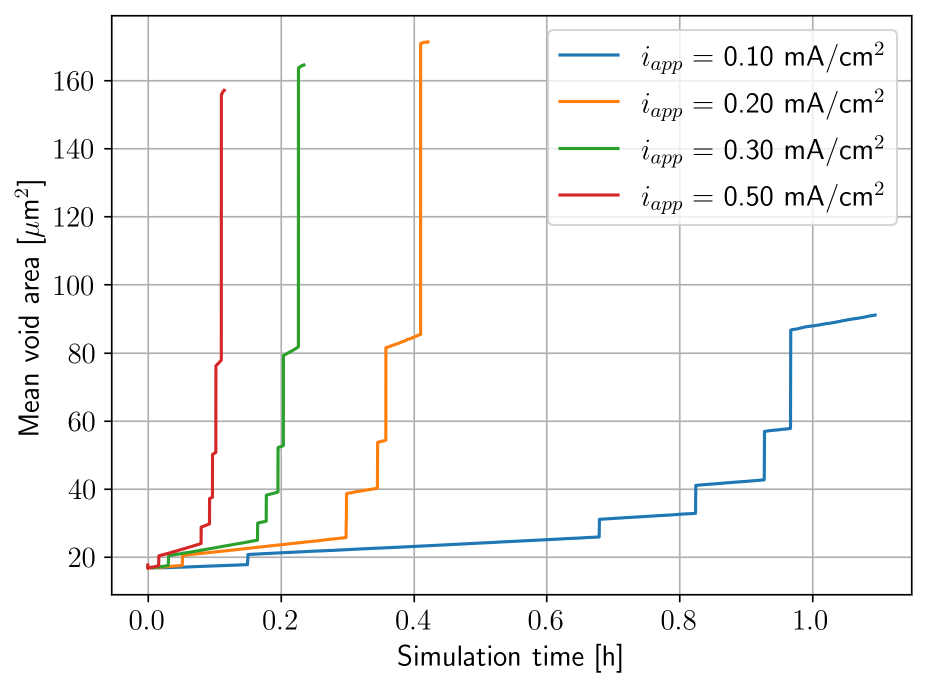}
	\subcaption{}
	\label{FigR10d}
\end{subfigure}
\begin{subfigure}{0.49\textwidth}
	\includegraphics[trim=0 0 0 0, clip, keepaspectratio,width=\linewidth]{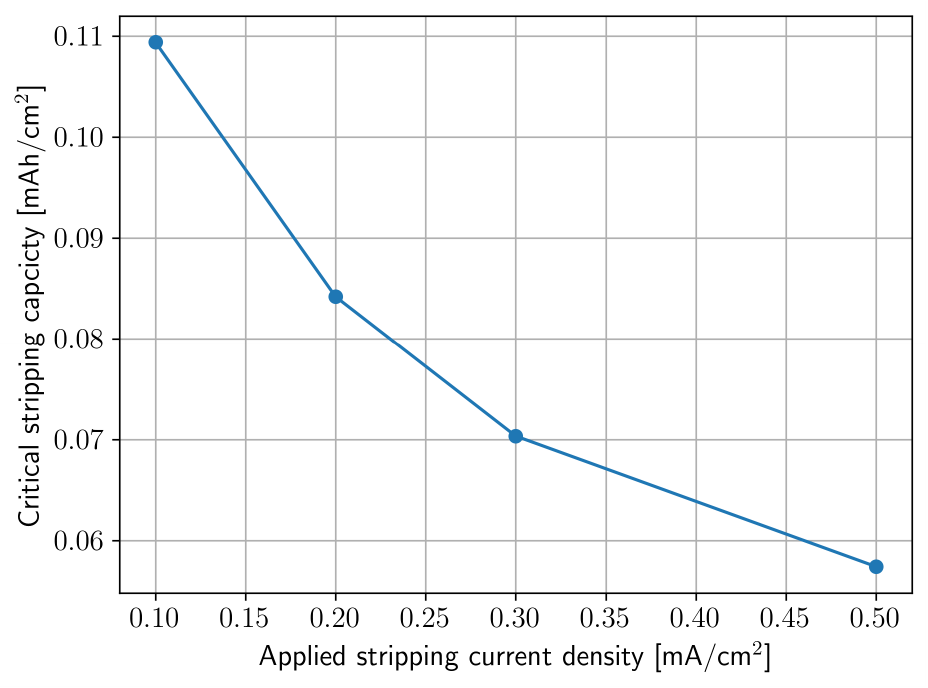}
	\subcaption{}
	\label{FigR10e}
\end{subfigure}
\caption{Stripping simulations with multiple interfacial voids at the electrode/electrolyte interface and a homogeneous solid electrolyte separator for different applied current densities. (a) Void morphology at three times with an applied current density of $0.2$ mA/cm$^2$. $\Omega_{ed}$ is shaded by the Na concentration and $\Omega_{el}$ by the normalized current density. (b) Void morphology after stripping for $0.23$ h with different applied current densities. Variation with time in the (c) number of voids and (d) mean void area. In (c), the black markers labelled $(i)$--$(vi)$ correspond to the microstructures shown in (a) and (b). (d) Critical stripping capacity as a function of applied current density.}
\end{figure}

It has been experimentally observed \cite{kasemchainan2019critical,spencer2019sodium} that contact loss between the electrode and SE during cycling occurs due to the growth, accumulation and coalescence of multiple interfacial voids. So, we investigate the time to contact loss and the coalescence rate in simulations of multiple interfacial voids during stripping. Similar to Zhao \textit{et al.} \cite{zhao2022phase}, we initialize our simulations with seven semi-circular-shaped voids of different radii at the electrode/electrolyte interface, as shown in Fig.~\ref{FigR10a}. We assume an initial negative electrode thickness of $28.6$ $\mu$m. The simulations run until complete contact loss occurs between the electrode and the SE separator.

Figure \ref{FigR10a} shows the evolution of interfacial voids during stripping with an applied current density of $0.2$ mA/cm$^2$. As stripping progresses, the void area increases while the number of voids decreases due to coalescence until complete contact loss occurs. Multiple current hot spots develop in the SE separator near the void edges with increasing contact loss. Once complete contact loss occurs the flux of Na/Na$^{+}$ leaving the electrode/electrolyte interface becomes effectively zero, which hinders further shrinkage of the negative electrode. 

The applied current density directly impacts the void coalescence rate during stripping. Figure \ref{FigR10b} shows the void morphology after stripping for $0.23$ h with different applied current densities. Although the voids appear qualitatively similar after stripping for the same duration, the coalescence rate and void area increase with increasing applied current density.

To quantify the coalescence rate, we calculate the number of voids and the mean void area, i.e., the total void area divided by the number of voids, as a function of time, as shown in Figs.~\ref{FigR10c} and \ref{FigR10d}, respectively. The decrease in the number of voids due to coalescence accelerates with increasing applied current density, and the growth of the void area increases with increasing applied current density. Thus, the time to complete contact loss decreases with increasing applied current density. To quantify this, following Agier \textit{et al.} \cite{agier2022void}, we calculate the critical stripping capacity $C_{crit} = i_{app}t_{f}$, which is the product of applied current density and the time to contact loss $t_f$ (see Fig.~\ref{FigR10d}). The critical stripping capacity decreases with increasing applied current density, which is in qualitative agreement with experiments \cite{krauskopf2019toward} and simulations in Li/LLZO systems \cite{agier2022void,barai2024study}. However, the predicted critical stripping capacity is almost one-tenth of the measured critical cell capacity, which is generally around $\sim 1$ mAh/cm$^2$ for a Li/LLZO system when operated under a stack pressure of $0.7$ MPa and an applied current density of $0.1$ mA/cm$^2$ \cite{krauskopf2019toward}, \cite{barai2024study}. Two reasons for this discrepancy are that i) our simulations begin with voids already present at the interface, meaning that we have effectively skipped past cycles during which voids nucleated at the interface, and ii) our model does not consider the effect of the stack pressure, which suppresses void growth and thus increases the time to failure

\subsubsection{Stripping simulations with a polycrystalline solid electrolyte (SE)}
\label{secR3.2}
\begin{figure}
\centering
\begin{subfigure}{0.47\textwidth}
	\includegraphics[trim=0 0 0 0, clip, keepaspectratio,width=\linewidth]{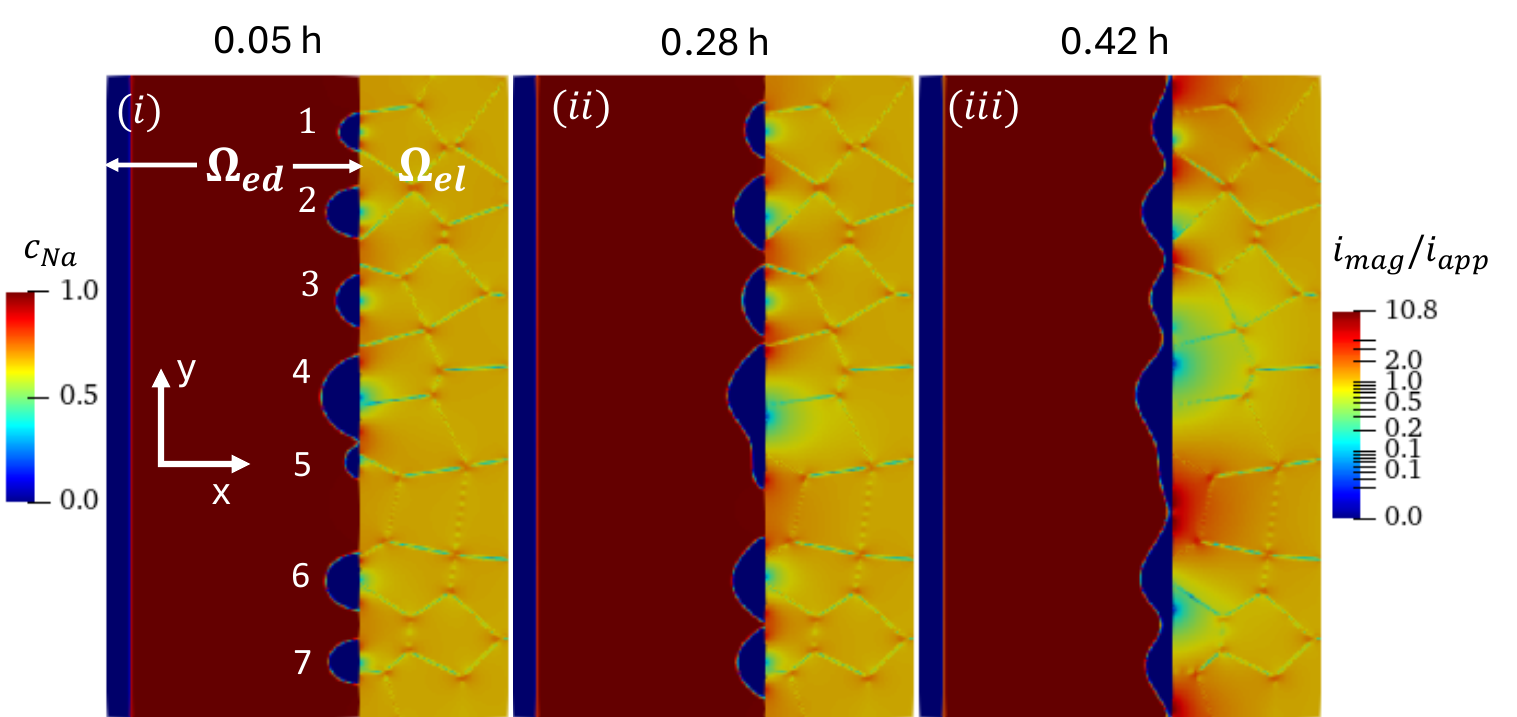}
	\subcaption{}
	\label{FigR11a}
\end{subfigure}
\begin{subfigure}{0.47\textwidth}
	\includegraphics[trim=0 0 0 0, clip, keepaspectratio,width=\linewidth]{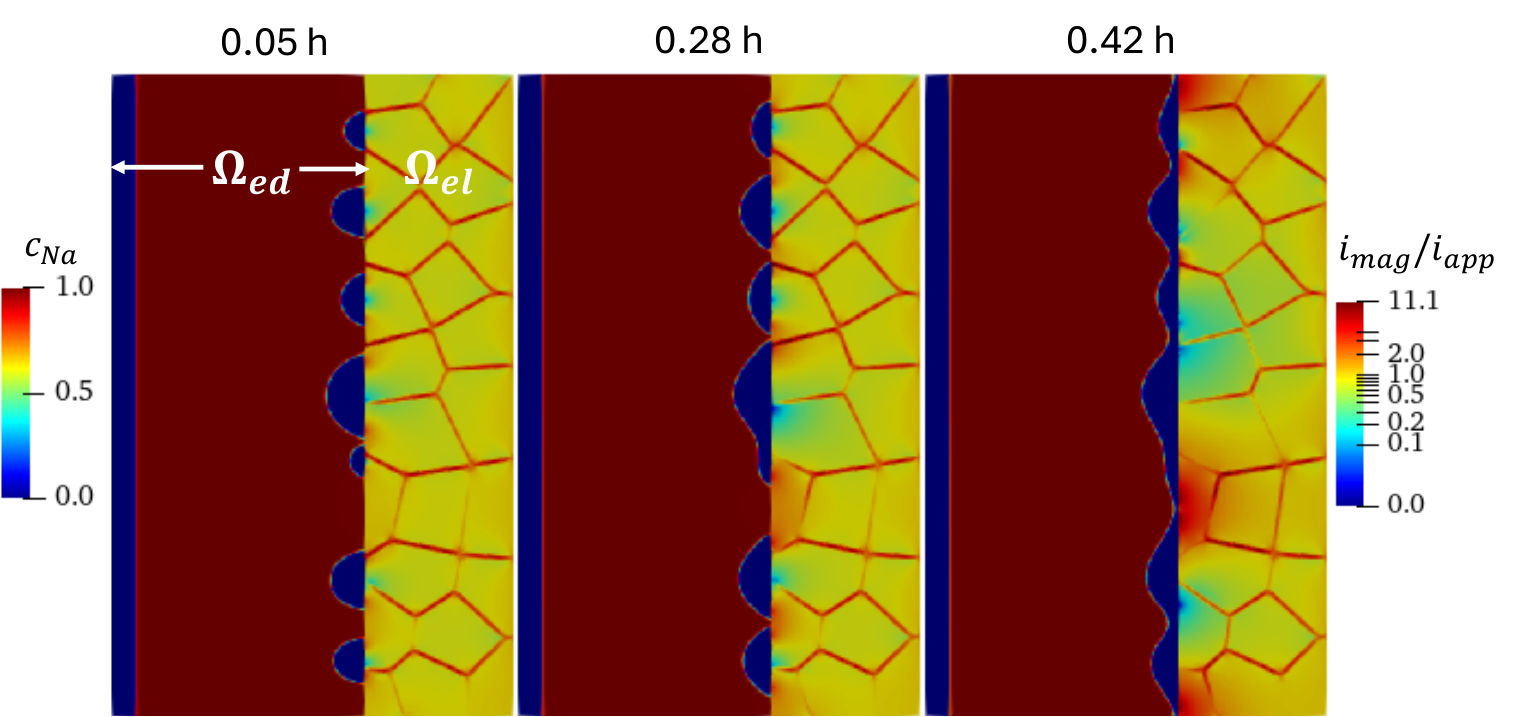}
	\subcaption{}
	\label{FigR11b}
\end{subfigure}
\begin{subfigure}{0.47\textwidth}
	\includegraphics[trim=0 0 0 0, clip, keepaspectratio,width=\linewidth]{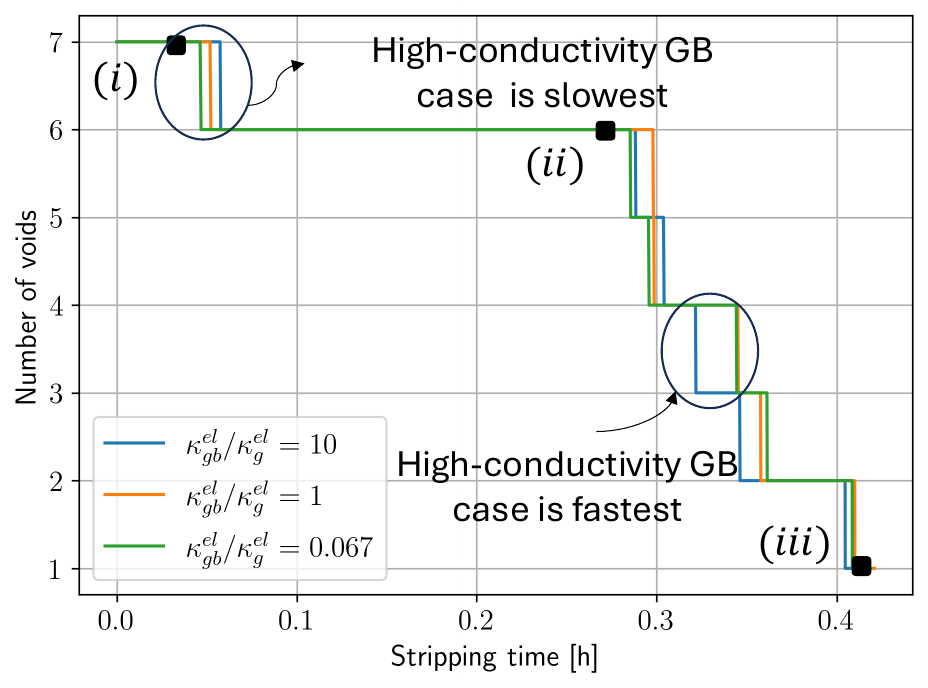}
	\subcaption{}
	\label{FigR11c}
\end{subfigure}
\begin{subfigure}{0.47\textwidth}
	\includegraphics[trim=0 0 0 0, clip, keepaspectratio,width=\linewidth]{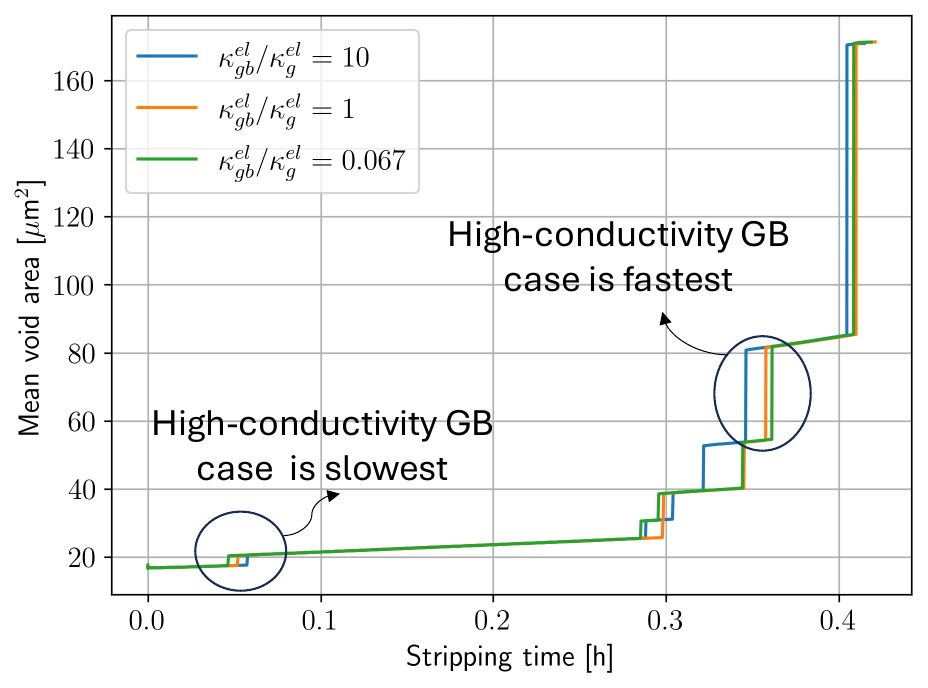}
	\subcaption{}
	\label{FigR11d}
\end{subfigure}
\begin{subfigure}{0.47\textwidth}
	\includegraphics[trim=0 0 0 0, clip, keepaspectratio,width=\linewidth]{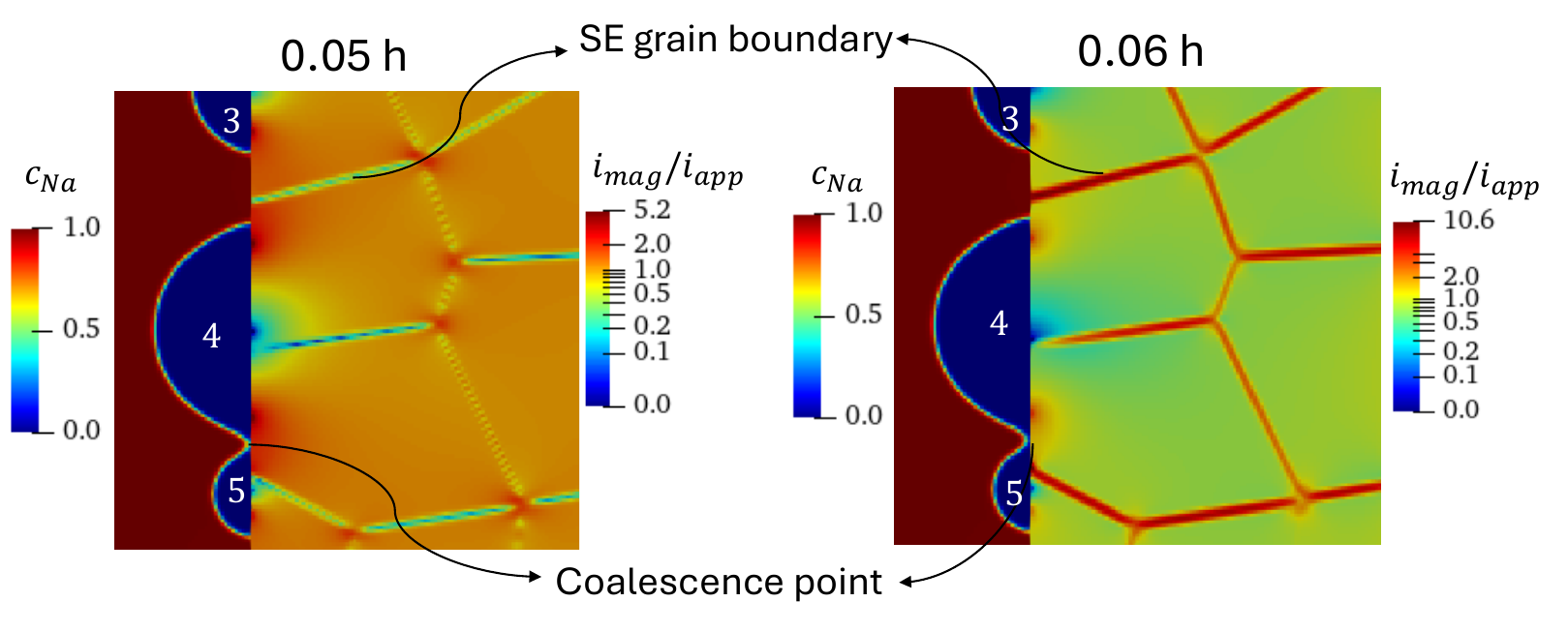}
	\subcaption{}
	\label{FigR11e}
\end{subfigure}
\begin{subfigure}{0.47\textwidth}
	\includegraphics[trim=0 0 0 0, clip, keepaspectratio,width=\linewidth]{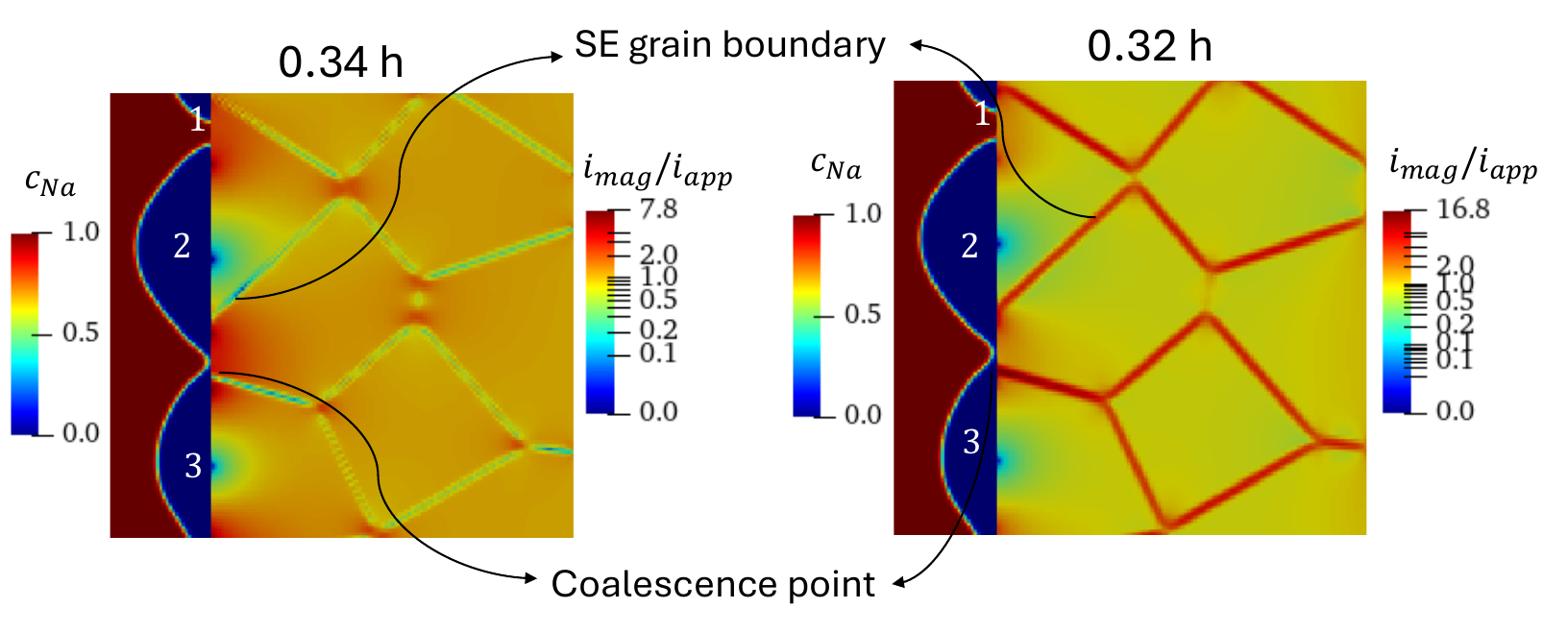}
	\subcaption{}
	\label{FigR11f}
\end{subfigure}
\caption{Stripping simulations with an applied current density of $0.2$ mA/cm$^2$, multiple interfacial voids, and a polycrystalline SE separator with different values of $\kappa_{gb}^{el}/\kappa_{g}^{el}$. Void morphology at three times with (a) low-conductivity and (b) high-conductivity GBs. The initial voids are numbered in (a). $\Omega_{ed}$ is shaded by the Na concentration and $\Omega_{el}$ by the normalized current density. Variation with time of the (c) number of voids and (d) mean void area. In (c), the black markers labelled $(i)-(iii)$ correspond to the microstructures shown in (a). Zoomed-in image of void morphologies when (e) voids 4 and 5 and (f) and 2 and 3 coalesce.}
\end{figure}

Finally, we perform simulations to investigate the role of SE GBs on void coalescence. The initial void radii and the electrode thickness are identical to Section \ref{secR3.1}. The only difference is that we include polycrystalline SEs with low-conductivity and high-conductivity GBs instead of a homogeneous SE, as in Section \ref{secR2.2}.  The simulations are performed under a fixed applied current density of $0.2$ mA/cm$^2$ and run until complete contact loss occurs between the electrode and the SE. For discussion purposes, we have labelled the voids as $1$ to $ 7$ starting from the top to bottom, as shown in Fig.~\ref{FigR11a}.

Figures \ref{FigR11a} and \ref{FigR11b} show the evolution of interfacial voids during stripping in the case of SEs with low-conductivity and high-conductivity GBs, respectively. The void morphologies are almost identical, indicating that GB conductivities only have small effects on the void evolution. Similar to Section \ref{secR3.1}, multiple current density hot spots occur in the SE separator near the void edges due to contact loss. We again plot the number of voids and the mean void area versus time, as shown in Figs.~\ref{FigR11c} and \ref{FigR11d}. There are noticeable differences in the void coalescence times between the low-conductivity, homogeneous, and high-conductivity GB cases. 
%
%However, in Section \ref{secR2.2}, we also demonstrated marginal differences in the migration rate of a single void in polycrystalline SEs compared to the homogenous SE case.
%We then attributed these differences to the interaction between the void edge and the intersecting SE GBs. Since the void migration and coalescence rates are related, these minor differences are reflected in the rate at which the number of voids decreases (Fig \ref{FigR11c}) and the mean void area increases (Fig.~\ref{FigR11d}) during stripping. Broadly speaking, these figures show that the void coalescence rates of SEs with low-conductive and high-conductive GBs, i.e., $\kappa_{gb}^{el}/\kappa_{g}^{el}=\{0.067, 10\}$, differ marginally from the homogeneous SE case, i.e., $\kappa_{gb}^{el}/\kappa_{g}^{el}=1$. For instance, in the case of SE with high-conductive GBs, the time at which voids labelled as $4$ and $5$ coalesce is approximately $0.06$ h (Fig.~\ref{FigR11e}), which is marginally longer than the low-conductive GB case and the homogenous SE case, as indicated in Fig.~\ref{FigR11c}. However, it is also evident from this figure that if voids labelled as $2$ and $3$ are considered (Fig.~\ref{FigR11f}), the time to coalesce is shortest, approximately $0.32$ h, for SE with high-conductive GBs. 
%
We believe the time to coalescence depends on the location at which the voids coalesce relative to the intersecting SE GB along the electrode/electrolyte interface. If the location of void coalescence lies near a SE GB, as in the case of the coalescence of Voids 2 and 3 (Fig.~\ref{FigR11f}), then the coalescence rate is faster with high-conductivity GBs. This is because the void edge velocity is faster near a high-conductivity GB, as shown in Section \ref{secR2.2}. If the location of void coalescence lies away from an SE GB, as in the case of the coalescence of Voids 4 and 5 (Fig.~\ref{FigR11e}), then the coalescence rate is slower with high-conductivity GBs.

We have performed additional void coalescence simulations with bi-crystal SEs that conclusively show that void coalescence in SEs with high-conductivity GBs is faster if the coalescence location is near a GB and may be slower if the coalescence location is not near a GB. These results are presented in Section \ref{SI_secS8} of the Supporting Information. 
\section{Conclusions}
\label{sec:conclusions}
We have presented an electrochemical phase-field model to simulate the evolution of a metallic negative electrode during the stripping and plating of an all-solid-state battery, first assuming a perfect contact between the electrode and the solid-electrolyte separator. We extended this model to simultaneously simulate the evolution of single and multiple interfacial voids at the electrode/solid-electrolyte separator interface during stripping and plating. The model was numerically solved using the finite-element method. For the first time, the stripping and plating processes in a solid-state cell with a metallic Na electrode in contact with a Na-$\beta^{\prime\prime}$-alumina (NBA) ceramic solid electrolyte separator were simulated. 

Both homogeneous and polycrystalline solid electrolytes were considered in this work. We demonstrated the impact of applied current density on electrode deposition/depletion and void edge velocities. The unique contribution of this work compared to previous works \cite{zhao2022phase,narayan2020modeling,jang2021towards,agier2022void,barai2024study} is that the role of the solid electrolyte grain boundary properties on deposition/depletion rates and void edge migration rates were directly considered.  Our results show that:
\begin{itemize}
\item The Na negative electrode shrinks and grows during stripping and plating, respectively. The depletion and deposition rates are proportional to the applied current density and consistent with sharp-interface electrochemical models \cite{mishra2021perspective}, assuming a perfect electrode/electrolyte interface.
\item For a perfect electrode/electrolyte interface, the depletion and deposition rates during the cycling of a solid-state cell are not impacted by the solid electrolyte grain boundary properties.
\item Interfacial voids preferentially grow and shrink along the electrode/electrolyte interface during stripping and plating. Their migration rate is proportional to the flux of Na/Na${^+}$ at the void edge, which depends on the applied current density. These voids block the shrinkage and growth of the Na-negative electrode during stripping and plating. 
\item SE GB properties impact the local interfacial void migration rate and the void coalescence rate, though the effect is small. 
\end{itemize} 

Though this work has made significant progress in modeling solid-state Na batteries, additional work is needed. First, the charge-transfer kinetics at the electrode/electrolyte interface are currently ignored but should be included in the future model. Second, the effect of applied stack pressure on void growth should be incorporated into this model. Third, multiple charge/discharge cycles should be simulated to predict cell failure accurately. Finally, 3D results should be carried out to ensure that the general trends observed here do not change.

\section*{Declaration of Competing Interest}
The authors declare that they have no known competing financial interests or personal relationships that could have appeared to influence the work reported in this paper.

\section*{Acknowledgements}
This work was funded by a subcontract from the company QuantumScape. The authors would like to thank Chaitanya Bhave for his initial work modeling batteries with his electrochemical phase-field model developed for molten salt corrosion. Part of this work was performed under the auspices of the US Department of Energy by Lawrence Livermore National Laboratory under Contract DE-AC52-07NA27344 (Sourav Chatterjee).

\section*{Data Availability}
The data that support the findings of this study are available from the corresponding authors (SC and MT) upon reasonable request.

\newpage
\section*{Appendix A}
\label{Appendix_A}

\setcounter{equation}{0} % Reset equation counter in the appendix
\renewcommand{\theequation}{A.\arabic{equation}}

\setcounter{subsection}{0} 
\renewcommand{\thesubsection}{A.\arabic{subsection}}

\subsection{Reduction of Eq.~\eqref{Eqn6} to Fick's law}
\label{AppendixA1}
It is important to note that in a grand-potential-based formulation, the overall concentration is a function of the phase-field variables and the diffusion potential of the species. This is because by definition, $c_{M}^{ed}(\{\xi\}, \mu_{M}^{ed})= -\delta \Psi/\delta \mu_{M}^{ed}$, where $\{\xi\}$ represents the set of three phase-field variables $\xi_a$, $\xi_m$ and $\xi_v$. Consequently, its gradient may be written as:
\begin{align}
\nabla c_{M}^{ed} =  \sum_{p=a,m,v}\frac{\partial c_{M}^{ed}}{\partial \xi_{p}}\nabla \xi_{p} + \frac{\partial c_{M}^{ed}}{\partial \mu_{M}^{ed}}\nabla \mu_{M}^{ed}.
\label{EqnA1}
\end{align}
Moreover, using the definition of $c_{M}^{ed}$ and Eq.~\eqref{Eqn1}, it follows that 
\begin{align}
c_{M}^{ed}(\{\xi\}, \mu_{M}^{ed}) = \sum_{p=a,m,v}h_{p}^{ed}c_{M}^{p},
\label{EqnA2}
\end{align}
where we have used the relation $c_{M}^{p} =-\partial \omega_{p}/\partial \mu_{M}^{ed}$ that follows from Eq.~\eqref{Eqn3}. Using Eq.~\eqref{EqnA2}, Eq.~\eqref{EqnA1} may be equivalently written as:
\begin{align}
\nabla c_{M}^{ed} =  \sum_{p=a,m,v}\left(\sum_{r=a,m,v}\frac{\partial h_{r}^{ed}}{\partial \xi_{p}}c_{M}^{r}\right)\nabla \xi_{p} + \frac{\partial c_{M}^{ed}}{\partial \mu_{M}^{ed}}\nabla \mu_{M}^{ed}.
\label{EqnA3}
\end{align}
Note that the first term in Eq.~\eqref{EqnA3} vanishes in the bulk regions of the phases since $\nabla \xi_{p}=0$ and $\partial h_{r}^{ed}/\partial \xi_{p}=0$. We can, therefore, say that in the bulk regions $\left.\nabla c_{M}^{ed}\right|_{\xi_p=1} = (\partial c_{M}^{ed}/\partial \mu_{M}^{ed})\nabla \mu_{M}^{ed}$, where the subscript $\xi_p=1$ indicates the bulk regions of either the auxiliary phase, metallic negative electrode or the void phase. Thus, using $M_{p} = D_{M}^{p}dc_{M}^{p}/d\mu_{M}^{ed}$, Eq.~\eqref{Eqn6} in the bulk regions reduces to:
\begin{align}
\left.\boldsymbol{j}_{M}^{ed}\right|_{\xi_{p}=1} = -\frac{D_{M}^{p} }{v_{m}}\left.\nabla c_{M}^{ed}\right|_{\xi_{p=1}}, \quad p =a,m,v.
\label{EqnA4}
\end{align}

\subsection{Derivation of Eq.~\eqref{Eqn8}}
\label{AppendixA2}
To derive Eq.~\eqref{Eqn8}, we start by taking the time derivative of Eq.~\eqref{EqnA2}. This yields:
\begin{align}
\frac{\partial c_{M}^{ed}}{\partial t} = \sum_{p=a,m,v} \frac{\partial h_{p}}{\partial t}c_{M}^{p} + \sum_{p=a,m,v} h_{p}\frac{\partial c_{M}^{p}}{\partial t}.
\label{EqnA5}
\end{align}
Further, it can be shown that:
\begin{align}
 \frac{\partial h_{p}}{\partial t} &= \sum_{q=a,m,v}\frac{\partial h_{p}}{\partial \xi_{q}}\frac{\partial \xi_{q}}{\partial t}\label{EqnA6}\\
\frac{\partial c_{M}^{p}}{\partial t} &= \frac{\partial c_{M}^{p}}{\partial \mu_{M}^{ed}}\frac{\partial \mu_{M}^{ed}}{\partial t}\label{EqnA7}.
\end{align} 
Substituting Eqs.~\eqref{EqnA6} and \eqref{EqnA7} in \eqref{EqnA5} yields:
\begin{align}
\frac{\partial c_{M}^{ed}}{\partial t} = \sum_{p=a,m,v}\left(\sum_{q=a,m,v}\frac{\partial h_{p}}{\partial \xi_{q}}\frac{\partial \xi_{q}}{\partial t}\right)  c_{M}^{p} + \sum_{p=a,m,v} h_{p}\frac{\partial c_{M}^{p}}{\partial \mu_{M}^{ed}}\frac{\partial \mu_{M}^{ed}}{\partial t}.
\label{EqnA8}
\end{align}
Using Eq.~\eqref{Eqn6}  in Eq. \ref{EqnA5} yields Eq.~\eqref{Eqn8}, i.e.,
\begin{align}
\left[\sum_{\substack{p=\\a,m,v}} \frac{h_{p}}{v_{m}}\frac{\partial c_{M}^{p}}{\partial \mu_{M}^{ed}}\right]\frac{\partial \mu_{M}^{ed}}{\partial t} 
+ \nabla \boldsymbol{j}_{M}^{ed}
+\sum_{\substack{p=\\a,m,v}}\left(\sum_{q=a,m,v}\frac{\partial h_{p}}{\partial \xi_{q}}\frac{\partial \xi_{q}}{\partial t}\right)  \frac{c_{M}^{p}}{v_{m}}
= 0.
\label{EqnA9}
\end{align}

\subsection{Derivation of the grand-potential expressions}
\label{AppendixA3}
As discussed in Section \ref{numerical_method}, we assume that the free energy densities of the auxiliary phase, metallic negative electrode, and void are parabolic functions of the Na mole fractions. This yields:
\begin{align}
f_{p}\left(c_{Na}^{p}\right) = \frac{A^{p}}{2}(c_{Na}^{p}- c_{Na}^{eq,p})^2 \quad p=a,m,v\quad \text{in} \quad \Omega_{ed}.
\label{EqnA10}
\end{align}
The diffusion potential of Na in the region $\Omega^{ed}$ is then obtained from Eq.~\eqref{EqnA10} using $\mu_{Na}^{ed}=v_{m}\left(\partial f_{p}/\partial c_{Na}^{p}\right)$, which yields: $\mu_{Na}^{ed}\left(c_{Na}^{p}\right) = v_{m}A^{p}(c_{Na}^{p}- c_{Na}^{eq,p})$. This relation can be inverted to obtain the phase mole fractions as functions of Na diffusion potential, which is the independent variable in our model:
\begin{align}
c_{Na}^{p}\left(\mu_{Na}^{ed}\right) = \frac{\mu_{Na}^{ed}}{A^{p}v_{m}} +  c_{Na}^{eq,p} \quad p=a,m,v\quad \text{in} \quad \Omega_{ed}.
\label{EqnA11}
\end{align}
Substituting Eq.~\eqref{EqnA11} in Eq.~\eqref{Eqn3} yields the grand-potential densities of the auxiliary phase, metallic negative electrode, and void as functions of Na diffusion potential:
\begin{align}
\omega_{p}\left(\mu_{Na}^{ed}\right)  = -\left[\frac{\left(\mu_{Na}^{ed}\right)^2}{v_{m}^2A^{p}} + \frac{\mu_{Na}^{ed}c_{Na}^{eq,p}}{v_{m}}\right]\quad p=a,m,v\quad \text{in} \quad \Omega_{ed}.
\label{EqnA12}
\end{align}

%%%%%
\subsection{Non-dimensionalization of the governing equations}
\label{nondimensional_form}
This section provides the non-dimensional form of the governing equations assuming Na is the diffusing species in the auxiliary-electrode-void region, i.e., $\Omega_{ed}$ and Na$^{+}$ is the diffusing species in the electrolyte region, i.e., $\Omega_{el}$. For numerical convergence reasons, we scale the gradient operator, time, energy density and the field variables as follows:
\begin{align}
\widetilde{\nabla } = \nabla l_{c},\quad 
\widetilde{t} = \frac{t}{\tau},\quad 
\widetilde{\omega}_{p}=  \frac{\omega_{p}}{f_{c}}, \quad
\widetilde{\mu}^{ed}_{Na} = \frac{\mu_{Na}^{ed}}{RT}, \quad 
\widetilde{\mu}^{el}_{Na^{^{+}}} = \frac{\mu_{Na^{^{+}}}^{el}}{RT},\quad 
\widetilde{\phi}^{el} = \frac{F\phi^{el}}{RT},\quad
\widetilde{\phi}^{ed} = \frac{F\phi^{ed}}{RT},
\label{EqnA13}
\end{align}
where tilde $\widetilde{\left(\cdot\right)}$ indicates a non-dimensional quantity; $l_{c}$ is a characteristic length; $\tau$ is a characteristic time; $R$ is universal gas constant, $T$ is temperature; $F$ is Faraday constant; and $f_c$ is characteristic energy density. For our simulations, we assumed $\tau= l_{c}^2/D_{Na}^{m}$ and $f_{c} = RT/v_{m}$. These parameters are listed in  Table \ref{tab:my_table}. Using the variables in Eq. (\ref{EqnA10}), Eq. (\ref{Eqn5}) may be written as:
\begin{align}
\begin{split}
\frac{\partial \xi_{p}}{\partial \widetilde{t}}  + \widetilde{L}_{\phi}\left[\left\{\xi_{p}^{3} - \xi_{p}  + 3\xi_{p}\sum_{\substack{q=a,m,v \\ q\neq p}} \xi_{q}^{2}\right\} 
- \widetilde{\kappa}\widetilde{\nabla}^2 \xi_{p} + 
\frac{f_{c}}{m}\sum_{\substack{q=a,m,v}}\frac{\partial h^{ed}_{q}}{\partial \xi_{p}}\widetilde{\omega}_{q}\right] =0 \quad p = a,m,v, \quad \text{in}\quad  \Omega_{ed},
\label{EqnA14}
\end{split}
\end{align}
where $\widetilde{L}_{\phi} = L_{\phi} \tau m$ and  $\widetilde{\kappa} = \kappa/\left(ml_{c}^{2}\right)$. It should be noted that the dimensionless grand potential densities in Eq.~\eqref{EqnA14} are related to Eq.~\eqref{EqnA12} via: 
\begin{align}
\widetilde{\omega}_{p}\left(\widetilde{\mu}_{Na}^{ed}\right)  = -\left[\frac{\left(\widetilde{\mu}_{Na}^{ed}\right)^2}{\widetilde{A}^{p}} + \widetilde{\mu}_{Na}^{ed}c_{Na}^{eq,p}\right] = \frac{\omega_{p}}{f_{c}}\quad p=a,m,v\quad \text{in} \quad \Omega_{ed},
\end{align}
where $\widetilde{A}^{p} = A^{p}/f_{c}$ for $p=\{a,m,v\}$ denote the non-dimensional parabolic coefficients of the auxiliary phase, negative electrode and void. These values are provided in Table \ref{tab:my_table}.
Similarly, substituting Eq.~\eqref{Eqn6} in Eq.~\eqref{Eqn8} and using the variables in Eq.~\eqref{EqnA13} yields:
\begin{align}
\begin{split}
\left[\sum_{\substack{p=\\a,m,v}}\frac{dc_{Na}^{p}}{d \widetilde{\mu}_{Na}^{ed}}h_{p}^{ed}\right]\frac{\partial \widetilde{\mu}^{ed}_{Na}}{\partial \widetilde{t}}
-\widetilde{\nabla}\left(h_{a}^{ed}{\widetilde{M}}_{a} + h_{m}^{ed}\widetilde{M}_{m} + h_{v}^{ed}{\widetilde{M}}_{v} \right)\widetilde{\nabla}\widetilde{\mu}_{Na}^{ed}\\
+\sum_{\substack{p=\\a,m,v}}\left(\sum_{q=a,m,v}\frac{\partial h_{p}}{\partial \xi_{q}}\frac{\partial \xi_{q}}{\partial \widetilde{t}}\right) c_{Na}^{p}
= 0
\quad  \text{in}  \quad \Omega_{ed},
\end{split}
\label{EqnA15}
\end{align}
where $\widetilde{M}_{p} = \left(M_{p}\tau RT\right)/l_{c}^{2}$ for $p=\{a,m,v\}$ denote the non-dimensional mobilities in the auxiliary phase, negative electrode and void. The phase mole fractions of Na in these phases, which are required in Eq.~\eqref{EqnA15}, are obtained from Eq.~\eqref{EqnA11} and can be written in a non-dimensional form as
\begin{align}
c_{Na}^{p}\left(\widetilde{\mu}_{Na}^{ed}\right) = \frac{\widetilde{\mu}_{Na}^{ed}}{\widetilde{A}^{p}} +  c_{Na}^{eq,p} \quad p=a,m,v\quad \text{in} \quad \Omega_{ed}.
\label{EqnA15b}
\end{align}
Using Eqn \eqref{EqnA15b}, the inverse of thermodynamic factors, which are required in Eq.~\eqref{EqnA15}, can be written in a non-dimensional form according to
\begin{align}
\frac{d c_{Na}^{p}}{d \widetilde{\mu}_{Na}^{ed}} = \frac{1}{\widetilde{A}^{p}}\quad p=a,m,v\quad \text{in} \quad \Omega_{ed}.
\label{EqnA15c}
\end{align}
Likewise, substituting Eq.~\eqref{Eqn9} in Eq.~\eqref{Eqn10} and using the variables in Eq.~\eqref{EqnA13} yields:
\begin{align}
\begin{split}
-\widetilde{\nabla}\left[\left(\widetilde{\sigma}_{a}h_{a}^{ed}  + \widetilde{\sigma}_{m}h_{m}^{ed} +  \widetilde{\sigma}_{v}h_{v}^{ed}\right)\widetilde{\nabla} \widetilde{\phi}^{ed}\right]=0\quad  \text{in}  \quad \Omega_{ed},
\end{split}
\label{EqnA16}
\end{align}
where $\widetilde{\sigma}_{p} = \left(\sigma_{p}RT\tau v_{m}\right)/\left(F^{2}l_{c}^{2}\right)$ for $p=\{a,m,v\}$ denote the non-dimensional electronic conductivities of the auxiliary phase, negative electrode and void. 

Following a similar procedure, we can obtain the non-dimensional form of the equations in the region $\Omega_{el}$. Specifically, combining Eq.~\eqref{Eqn11} with Eq.~\eqref{Eqn14} and  Eq.~\eqref{Eqn15} with Eq.~\eqref{Eqn16} yields:
\begin{align}
\frac{\partial \hat{c}}{\partial \widetilde{\mu}^{el}_{Na^+}}\frac{\partial \widetilde{\mu}^{el}_{Na^{^+}}}{\partial \widetilde{t}} -
\widetilde{\nabla} \left[\widetilde{M}^{g,el}_{Na^{^+}} (h^{el}_{g} - h^{el}_{gb})  + \widetilde{M}^{gb,el}_{Na^{^+}}h^{el}_{gb}\right]\left(\widetilde{\nabla} \widetilde{\mu}^{el}_{Na^{^+}} + \widetilde{\nabla} \widetilde{\phi}^{el}\right) &= 0 \quad  \text{in}  \quad \Omega_{el}\label{EqnA17a}\\
-\widetilde{\nabla}\left[\left\{\widetilde{\kappa}_{g}^{el}(h_{g}^{el} -h_{gb}^{el}) + \widetilde{\kappa}_{gb}^{el}h_{gb}^{el}\right\}\widetilde{\nabla} \widetilde{\phi}^{el}\right]&=0\quad  \text{in}  \quad \Omega_{el}\label{EqnA17b},
\end{align}
where $\widetilde{M}^{p,el}_{Na^+} = \left(M^{p,el}_{Na^+}\tau RT\right)/l_{c}^{2}$ for $p=\{g, gb\}$ are the non-dimensional mobilities of the Na$^+$ in the solid electrolyte grains and grain boundaries, and $\widetilde{\kappa}_{p}^{el} = \left(\kappa_{p}^{el}RT\tau v_{m}\right)/\left(F^{2}l_{c}^{2}\right)$ for $p=\{g, gb\}$  are the non-dimensional ionic conductivities of the solid electrolyte grains and grain boundaries. Like Eq.~\eqref{EqnA15c}, the inverse of the thermodynamic factor in the solid electrolyte is required in Eq.~\eqref{EqnA17a}, and is obtained as follows:
\begin{align}
\frac{d \hat{c}}{d \widetilde{\mu}_{Na^+}^{el}} = \frac{1}{\widetilde{A}^{el}}\quad \text{in} \quad \Omega_{el}
\label{EqnA17c},
\end{align}
where $\widetilde{A}^{el} = A^{el}/f_{c}$ is the non-dimensional parabolic coefficient of the free energy density of the solid electrolyte phase (Table \ref{tab:my_table}).

%References%%%%%%%%%%%%%%%%%%%%%%%%%%%%%%%%%%%%%%%
% Suppress the bibliography section
\newpage
\bibliographystyle{elsarticle-num.bst}
\bibliography{paper4.bib}

\begin{thebibliography}{10}
\expandafter\ifx\csname url\endcsname\relax
  \def\url#1{\texttt{#1}}\fi
\expandafter\ifx\csname urlprefix\endcsname\relax\def\urlprefix{URL }\fi
\expandafter\ifx\csname href\endcsname\relax
  \def\href#1#2{#2} \def\path#1{#1}\fi

\bibitem{takada2013progress}
K.~Takada, Progress and prospective of solid-state lithium batteries, Acta
  Materialia 61~(3) (2013) 759--770.

\bibitem{wang2016general}
B.~Wang, X.-Y. Lu, Y.~Tang, W.~Ben, General polyethyleneimine-mediated
  synthesis of ultrathin hexagonal co3o4 nanosheets with reactive facets for
  lithium-ion batteries, ChemElectroChem 3~(1) (2016) 55--65.

\bibitem{janek2016solid}
J.~Janek, W.~G. Zeier, A solid future for battery development, Nature energy
  1~(9) (2016) 1--4.

\bibitem{barchasz2012lithium}
C.~Barchasz, F.~Molton, C.~Duboc, J.-C. Lepr{\^e}tre, S.~Patoux, F.~Alloin,
  Lithium/sulfur cell discharge mechanism: an original approach for
  intermediate species identification, Analytical chemistry 84~(9) (2012)
  3973--3980.

\bibitem{liu2019pathways}
J.~Liu, Z.~Bao, Y.~Cui, E.~J. Dufek, J.~B. Goodenough, P.~Khalifah, Q.~Li,
  B.~Y. Liaw, P.~Liu, A.~Manthiram, et~al., Pathways for practical high-energy
  long-cycling lithium metal batteries, Nature Energy 4~(3) (2019) 180--186.

\bibitem{feng2022review}
X.~Feng, H.~Fang, N.~Wu, P.~Liu, P.~Jena, J.~Nanda, D.~Mitlin, Review of
  modification strategies in emerging inorganic solid-state electrolytes for
  lithium, sodium, and potassium batteries, Joule 6~(3) (2022) 543--587.

\bibitem{park2017dendrite}
K.~Park, J.~B. Goodenough, Dendrite-suppressed lithium plating from a liquid
  electrolyte via wetting of li3n, Advanced Energy Materials 7~(19) (2017)
  1700732.

\bibitem{vaalma2018cost}
C.~Vaalma, D.~Buchholz, M.~Weil, S.~Passerini, A cost and resource analysis of
  sodium-ion batteries, Nature reviews materials 3~(4) (2018) 1--11.

\bibitem{battistel_electrochemical_2020}
A.~Battistel, M.~S. Palagonia, D.~Brogioli, F.~L. Mantia, R.~Trócoli,
  Electrochemical {Methods} for {Lithium} {Recovery}: {A} {Comprehensive} and
  {Critical} {Review}, Adv. Mater. (2020).

\bibitem{zhao2018solid}
C.~Zhao, L.~Liu, X.~Qi, Y.~Lu, F.~Wu, J.~Zhao, Y.~Yu, Y.-S. Hu, L.~Chen,
  Solid-state sodium batteries, Advanced Energy Materials 8~(17) (2018)
  1703012.

\bibitem{lee2019sodium}
B.~Lee, E.~Paek, D.~Mitlin, S.~W. Lee, Sodium metal anodes: emerging solutions
  to dendrite growth, Chemical reviews 119~(8) (2019) 5416--5460.

\bibitem{han2019high}
F.~Han, A.~S. Westover, J.~Yue, X.~Fan, F.~Wang, M.~Chi, D.~N. Leonard, N.~J.
  Dudney, H.~Wang, C.~Wang, High electronic conductivity as the origin of
  lithium dendrite formation within solid electrolytes, Nature Energy 4~(3)
  (2019) 187--196.

\bibitem{lu_electrolyte_2018}
Y.~Lu, L.~Li, Q.~Zhang, Z.~Niu, J.~Chen, Electrolyte and {Interface}
  {Engineering} for {Solid}-{State} {Sodium} {Batteries}, Joule 2~(9) (2018)
  1747--1770.
\newblock \href {https://doi.org/10.1016/j.joule.2018.07.028}
  {\path{doi:10.1016/j.joule.2018.07.028}}.

\bibitem{kazyak2020li}
E.~Kazyak, R.~Garcia-Mendez, W.~S. LePage, A.~Sharafi, A.~L. Davis, A.~J.
  Sanchez, K.-H. Chen, C.~Haslam, J.~Sakamoto, N.~P. Dasgupta, Li penetration
  in ceramic solid electrolytes: operando microscopy analysis of morphology,
  propagation, and reversibility, Matter 2~(4) (2020) 1025--1048.

\bibitem{spencer2022high}
D.~Spencer~Jolly, J.~Perera, S.~D. Pu, D.~L. Melvin, P.~Adamson, P.~G. Bruce,
  High critical currents for dendrite penetration and voiding in potassium
  metal anode solid-state batteries, Journal of Solid State Electrochemistry
  26~(9) (2022) 1961--1968.

\bibitem{ECheng2017}
E.~J. Cheng, A.~Sharafi, J.~Sakamoto, Intergranular li metal propagation
  through polycrystalline li6. 25al0. 25la3zr2o12 ceramic electrolyte,
  Electrochimica Acta 223 (2017) 85--91.

\bibitem{sharafi2017controlling}
A.~Sharafi, C.~G. Haslam, R.~D. Kerns, J.~Wolfenstine, J.~Sakamoto, Controlling
  and correlating the effect of grain size with the mechanical and
  electrochemical properties of li 7 la 3 zr 2 o 12 solid-state electrolyte,
  Journal of Materials Chemistry A 5~(40) (2017) 21491--21504.

\bibitem{wang2019characterizing}
M.~J. Wang, R.~Choudhury, J.~Sakamoto, Characterizing the li-solid-electrolyte
  interface dynamics as a function of stack pressure and current density, Joule
  3~(9) (2019) 2165--2178.

\bibitem{kasemchainan2019critical}
J.~Kasemchainan, S.~Zekoll, D.~Spencer~Jolly, Z.~Ning, G.~O. Hartley,
  J.~Marrow, P.~G. Bruce, Critical stripping current leads to dendrite
  formation on plating in lithium anode solid electrolyte cells, Nature
  materials 18~(10) (2019) 1105--1111.

\bibitem{XKe2020}
X.~Ke, Y.~Wang, L.~Dai, C.~Yuan, Cell failures of all-solid-state lithium metal
  batteries with inorganic solid electrolytes: Lithium dendrites, Energy
  Storage Materials 33 (2020) 309--328.

\bibitem{raj2022direct}
V.~Raj, V.~Venturi, V.~R. Kankanallu, B.~Kuiri, V.~Viswanathan, N.~P.~B.
  Aetukuri, Direct correlation between void formation and lithium dendrite
  growth in solid-state electrolytes with interlayers, Nature Materials 21~(9)
  (2022) 1050--1056.

\bibitem{hong2018operando}
Y.-S. Hong, N.~Li, H.~Chen, P.~Wang, W.-L. Song, D.~Fang, In operando
  observation of chemical and mechanical stability of li and na dendrites under
  quasi-zero electrochemical field, Energy Storage Materials 11 (2018)
  118--126.

\bibitem{bay2020sodium}
M.-C. Bay, M.~Wang, R.~Grissa, M.~V. Heinz, J.~Sakamoto, C.~Battaglia, Sodium
  plating from na-$\beta^{\prime\prime}$-alumina ceramics at room temperature,
  paving the way for fast-charging all-solid-state batteries, Advanced Energy
  Materials 10~(3) (2020) 1902899.

\bibitem{koshikawa2018dynamic}
H.~Koshikawa, S.~Matsuda, K.~Kamiya, M.~Miyayama, Y.~Kubo, K.~Uosaki,
  K.~Hashimoto, S.~Nakanishi, Dynamic changes in charge-transfer resistance at
  li metal/li7la3zr2o12 interfaces during electrochemical li
  dissolution/deposition cycles, Journal of power sources 376 (2018) 147--151.

\bibitem{manalastas2019mechanical}
W.~Manalastas~Jr, J.~Rikarte, R.~J. Chater, R.~Brugge, A.~Aguadero, L.~Buannic,
  A.~Llord{\'e}s, F.~Aguesse, J.~Kilner, Mechanical failure of garnet
  electrolytes during li electrodeposition observed by in-operando microscopy,
  Journal of Power Sources 412 (2019) 287--293.

\bibitem{devaux2015failure}
D.~Devaux, K.~J. Harry, D.~Y. Parkinson, R.~Yuan, D.~T. Hallinan, A.~A.
  MacDowell, N.~P. Balsara, Failure mode of lithium metal batteries with a
  block copolymer electrolyte analyzed by x-ray microtomography, Journal of The
  Electrochemical Society 162~(7) (2015) A1301.

\bibitem{krauskopf2020physicochemical}
T.~Krauskopf, F.~H. Richter, W.~G. Zeier, J.~Janek, Physicochemical concepts of
  the lithium metal anode in solid-state batteries, Chemical reviews 120~(15)
  (2020) 7745--7794.

\bibitem{spencer2019sodium}
D.~Spencer~Jolly, Z.~Ning, J.~E. Darnbrough, J.~Kasemchainan, G.~O. Hartley,
  P.~Adamson, D.~E. Armstrong, J.~Marrow, P.~G. Bruce, Sodium/na
  $\beta^{\prime\prime}$ alumina interface: effect of pressure on voids, ACS
  applied materials \& interfaces 12~(1) (2019) 678--685.

\bibitem{krauskopf2019toward}
T.~Krauskopf, H.~Hartmann, W.~G. Zeier, J.~Janek, Toward a fundamental
  understanding of the lithium metal anode in solid-state batteries—an
  electrochemo-mechanical study on the garnet-type solid electrolyte li6.
  25al0. 25la3zr2o12, ACS applied materials \& interfaces 11~(15) (2019)
  14463--14477.

\bibitem{lu2022void}
Y.~Lu, C.-Z. Zhao, J.-K. Hu, S.~Sun, H.~Yuan, Z.-H. Fu, X.~Chen, J.-Q. Huang,
  M.~Ouyang, Q.~Zhang, The void formation behaviors in working solid-state li
  metal batteries, Science Advances 8~(45) (2022) eadd0510.

\bibitem{krauskopf2019diffusion}
T.~Krauskopf, B.~Mogwitz, C.~Rosenbach, W.~G. Zeier, J.~Janek, Diffusion
  limitation of lithium metal and li--mg alloy anodes on llzo type solid
  electrolytes as a function of temperature and pressure, Advanced Energy
  Materials 9~(44) (2019) 1902568.

\bibitem{park2021semi}
R.~J.-Y. Park, C.~M. Eschler, C.~D. Fincher, A.~F. Badel, P.~Guan, M.~Pharr,
  B.~W. Sheldon, W.~C. Carter, V.~Viswanathan, Y.-M. Chiang, Semi-solid alkali
  metal electrodes enabling high critical current densities in solid
  electrolyte batteries, Nature Energy 6~(3) (2021) 314--322.

\bibitem{MYang2021}
M.~Yang, Y.~Liu, A.~M. Nolan, Y.~Mo, Interfacial atomistic mechanisms of
  lithium metal stripping and plating in solid-state batteries, Advanced
  Materials 33 (3 2021).
\newblock \href {https://doi.org/10.1002/adma.202008081}
  {\path{doi:10.1002/adma.202008081}}.

\bibitem{yang2021maintaining}
C.-T. Yang, Y.~Qi, Maintaining a flat li surface during the li stripping
  process via interface design, Chemistry of Materials 33~(8) (2021)
  2814--2823.

\bibitem{agier2022void}
J.~Agier, S.~Shishvan, N.~Fleck, V.~Deshpande, Void growth within li electrodes
  in solid electrolyte cells, Acta Materialia 240 (2022) 118303.

\bibitem{zhao2022phase}
Y.~Zhao, R.~Wang, E.~Mart{\'\i}nez-Pa{\~n}eda, A phase field
  electro-chemo-mechanical formulation for predicting void evolution at the
  li--electrolyte interface in all-solid-state batteries, Journal of the
  Mechanics and Physics of Solids 167 (2022) 104999.

\bibitem{barai2024study}
P.~Barai, T.~Fuchs, E.~Trevisanello, F.~H. Richter, J.~Janek, V.~Srinivasan,
  Study of void formation at the lithium| solid electrolyte interface,
  Chemistry of Materials 36~(5) (2024) 2245--2258.

\bibitem{milan2022role}
E.~Milan, M.~Pasta, The role of grain boundaries in solid-state li-metal
  batteries, Materials Futures 2~(1) (2022) 013501.

\bibitem{dawson2018atomic}
J.~A. Dawson, P.~Canepa, T.~Famprikis, C.~Masquelier, M.~S. Islam, Atomic-scale
  influence of grain boundaries on li-ion conduction in solid electrolytes for
  all-solid-state batteries, Journal of the American Chemical Society 140~(1)
  (2018) 362--368.

\bibitem{breuer2015separating}
S.~Breuer, D.~Prutsch, Q.~Ma, V.~Epp, F.~Preishuber-Pfl{\"u}gl, F.~Tietz,
  M.~Wilkening, Separating bulk from grain boundary li ion conductivity in the
  sol--gel prepared solid electrolyte li 1.5 al 0.5 ti 1.5 (po 4) 3, Journal of
  Materials Chemistry A 3~(42) (2015) 21343--21350.

\bibitem{yamada2015reduced}
H.~Yamada, D.~Tsunoe, S.~Shiraishi, G.~Isomichi, Reduced grain boundary
  resistance by surface modification, The Journal of Physical Chemistry C
  119~(10) (2015) 5412--5419.

\bibitem{tian2019interfacial}
H.-K. Tian, Z.~Liu, Y.~Ji, L.-Q. Chen, Y.~Qi, Interfacial electronic properties
  dictate li dendrite growth in solid electrolytes, Chemistry of Materials
  31~(18) (2019) 7351--7359.

\bibitem{tantratian2021unraveling}
K.~Tantratian, H.~Yan, K.~Ellwood, E.~T. Harrison, L.~Chen, Unraveling the li
  penetration mechanism in polycrystalline solid electrolytes, Advanced Energy
  Materials 11~(13) (2021) 2003417.

\bibitem{mishra2021perspective}
L.~Mishra, A.~Subramaniam, T.~Jang, K.~Shah, M.~Uppaluri, S.~A. Roberts, V.~R.
  Subramanian, Perspective—mass conservation in models for
  electrodeposition/stripping in lithium metal batteries, Journal of The
  Electrochemical Society 168~(9) (2021) 092502.

\bibitem{narayan2020modeling}
S.~Narayan, L.~Anand, On modeling the detrimental effects of inhomogeneous
  plating-and-stripping at a lithium-metal/solid-electrolyte interface in a
  solid-state-battery, Journal of The Electrochemical Society 167~(4) (2020)
  040525.

\bibitem{jang2021towards}
T.~Jang, L.~Mishra, K.~Shah, A.~Subramaniam, M.~Uppaluri, S.~A. Roberts, V.~R.
  Subramanian, Towards real-time simulation of two-dimensional models for
  electrodeposition/stripping in lithium-metal batteries, ECS Transactions
  104~(1) (2021) 131.

\bibitem{jang2022battphase}
T.~Jang, L.~Mishra, S.~A. Roberts, B.~Planden, A.~Subramaniam, M.~Uppaluri,
  D.~Linder, M.~P. Gururajan, J.-G. Zhang, V.~R. Subramanian, Battphase—a
  convergent, non-oscillatory, efficient algorithm and code for predicting
  shape changes in lithium metal batteries using phase-field models: Part i.
  secondary current distribution, Journal of The Electrochemical Society
  169~(8) (2022) 080516.

\bibitem{plapp2011unified}
M.~Plapp, Unified derivation of phase-field models for alloy solidification
  from a grand-potential functional, Physical Review E 84~(3) (2011) 031601.

\bibitem{Choudhary2012}
A.~Choudhury, B.~Nestler,
  \href{https://link.aps.org/doi/10.1103/PhysRevE.85.021602}{Grand-potential
  formulation for multicomponent phase transformations combined with
  thin-interface asymptotics of the double-obstacle potential}, Phys. Rev. E
  85~(2) (2012) 021602--1--16.
\newblock \href {https://doi.org/10.1103/PhysRevE.85.021602}
  {\path{doi:10.1103/PhysRevE.85.021602}}.
\newline\urlprefix\url{https://link.aps.org/doi/10.1103/PhysRevE.85.021602}

\bibitem{bhave2023electrochemical}
C.~V. Bhave, G.~Zheng, K.~Sridharan, D.~Schwen, M.~R. Tonks, An electrochemical
  mesoscale tool for modeling the corrosion of structural alloys by molten
  salt, Journal of Nuclear Materials 574 (2023) 154147.

\bibitem{kim1999phase}
S.~G. Kim, W.~T. Kim, T.~Suzuki, Phase-field model for binary alloys, Physical
  review e 60~(6) (1999) 7186.

\bibitem{moelans2011}
N.~Moelans, A quantitative and thermodynamically consistent phase-field
  interpolation function for multi-phase systems, Acta Materialia 59~(3) (2011)
  1077--1086.

\bibitem{allen1979microscopic}
S.~M. Allen, J.~W. Cahn, A microscopic theory for antiphase boundary motion and
  its application to antiphase domain coarsening, Acta metallurgica 27~(6)
  (1979) 1085--1095.

\bibitem{fabre2011charge}
S.~D. Fabre, D.~Guy-Bouyssou, P.~Bouillon, F.~Le~Cras, C.~Delacourt,
  Charge/discharge simulation of an all-solid-state thin-film battery using a
  one-dimensional model, Journal of The Electrochemical Society 159~(2) (2011)
  A104.

\bibitem{tian2017simulation}
H.-K. Tian, Y.~Qi, Simulation of the effect of contact area loss in
  all-solid-state li-ion batteries, Journal of The Electrochemical Society
  164~(11) (2017) E3512.

\bibitem{schmidt2023three}
C.~P. Schmidt, S.~Sinzig, V.~Gravemeier, W.~A. Wall, A three-dimensional finite
  element formulation coupling electrochemistry and solid mechanics on resolved
  microstructures of all-solid-state lithium-ion batteries, Computer Methods in
  Applied Mechanics and Engineering 417 (2023) 116468.

\bibitem{Cogswell2015}
D.~A. Cogswell,
  \href{https://link.aps.org/doi/10.1103/PhysRevE.92.011301}{Quantitative
  phase-field modeling of dendritic electrodeposition}, Phys. Rev. E 92 (2015)
  011301.
\newblock \href {https://doi.org/10.1103/PhysRevE.92.011301}
  {\path{doi:10.1103/PhysRevE.92.011301}}.
\newline\urlprefix\url{https://link.aps.org/doi/10.1103/PhysRevE.92.011301}

\bibitem{hong2018phase}
Z.~Hong, V.~Viswanathan, Phase-field simulations of lithium dendrite growth
  with open-source software, ACS Energy Letters 3~(7) (2018) 1737--1743.

\bibitem{jeon2022phase}
J.~Jeon, G.~H. Yoon, T.~Vegge, J.~H. Chang, Phase-field investigation of
  lithium electrodeposition at different applied overpotentials and operating
  temperatures, ACS applied materials \& interfaces 14~(13) (2022)
  15275--15286.

\bibitem{muntaha2023impact}
M.~A. Muntaha, S.~Chatterjee, S.~Blondel, L.~Aagesen, D.~Andersson, B.~Wirth,
  M.~Tonks, Impact of grain boundary and surface diffusion on predicted fission
  gas bubble behavior and release in uo$_2$ fuel (2023).
\newblock \href {http://arxiv.org/abs/2310.06795} {\path{arXiv:2310.06795}}.

\bibitem{balluffi2005kinetics}
R.~W. Balluffi, S.~M. Allen, W.~C. Carter, Kinetics of materials, John Wiley \&
  Sons, 2005.

\bibitem{giudicelli2024moose}
G.~Giudicelli, A.~Lindsay, L.~Harbour, C.~Icenhour, M.~Li, J.~E. Hansel,
  P.~German, P.~Behne, O.~Marin, R.~H. Stogner, J.~M. Miller, D.~Schwen,
  Y.~Wang, L.~Munday, S.~Schunert, B.~W. Spencer, D.~Yushu, A.~Recuero, Z.~M.
  Prince, M.~Nezdyur, T.~Hu, Y.~Miao, Y.~S. Jung, C.~Matthews, A.~Novak,
  B.~Langley, T.~Truster, N.~Nobre, B.~Alger, D.~Andr{\v{s}}, F.~Kong,
  R.~Carlsen, A.~E. Slaughter, J.~W. Peterson, D.~Gaston, C.~Permann,
  \href{https://www.sciencedirect.com/science/article/pii/S235271102400061X}{3.0
  - {MOOSE}: Enabling massively parallel multiphysics simulations}, {SoftwareX}
  26 (2024) 101690.
\newblock \href {https://doi.org/https://doi.org/10.1016/j.softx.2024.101690}
  {\path{doi:https://doi.org/10.1016/j.softx.2024.101690}}.
\newline\urlprefix\url{https://www.sciencedirect.com/science/article/pii/S235271102400061X}

\bibitem{tonks2012object}
M.~R. Tonks, D.~Gaston, P.~C. Millett, D.~Andrs, P.~Talbot, An object-oriented
  finite element framework for multiphysics phase field simulations,
  Computational Materials Science 51~(1) (2012) 20--29.

\bibitem{schwen2017rapid}
D.~Schwen, L.~K. Aagesen, J.~W. Peterson, M.~R. Tonks, Rapid multiphase-field
  model development using a modular free energy based approach with automatic
  differentiation in moose/marmot, Computational Materials Science 132 (2017)
  36--45.

\bibitem{hu2007thermodynamic}
S.~Hu, J.~Murray, H.~Weiland, Z.~Liu, L.~Chen, Thermodynamic description and
  growth kinetics of stoichiometric precipitates in the phase-field approach,
  Calphad 31~(2) (2007) 303--312.

\bibitem{heinz2021grain}
M.~V. Heinz, M.-C. Bay, U.~F. Vogt, C.~Battaglia, Grain size effects on
  activation energy and conductivity: Na-$\beta^{\prime\prime}$-alumina
  ceramics and ion conductors with highly resistive grain boundary phases, Acta
  Materialia 213 (2021) 116940.

\bibitem{kellner2022modeling}
M.~Kellner, S.~N. Enugala, B.~Nestler, Modeling of stoichiometric phases in
  off-eutectic compositions of directional solidifying nbsi-10ti for
  phase-field simulations, Computational Materials Science 203 (2022) 111046.

\bibitem{yu2012extended}
H.-C. Yu, H.-Y. Chen, K.~Thornton, Extended smoothed boundary method for
  solving partial differential equations with general boundary conditions on
  complex boundaries, Modelling and Simulation in Materials Science and
  Engineering 20~(7) (2012) 075008.

\bibitem{wang2022optimizing}
C.~Wang, C.~Sun, Z.~Sun, B.~Wang, T.~Song, Y.~Zhao, J.~Li, H.~Jin, Optimizing
  the na metal/solid electrolyte interface through a grain boundary design,
  Journal of Materials Chemistry A 10~(10) (2022) 5280--5286.

\bibitem{jain2013commentary}
A.~Jain, S.~P. Ong, G.~Hautier, W.~Chen, W.~D. Richards, S.~Dacek, S.~Cholia,
  D.~Gunter, D.~Skinner, G.~Ceder, et~al., Commentary: The materials project: A
  materials genome approach to accelerating materials innovation, APL materials
  1~(1) (2013).

\bibitem{sullivan1964measurement}
G.~A. Sullivan, J.~W. Weymouth, Measurement of the equilibrium net vacancy
  concentration in sodium, Physical Review 136~(4A) (1964) A1141.

\bibitem{mundy1971effect}
J.~N. Mundy, Effect of pressure on the isotope effect in sodium self-diffusion,
  Physical Review B 3~(8) (1971) 2431.

\bibitem{west2022solid}
A.~R. West, Solid state chemistry and its applications, John Wiley \& Sons,
  2022.

\bibitem{kim2023pubchem}
S.~Kim, J.~Chen, T.~Cheng, A.~Gindulyte, J.~He, S.~He, Q.~Li, B.~A. Shoemaker,
  P.~A. Thiessen, B.~Yu, et~al., Pubchem 2023 update, Nucleic acids research
  51~(D1) (2023) D1373--D1380.

\end{thebibliography}

\newpage
% Large and centered text
\begin{center}
    {\LARGE \underline{Supporting Information}}
\end{center}
\section*{}
\setcounter{subsection}{0} 
\renewcommand{\thesubsection}{S.\arabic{subsection}}

\renewcommand{\thefigure}{S\arabic{figure}}
\setcounter{figure}{0}% Reset the figure counter to start from -1

\setcounter{equation}{0} % Reset equation counter in the appendix
\renewcommand{\theequation}{S.\arabic{equation}}

\subsection{Section S1}
\label{SI_secS1}

\begin{figure}[H]
\centering
\includegraphics[trim=0 0 0 0, clip, keepaspectratio,width=\linewidth]{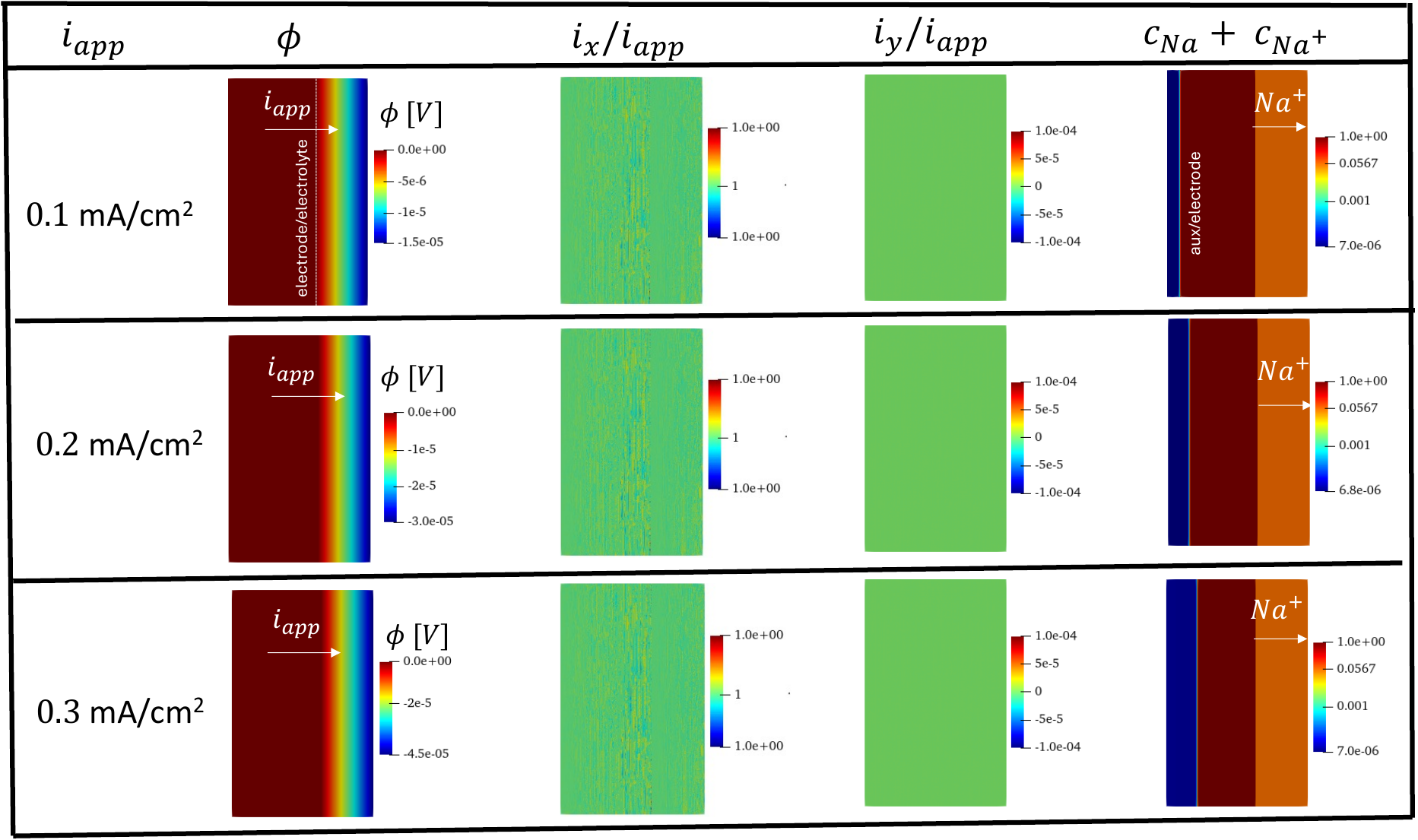}
\caption{Results of the stripping simulation with a perfect electrode/electrolyte interface and a homogeneous SE separator after 3 h. The first column shows the applied stripping current densities, $i_{app}$ in units of mA/cm$^2$; the second column shows the spatial distribution of electric potential $\phi$ across the domain in units of volts; the third column shows the spatial distribution of current density along the x-axis normalized by the applied current density, $i_{x}/ i_{app}$; the fourth column shows the spatial distribution of current density along the $y$-axis normalized by the applied current density, $i_{y}/ i_{app}$; and the fourth column shows the spatial distribution in the auxiliary phase, electrode and solid electrolyte separator, $c_{Na} + c_{Na^+}$}
\label{Fig_SI_secS1}
\end{figure}
Figure \ref{Fig_SI_secS1} shows that the spatial distributions of electric potential and current density along the y-axis are uniform, assuming a perfect electrode/electrolyte interface. Thus, the electrostatic problem can be reduced to a one-dimensional problem. Since the electronic conductivity of the auxiliary phase equals the metallic negative electrode, the electric current in the region $\Omega_{ed}$ may be written as $i^{ed} = -\sigma_{m}\left(d\phi^{ed}/dx\right)$. Consequently, Eq.~\eqref{Eqn10} yields:
\begin{align}
\sigma_{m}\frac{\partial^{2} \phi^{ed}}{\partial x^2} = 0  \quad \text{in}\quad \Omega^{ed}.
\label{EqnS1}
\end{align}
Likewise, since the solid electrolyte separator is homogeneous, the ionic current in the region $\Omega_{ed}$ may be written as $i^{el} = -\kappa_{g}^{el}\left(d\phi^{el}/dx\right)$. Consequently, Eq.~\eqref{Eqn15} yields:
\begin{align}
\kappa_{g}^{el}\frac{\partial^{2} \phi^{el}}{\partial x^2} = 0 \quad \text{in}\quad \Omega^{el}.
\label{EqnS2}
\end{align}
The general form of electric potentials that satisfy Eqs.~\eqref{EqnS1} and \eqref{EqnS2} are:
\begin{align}
\phi^{ed}(x) &= \mathcal{A}^{ed}x + \mathcal{B}^{ed},\label{EqnS3}\\
\phi^{el}(x) &= \mathcal{A}^{el}x + \mathcal{B}^{el},\label{EqnS4}
\end{align}
where $\mathcal{A}^{ed}$, $\mathcal{B}^{ed}$, $\mathcal{A}^{el}$,  and $\mathcal{B}^{el}$ are four unknown constants. These four unknowns can be determined from the four boundary conditions. Two of these boundary conditions follow from the interfacial conditions, i.e., Eqs.~\eqref{Eqn19} and  \eqref{Eqn20}. Thus, using Eqs.~\eqref{EqnS3} and \eqref{EqnS4} in Eqs. \eqref{Eqn19} and \eqref{Eqn20}, we obtain
\begin{align}
\left.\phi^{ed}\right|_{x=x_{in}} &= \left.\phi^{el}\right|_{x=x_{in}} \implies  \mathcal{A}^{ed}x_{in} + \mathcal{B}^{ed} =  \mathcal{A}^{el}x_{in} + \mathcal{B}^{el}, \label{EqnS5}\\
\left.i^{ed}\right|_{x=x_{in}} &= \left.i^{el}\right|_{x=x_{in}}\implies \sigma_{m}\mathcal{A}^{ed} =  \kappa_{g}^{el}\mathcal{A}^{el}, \label{EqnS6}
\end{align}
where $x_{in}$ is the position of the electrode/electrolyte interface. Note that the electrode/electrolyte interface does not vary with time in our simulations. The remaining two boundary conditions are at the left and right boundaries. At the left boundary, i.e., $x=0$, Eq.~\eqref{Eqn21} yields:
\begin{align}
\left.\phi^{ed}\right|_{x=0} =0 \implies \mathcal{B}^{ed} = 0. \label{EqnS7}
\end{align}
Similarly, at the right boundary, i.e., $x=L$, where $L$ is the length of the domain, Eq. \eqref{Eqn22} yields 
\begin{align}
\left.i^{el}\right|_{x=L} =i_{app} \implies \mathcal{A}^{el} = \frac{i_{app}}{\kappa^{el}_{g}}. \label{EqnS8}
\end{align}
The remaining two unknowns are obtained by substituting Eqs. \eqref{EqnS7} and \eqref{EqnS8} into Eqs.~\eqref{EqnS5} and \eqref{EqnS6}, which gives:
\begin{align}
\mathcal{B}^{el} &= \left[\frac{\kappa_{g}^{el}}{\sigma_{m}} - 1\right]  i_{app} x_{in} \label{EqnS9}\\
\mathcal{A}^{ed} &= \frac{\kappa_{g}^{el}}{\sigma_{m}} \mathcal{A}^{el}=\frac{\kappa_{g}^{el}}{\sigma_{m}}\frac{i_{app}}{\kappa^{el}_{g}}\label{EqnS10}
\end{align}

\subsection{Section S2}
\label{SI_secS2}
In a sharp-interface electrochemical model, the velocity of the electrode/electrolyte interface, $v_{ed/el}$, is given by Faraday's law \cite{mishra2021perspective}:
\begin{align}
v_{ed/el}  = -\frac{M_{w}}{\rho}\frac{i_{ed/el}}{F},
\label{EqnS11}
\end{align}
where $M_{w}$ is the molar mass of the metal,  $\rho$ is the density of the metal, $F$ is Faraday constant, and $i_{ed/el}$ is the interfacial current at the electrode/electrolyte interface. It should be noted that the deposition/depletion rates depend on the velocity of the electrode/electrolyte interface in a sharp-interface model. Thus, we refer to Eq.~\eqref{EqnS11} as the sharp-interface deposition/depletion rates.

We make two assumptions to compare our simulated deposition/depletion rates with the sharp-interface rates. First, we assume the molar mass $M_{w}$ and density $\rho$ of the Na negative electrode to be $22.98$ g/mol and $0.9688$ g/cm$^3$ \cite{kim2023pubchem}, respectively. Second, we assume that the interfacial current density in Eq.~\eqref{EqnS11} equals the applied current density in our simulation, i.e., $i_{ed/el}=i_{app}$. This is reasonable because the electrode/electrolyte interface is perfect and planar, which leads to a homogeneous current density distribution, as shown in Fig. \ref{Fig_SI_secS1}.
\subsection{Section S3}
\label{SI_secS3}
\begin{figure}[tbh]
\centering
\includegraphics[trim=0 0 0 0, clip, keepaspectratio,width=0.5\linewidth]{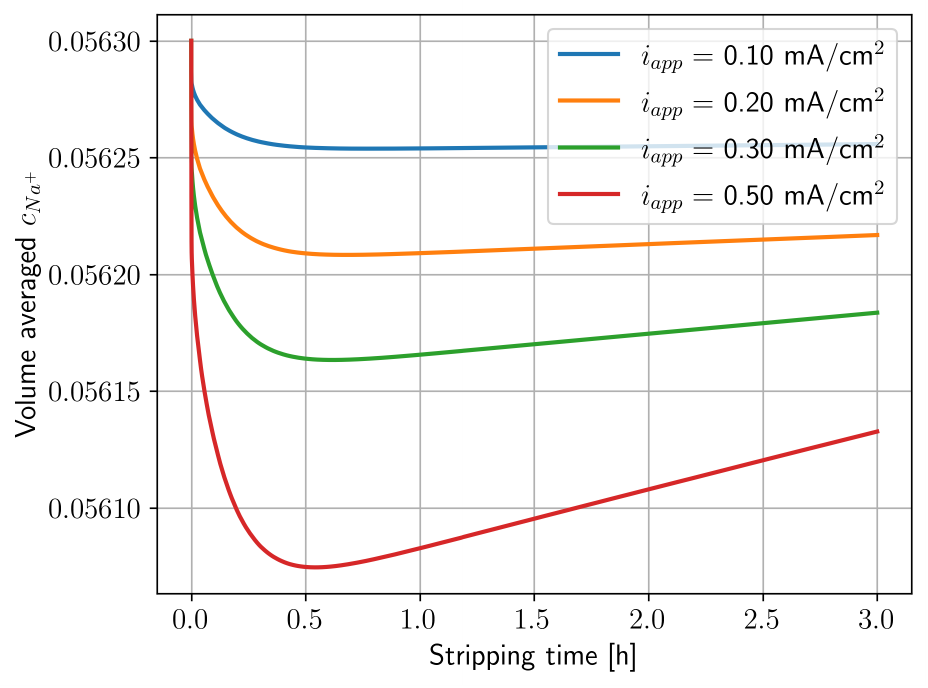}
\caption{Volume-averaged Na$^+$ concentration with time for different applied stripping current densities in the stripping simulations with a perfect electrode/electrolyte interface and a homogeneous SE separator.}
\label{Fig_SI_secS2a}
\end{figure}

Figure \ref{Fig_SI_secS2a} shows the volume-averaged Na$^+$ concentration in the SE separator with time in the stripping simulations with a perfect electrode/electrolyte interface and a homogeneous SE separator. The change in the volume-averaged concentration is very small during the simulation, indicating it can be assumed to be constant with time. This indicates that the flux of Na$^+$ in and out of the separator are equal.
\subsection{Section S4}
\label{SI_secS4}
This section shows how to calculate the Na loss or gain rates during stripping or plating analytically using the mass conservation equation, i.e., Eq.~\eqref{Eqn7}:
\begin{align}
\frac{1}{v_{m}}\frac{\partial c_{Na}^{ed}}{\partial t} +\nabla \cdot \boldsymbol{j}_{Na}^{ed}= 0, \quad \text{in}  \quad \Omega_{ed}.
\label{EqnS12}
\end{align}
Since the flux of Na is non-zero only at the electrode/electrolyte interface, taking the volume-integral of this equation and using the divergence theorem, Eq.~\eqref{EqnS12} may be written as:
\begin{align}
\int_{\Omega_{ed}}\frac{1}{v_{m}}\frac{\partial c_{Na}^{ed}}{\partial t} dv = -\int_{A_{ed}} \boldsymbol{j}_{Na}^{ed/el}\cdot \boldsymbol{n}_{ed/el}da,
\label{EqnS13}
\end{align}
where $\int_{A_{ed}}$ indicates the surface integral of the flux over the electrode/electrolyte interface, $\boldsymbol{j}_{Na}^{ed/el}$ is the flux at the electrode/electrolyte interface, and $\boldsymbol{n}_{ed/el}$ is the unit normal at the electrode/electrolyte interface (Fig.~\ref{Fig0a}). Further, the volume-averaged Na mole fraction, $\langle c_{Na} \rangle$, is defined as:
\begin{align}
\langle c_{Na} \rangle = \frac{\int_{\Omega_{ed}}c _{Na}^{ed}dv}{V_{ed}},
\label{EqnS14}
\end{align}
where $V_{ed}$ is the volume of the subregion $\Omega_{ed}$. Dividing Eq.~\eqref{EqnS13} by $V_{ed}$ and using Eq.~\eqref{EqnS14} yields:
\begin{align}
\frac{\partial \langle c _{Na}\rangle}{\partial t} = - \frac{v_{m}}{V_{ed}}\int_{A_{ed}} \boldsymbol{j}_{Na}^{ed/el}\cdot \boldsymbol{n}_{ed/el}da.
\label{EqnS15}
\end{align}
Morever, the area-averaged flux of Na$^{+}$ at the electrode/electrolyte interface is defined as:
\begin{align}
\langle j_{Na^+}^{ed/el} \rangle = \frac{1}{A_{ed}} \int_{A_{ed}}  \boldsymbol{j}_{Na^{+}}^{ed/el}\cdot \boldsymbol{n}_{ed/el}da.
\label{EqnS16a}
\end{align}
In Section \ref{RS1.1}, we showed that the area-averaged flux of Na$^{+}$ at the electrode/electrolyte interface was equal to the applied flux at the rightmost boundary, which is the ratio of applied current density divided by the Faraday constant once the simulations reach a steady state. Since the flux of Na$^{+}$ equals the flux of Na at the interface (see Eq.~\eqref{Eqn17}), this may be written as:
\begin{align}
\frac{1}{A_{ed}}\int_{A_{ed}} \boldsymbol{j}_{Na}^{ed/el}\cdot \boldsymbol{n}_{ed/el}da= \frac{1}{A_{ed}}\int_{A_{ed}} \boldsymbol{j}_{Na^{+}}^{ed/el}\cdot \boldsymbol{n}_{ed/el}da = \frac{i_{app}}{F},
\label{EqnS16}
\end{align}
where $A_{ed}$ is the area of the electrode/electrolyte interface. Note that the ratio of $A_{ed}/V_{ed}$ is simply the thickness of the region $\Omega_{ed}$, which is fixed in our simulation. This thickness is referred to as $L_{ed}=A_{ed}/V_{ed}$. Thus, substituting Eq.~\eqref{EqnS16} in Eq.~\eqref{EqnS15} yields:
\begin{align}
\frac{\partial \langle c _{Na}\rangle}{\partial t} = - \frac{v_{m}}{L_{ed}} \frac{i_{app}}{F},
\label{EqnS17}
\end{align}
where $\partial \langle c _{Na}\rangle/\partial t$ refers to the Na loss or gain rate. It is worth noting that, unlike Section \ref{SI_secS1}, this derivation does not assume that the electrode/electrolyte is perfect. Therefore, Eq.~\eqref{EqnS17} should also be valid for the cases with interfacial voids, and we have checked its validity for these cases. 

\subsection{Section S5}
\label{SI_secS5}
\begin{figure}[btp]
\centering
\includegraphics[trim=0 0 0 0, clip, keepaspectratio,width=\linewidth]{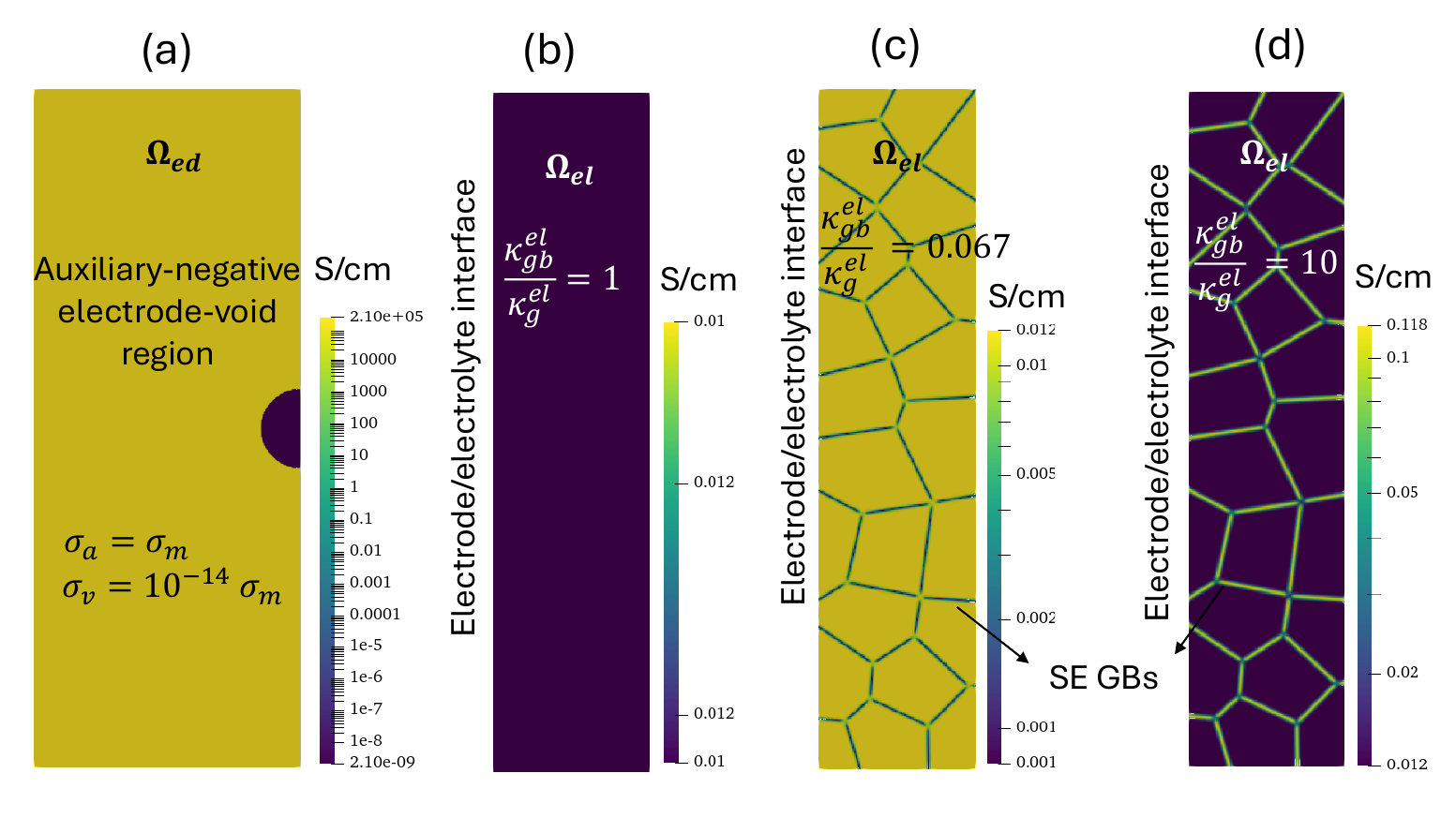}
\caption{Simulation domain for the simulations with polycrystal SE separators. (a) Spatial distribution of electronic conductivity in the region $\Omega_{ed}$. Spatial distribution of ionic conductivity in the region $\Omega_{el}$ for (b) homogeneous SE ($\kappa_{gb}^{el}/\kappa_g^{el}=1$), (c) low-conductivity SE ($\kappa_{gb}^{el}/\kappa_g^{el}=0.067$), and (d) high-conductivity SE ($\kappa_{gb}^{el}/\kappa_g^{el}=10$).}
\label{Fig_SI_secS5}
\end{figure}

In the simulations with polycrystalline SE, the grain structure is created using a Voronoi tesselation. The value of the ionic conductivity varies along the grain boundaries, as shown in Fig.~\ref{Fig_SI_secS5}.

\subsection{Section S6}
\label{SI_secS6}
In this subsection, we provide additional figures that support the arguments presented in Section \ref{secR2.2}. These supporting figures are briefly discussed in the following subsections.

\begin{figure}[tpb]
\centering
\begin{subfigure}{0.49\textwidth}
	\includegraphics[trim=0 0 0 0, clip, keepaspectratio,width=\linewidth]{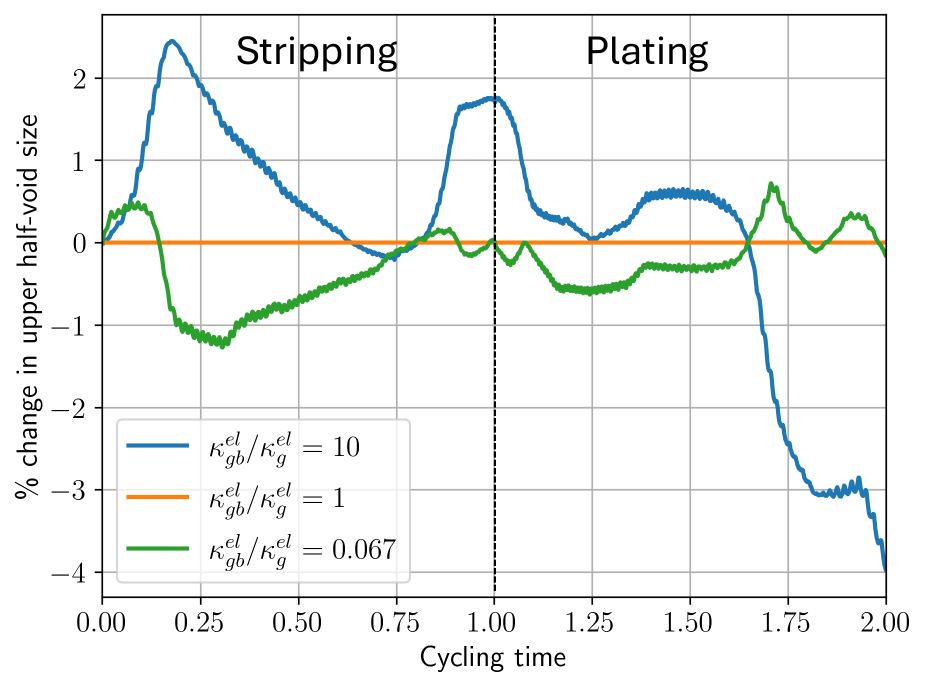}
	\subcaption{}
	\label{Fig1a_SI_secS6}
\end{subfigure}
\begin{subfigure}{0.49\textwidth}
	\includegraphics[trim=0 0 0 0, clip, keepaspectratio,width=\linewidth]{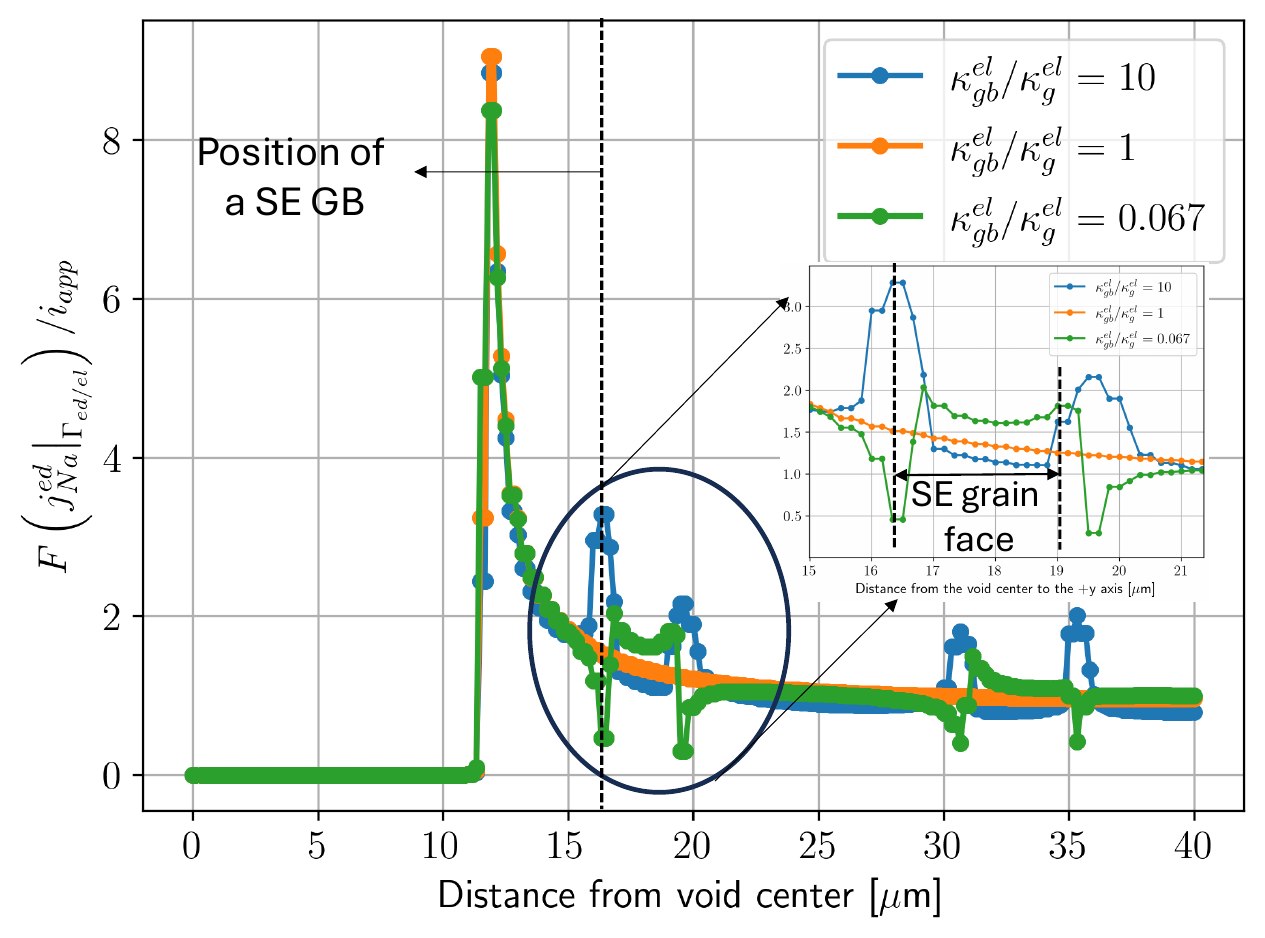}
	\subcaption{}
	\label{Fig1d_SI_secS6}
\end{subfigure}
\begin{subfigure}{0.42\textwidth}
	\includegraphics[trim=0 0 0 0, clip, keepaspectratio,width=\linewidth]{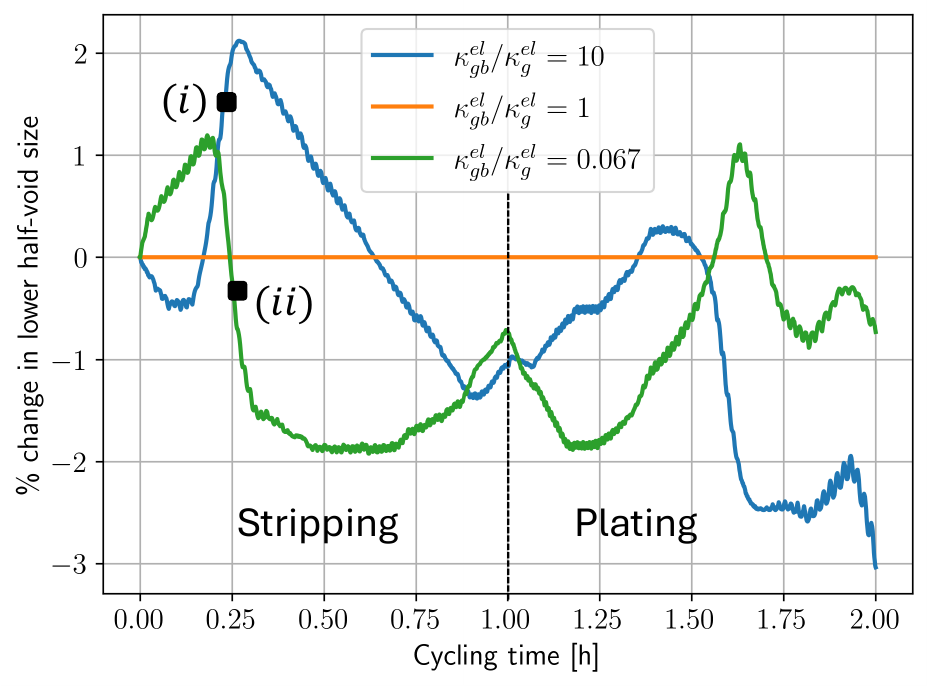}
	\subcaption{}
	\label{Fig1b_SI_secS6}
\end{subfigure}
\begin{subfigure}{0.57\textwidth}
	\includegraphics[trim=0 0 0 0, clip, keepaspectratio,width=\linewidth]{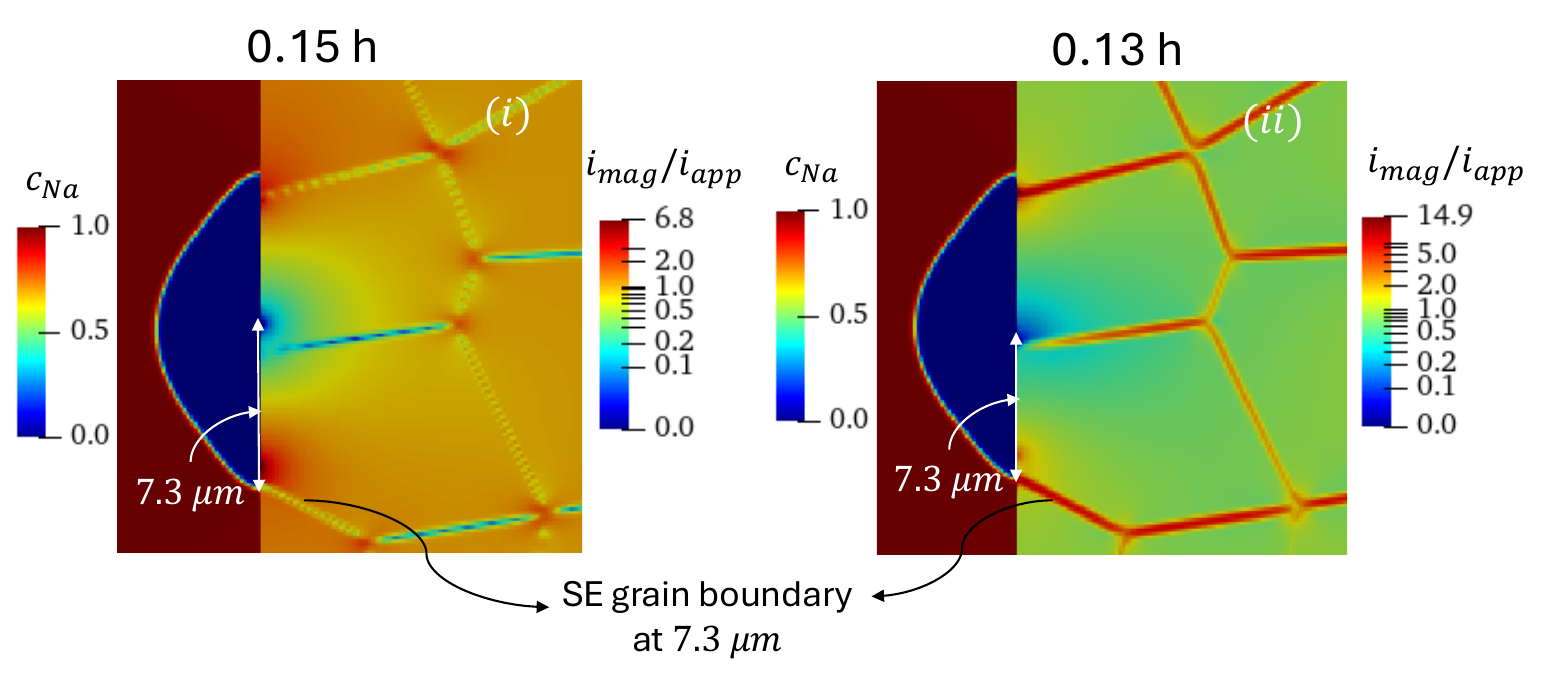}
	\subcaption{}
	\label{Fig1c_SI_secS6}
\end{subfigure}
\caption{(a) Percentage change in upper half-void size relative to the homogeneous solid electrolyte (SE) ($\kappa_{gb}^{el}/\kappa_{g}^{el}=1$) as a function of cycling time for low-conductivity ($\kappa_{gb}^{el}/\kappa_{g}^{el}=0.067$) and high-conductivity ($\kappa_{gb}^{el}/\kappa_{g}^{el}=10$) grain boundaries (GBs). (b) The normal component of the Na mass flux along the electrode/SE interface normalized with the applied flux, $\left.j_{Na}^{ed}\right|_{\Gamma_{ed/el}}/(i_{app}/F)$, with distance from the void center after $0.6$ h for $\kappa_{gb}^{el}/\kappa_{g}^{el}=0.067$, $1$ and $10$. (c) Percentage change in lower half-void size relative to the homogeneous SE with time for $\kappa_{gb}^{el}/\kappa_{g}^{el}=0.067$ and $10$. The black markers labeled $(i)$ and $(ii)$ correspond to the microstructures shown in (d). (d) Zoomed-in images of the simulation domains shown in Figs.~\ref{FigR8a} and \ref{FigR8b} when the lower void edge touches the SE GB at a distance of $7.3$ $\mu$m below the void center.}
\end{figure}

\subsubsection{Discussion for Fig.~\ref{Fig1a_SI_secS6}}
Figure \ref{Fig1a_SI_secS6} shows the percentage change calculated relative to the homogeneous solid electrolyte (SE) case for the entire cycle. The percentage changes in the entire cycle are less than $5\%$, which indicates that the local differences in void migration due to SE grain boundary conductivity are marginal. 

\subsubsection{Discussion for Fig.~\ref{Fig1d_SI_secS6}}
Figure \ref{Fig1d_SI_secS6} shows the spatial distribution of the interfacial flux of Na normalized by the applied flux along the electrode/electrolyte interface for SEs with low-conductivity, homogeneous and high-conductivity GBs after stripping for $0.6$ h. As in Section \ref{secR2.2}, in all cases, the interfacial flux is zero along the void,  jumps to a maximum at the void edge and gradually decreases until it equals the applied flux at the boundary. However, locally, near the SE GBs, the Na flux is lower and higher than the homogeneous case for SEs with low-conductivity and high-conductivity GBs, respectively. The inset of Fig.~\ref{Fig1d_SI_secS6} shows that, along the grains, the Na flux is marginally lower in the case of SE with high-conductivity GBs than in the other two cases.

\subsubsection{Discussion for Figs.~\ref{Fig1b_SI_secS6} and \ref{Fig1c_SI_secS6}}
Figure \ref{Fig1b_SI_secS6} shows the percentage change in the void edge position relative to the homogeneous solid electrolyte (SE) case in the direction of the $-y$-axis as a function of cycling time. Similar to Fig.~\ref{Fig1a_SI_secS6}, the percentage changes over the entire cycle are less than $3\%$, which indicates that the  SE grain boundary conductivity marginally impacts void migration. However, this marginal difference leads to a slightly asymmetric growth, as discussed in Section  \ref{secR2.2}. Note that the maximum deviation during stripping occurs around the instance when the void edge touches the SE GB located at a distance of $7.3$ $\mu$m below the void centre, as shown in Fig.~\ref{Fig1c_SI_secS6}. 

\subsection{Section S7}
\label{SI_secS7}
\begin{figure}[tbp]
\centering
\begin{subfigure}{0.47\textwidth}
	\includegraphics[trim=0 0 0 0, clip, keepaspectratio,width=\linewidth]{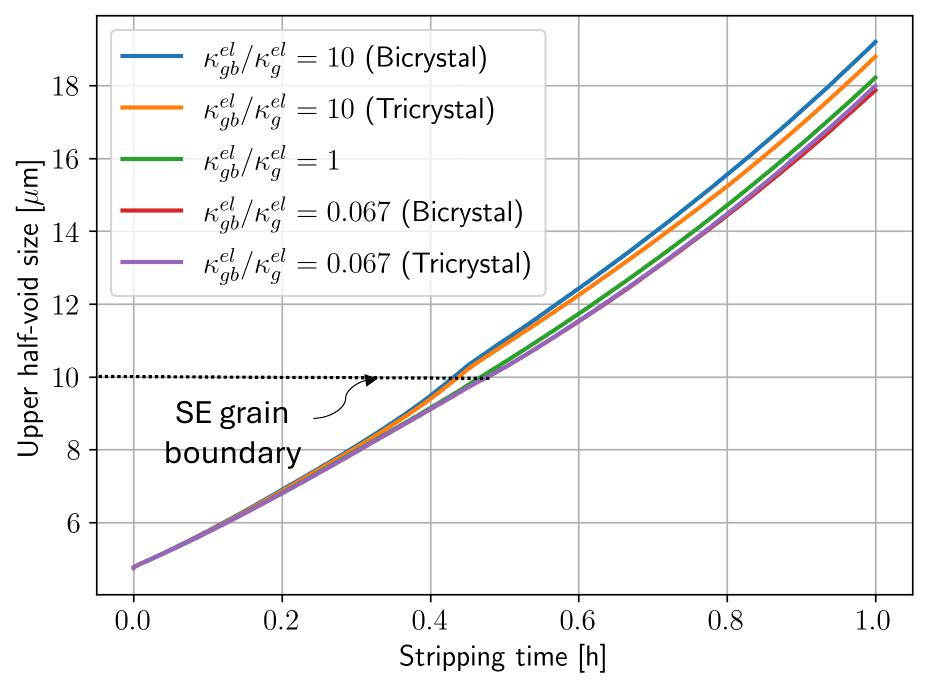}
	\subcaption{}
	\label{Fig1a_SI_secS7}
\end{subfigure}
\begin{subfigure}{0.47\textwidth}
	\includegraphics[trim=0 0 0 0, clip, keepaspectratio,width=\linewidth]{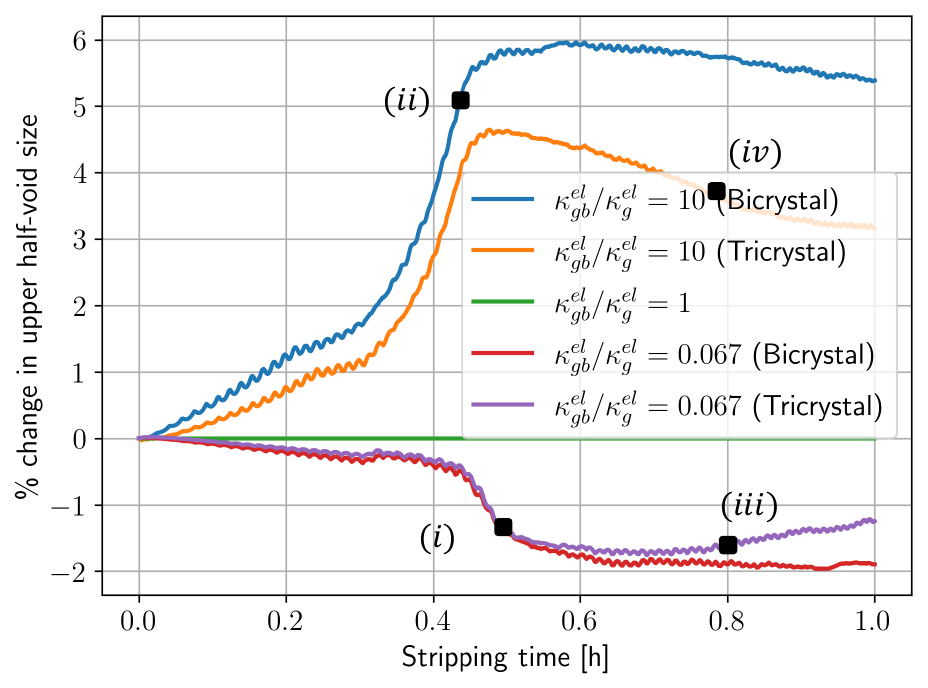}
	\subcaption{}
	\label{Fig1b_SI_secS7}
\end{subfigure}
\begin{subfigure}{0.47\textwidth}
	\includegraphics[trim=0 0 0 0, clip, keepaspectratio,width=\linewidth]{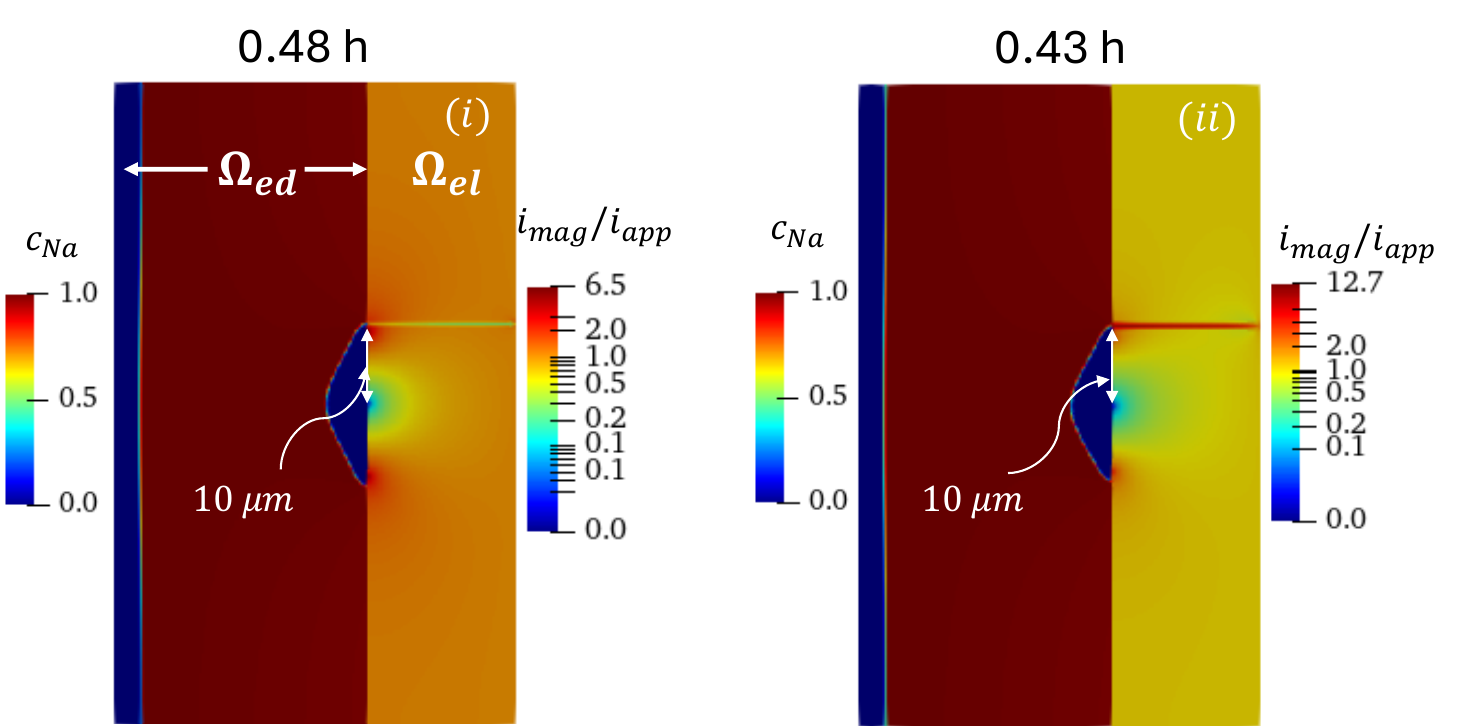}
	\subcaption{}
	\label{Fig1c_SI_secS7}
\end{subfigure}
\begin{subfigure}{0.47\textwidth}
	\includegraphics[trim=0 0 0 0, clip, keepaspectratio,width=\linewidth]{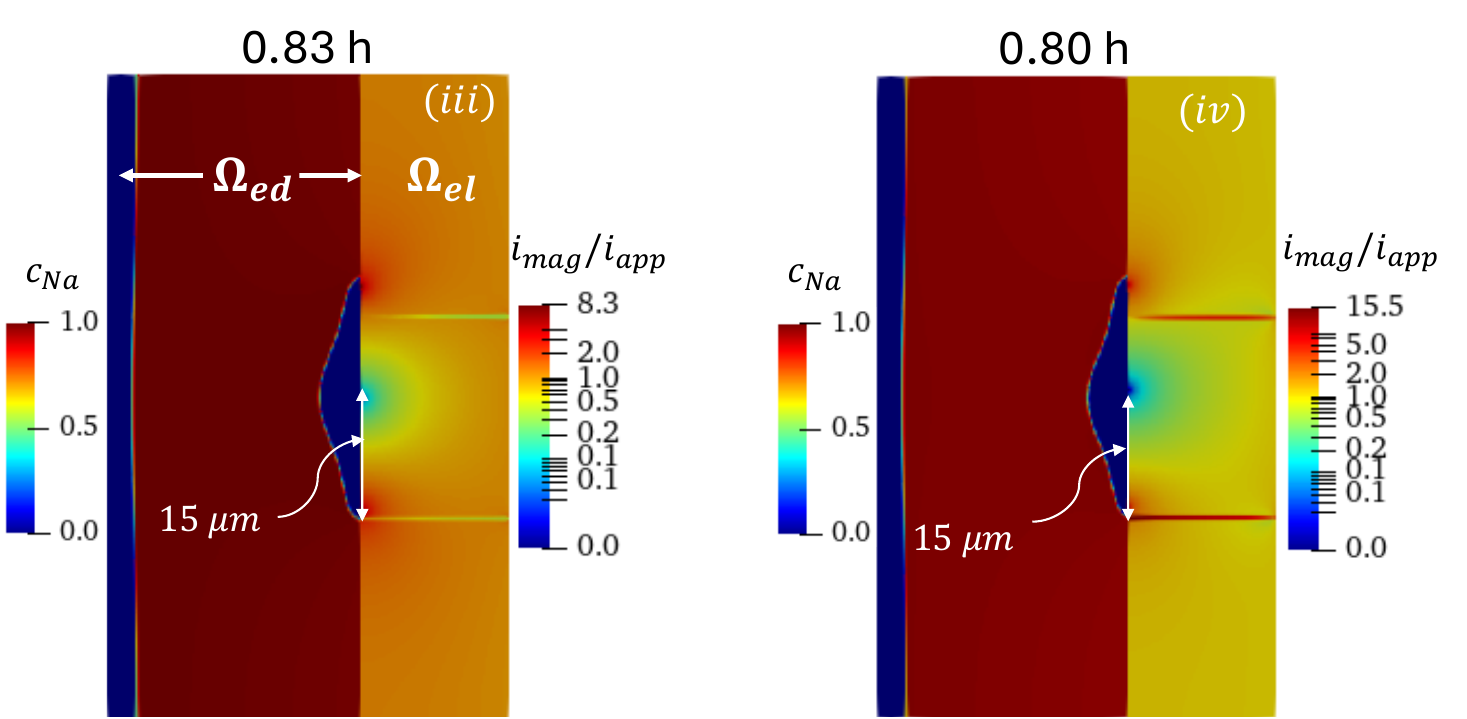}
	\subcaption{}
	\label{Fig1d_SI_secS7}
\end{subfigure}
\caption{Stripping simulation with a single interfacial void and bicrystal and tricrystal SEs under an applied current density of $0.2$ mA/cm$^2$. (a) Half-void size from the void center to the upper void edge for bicrystal and tricrystal polycrystalline SEs with low-conductivity ($\kappa_{gb}^{el}/\kappa_{g}^{el} = 0.067$) and high-conductivity ($\kappa_{gb}^{el}/\kappa_{g}^{el} = 10$), grain boundaries (GBs). The homogeneous SE case is indicated by $\kappa_{gb}^{el}/\kappa_{g}^{el} = 1$. The dotted black line indicates the SE GB at a distance of $10$ $\mu$m above the void center. (b) Percentage change in upper half-void size relative to the homogeneous SE as a function of time for bicrystal and tricrystal SEs with $\kappa_{gb}^{el}/\kappa_{g}^{el}=0.067$, $1$ and $10$. The black markers labelled $(i)$-$(iv)$ correspond to the microstructures shown in (c) and (d). Simulation domain when the void edge touches the SE GB in (c) a bicrystal SE, and (d) a tricrystal SE. $\Omega_{ed}$ is shaded by the Na concentration and $\Omega_{el}$ by the normalized current density.}
\end{figure}

To support our assertion that grain boundary (GB) conductivity locally impacts void migration, we present additional single void growth simulations in bicrystal and tricrystal solid electrolyte (SE) separators under an applied stripping current density of $0.2$ mA/cm$^2$. The initial void radius and Na negative electrode thickness are identical to the cases discussed in Section \ref{secR2.2}. Like Section \ref{secR2.2}, bicrystal and tricrystal SEs with low-conductivity and high-conductivity GBs are considered. In the case of the bicrystal SE, a diffuse horizontal GB is placed at a distance of $10$ $\mu$m above the void center (see Fig.~\ref{Fig1c_SI_secS7}). For the case of tricrystal SE, a horizontal GB is at a distance of $10$ $\mu$m above the void centre, and another at 15 $\mu$m below the void centre (Fig.~\ref{Fig1d_SI_secS7}). The GB-to-grain conductivity ratios in the high-conductivity and low-conductivity SE separators are identical to the polycrystalline SE cases discussed in Section \ref{secR2.2}.

Figure \ref{Fig1a_SI_secS7} shows the void size along the $+y$-axis as a function of stripping time in the case of bicrystal and tricrystal SEs with low-conductivity and high-conductivity GBs. The case with homogeneous SE (green curve) is also shown in this figure. As expected, the void edge grows along the electrode/electrolyte interface during stripping in all cases. However, the rate at which the void edge grows depends slightly on the SE GB conductivities. For instance, after stripping for $0.4$ h, the percentage change is less than $6\%$ and positive in the case of bicrystal and tricrystal SEs with high-conductivity GBs (see Fig. \ref{Fig1b_SI_secS7}).
In contrast, the percentage change is less than $2\%$ and negative for bicrystal and tricrystal SEs with low-conductivity GBs. As discussed in Section \ref{secR2.2}, a positive or negative percentage change implies a faster or slower void migration relative to the homogeneous SE case. Again, these deviations are due to the interaction between the void edge and the SE grain boundaries, as shown in Figs.~\ref{Fig1c_SI_secS7} and \ref{Fig1d_SI_secS7}. These figures show the instances when the void edge touches the bicrystal and tricrystal SE GBs, respectively.

\subsection{Section S8}
\label{SI_secS8}
\begin{figure}[tpb]
\centering
\begin{subfigure}{0.49\textwidth}
	\includegraphics[trim=0 0 0 0, clip, keepaspectratio,width=\linewidth]{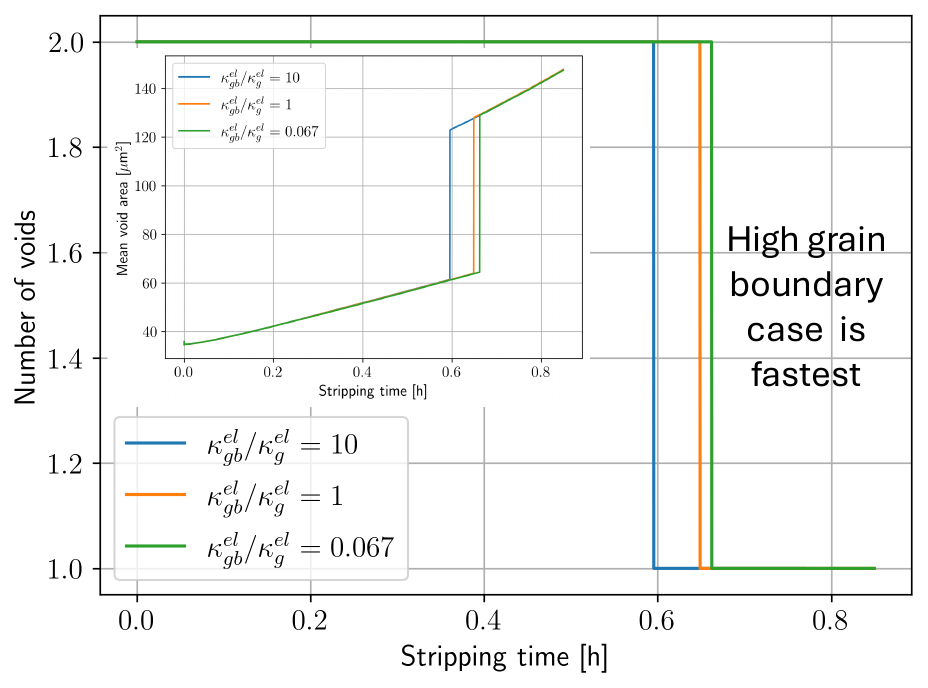}
	\subcaption{}
	\label{Fig1a_SI_secS8}
\end{subfigure}
\begin{subfigure}{0.49\textwidth}
	\includegraphics[trim=0 0 0 0, clip, keepaspectratio,width=\linewidth]{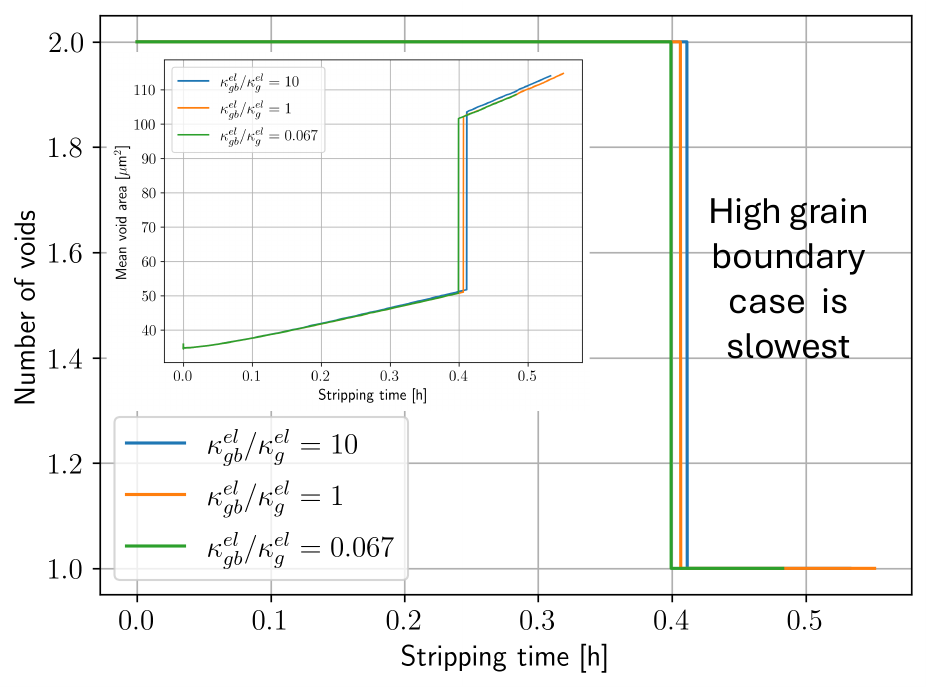}
	\subcaption{}
	\label{Fig1b_SI_secS8}
\end{subfigure}
\begin{subfigure}{0.49\textwidth}
	\includegraphics[trim=0 0 0 0, clip, keepaspectratio,width=\linewidth]{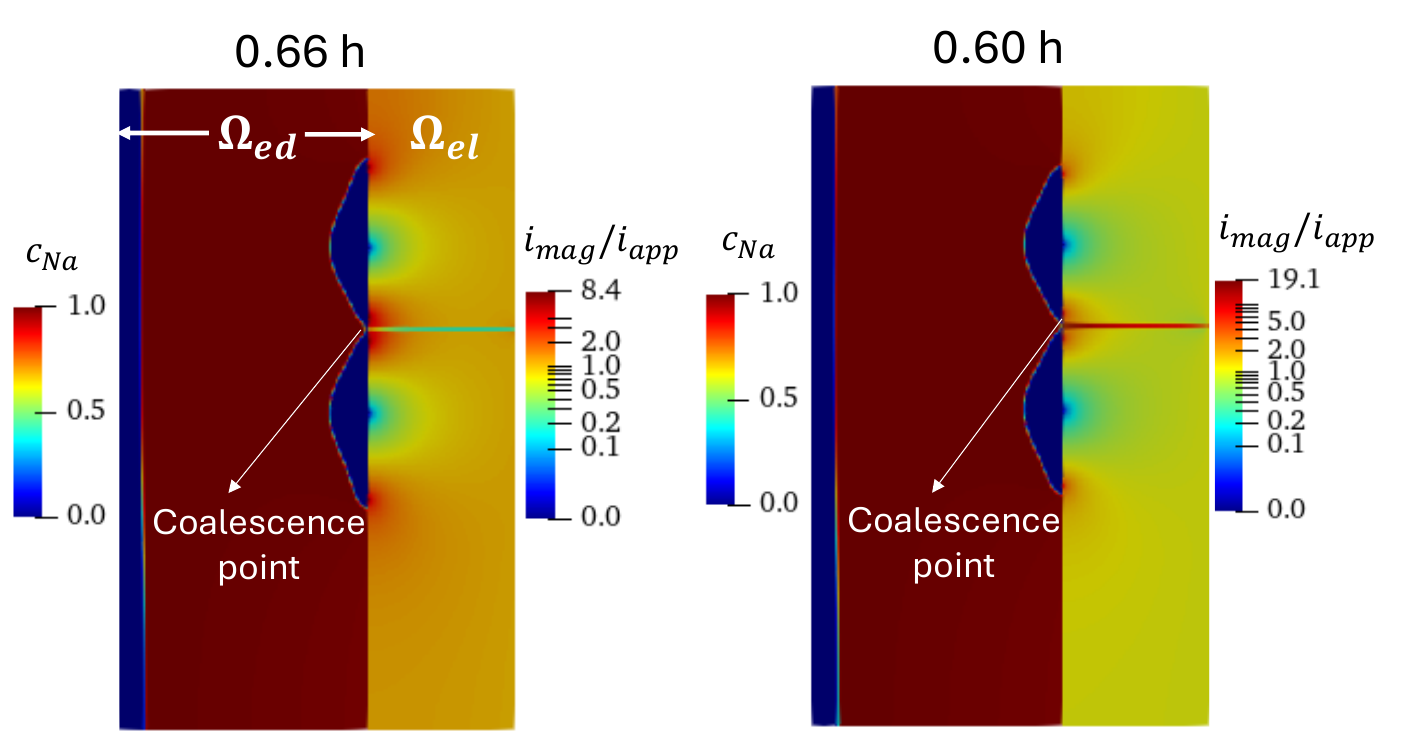}
	\subcaption{}
	\label{Fig1c_SI_secS8}
\end{subfigure}
\begin{subfigure}{0.49\textwidth}
	\includegraphics[trim=0 0 0 0, clip, keepaspectratio,width=\linewidth]{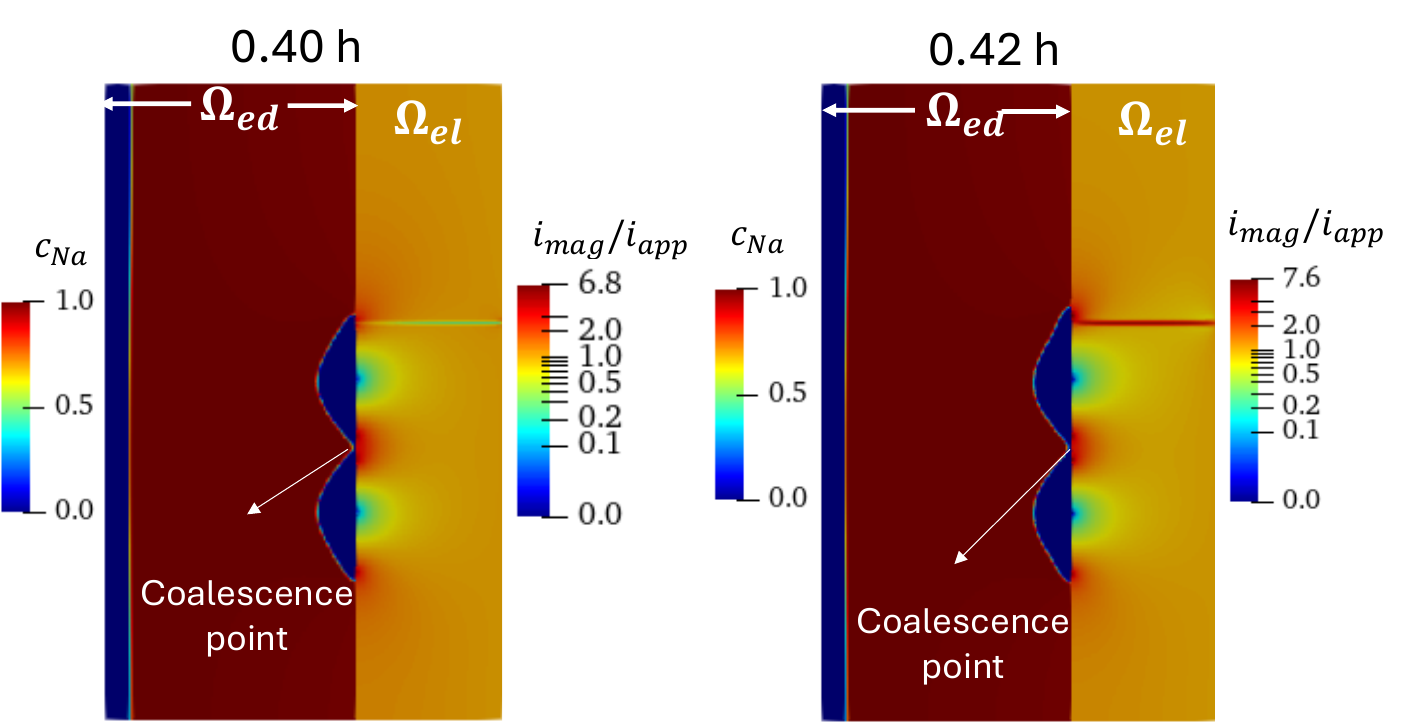}
	\subcaption{}
	\label{Fig1d_SI_secS8}
\end{subfigure}
\caption{Stripping simulations with two interfacial voids and a bicrystal solid electrolyte (SE) separator with low-conductivity and high-conductivity grain boundaries (GBs) under an applied current density of $0.2$ mA/cm$^2$. Number of voids as a function of stripping time when (a) the SE GB is near the void coalescence point and (b) when the SE GB is away from the void coalescence point. The insets show the mean void area with time for low-conductivity ($\kappa_{gb}^{el}/\kappa_{g}^{el}=0.067$), homogeneous ($\kappa_{gb}^{el}/\kappa_{g}^{el}=1$), and high-conductivity ($\kappa_{gb}^{el}/\kappa_{g}^{el}=10$) GBs in both cases. Simulation domain when the voids coalesce in (c) near and (d) away from the bicrystal SE GB. In (c) and (d), $\Omega_{ed}$ is shaded by the Na concentration and $\Omega_{el}$ by the ratio of the local current density to the applied current density.}
\end{figure}

To support our assertion that SE GB conductivity marginally affects void coalescence, we present additional multi-void simulations using bicrystal SEs with low-conductivity and high-conductivity GBs under an applied stripping current density of $0.2$ mA/cm$^2$. For the sake of simplicity, we consider only two interfacial voids located at the electrode/electrolyte interface. Initially, voids are semi-circular, each with a radius of $4.8$ $\mu$m. Two scenarios are simulated. First, the voids coalesce near a SE GB (Fig.~\ref{Fig1c_SI_secS7}); and in the second, the voids coalesce at a distance away from the SE GB (Fig.~\ref{Fig1d_SI_secS7}). Note that the intervoid spacings are different in the two scenarios. The void centers are $20$ $\mu$m apart in the first scenario, while they are $15.8$ $\mu$m apart in the second scenario. Consequently, it takes a slightly longer time for voids to coalesce in the former than in the latter.

Figure \ref{Fig1a_SI_secS8} shows the evolution of the number of voids as a function of stripping time when the SE GB is located near the void coalescence point.  The inset of Fig.~\ref{Fig1a_SI_secS8} shows the mean void area as a function of stripping time for low-conductivity, homogeneous and high-conductivity GB cases. As in Section \ref{secR3.2}, voids grow and eventually coalesce during stripping in all cases. However, the coalescence rate is marginally faster in the case of SE with high-conductivity GBs than in the other two cases (see Fig.~\ref{Fig1a_SI_secS8}). This is because, as discussed in Section \ref{secR2.2}, the void migration becomes marginally faster due to the presence of high-conductivity SE GBs. Figure \ref{Fig1c_SI_secS8} shows the instances when the voids coalesce near the low-conductivity and high-conductivity SE GBs.

Figure \ref{Fig1b_SI_secS8} shows the evolution of the number of voids as a function of stripping time when the SE GB is located away from the void coalescence point. The inset of Fig.~\ref{Fig1b_SI_secS8} shows the mean void area as a function of stripping time for low-conductivity, homogeneous and high-conductivity GB cases. Again, the voids grow and eventually coalesce in all cases. However, the coalescence rate is marginally faster in the case of SE with low-conductivity GBs than in the other two cases. This could be due to slightly faster void migration along the grains in the case of SEs with low-conductive GBs than in the other two cases. Figure \ref{Fig1d_SI_secS8} shows the instances when the voids coalesce away from the low-conductivity and high-conductivity SE GBs. These results indicate that the coalescence rate depends on the position of the SE GB from the void coalescence point.

\end{document}